\documentclass[a4paper,fleqn,dvipsnames,web,table]{cas-dc}

\usepackage{amsmath,amssymb,amsfonts}
\usepackage{framed,multirow}
\usepackage{xcolor}
\usepackage{mathrsfs}
\usepackage{bm}
\usepackage{tikz}
\usetikzlibrary{arrows.meta, spy, calc, arrows}
\usepackage{adjustbox}
\usepackage{floatrow}
\usepackage{array}
\usepackage{pgfplots}
\usepackage{tkz-kiviat} 
\usepackage{caption}
\usepackage[square,numbers]{natbib}
\usepackage{hyperref}
\usepackage{hhline}

\pgfplotsset{compat=1.8}
\usepgfplotslibrary{statistics}

\newcolumntype{P}[1]{>{\centering\arraybackslash}p{#1}}

\DeclareMathOperator*{\argmin}{\arg\min} 
\newcommand{\norm}[1]{\left\lVert#1\right\rVert}

\begin{document}
\shorttitle{Multi-Structure Bone Segmentation in Pediatric MR Images}
\shortauthors{A. Boutillon et~al.}

\title[mode=title]{Multi-Structure Bone Segmentation in Pediatric MR Images with Combined Regularization from Shape Priors and Adversarial Network}

\author[1,2]{Arnaud Boutillon}[orcid=0000-0001-5855-2770]
\cormark[1]

\author[2,3,4]{Bhushan Borotikar}[orcid=0000-0002-3404-6547]

\author[1,2]{Valérie Burdin}[orcid=0000-0001-6012-9883]

\author[1,2]{Pierre-Henri Conze}[orcid=0000-0003-2214-3654]

\address[1]{IMT Atlantique, Brest, France}
\address[2]{LaTIM UMR 1101, Inserm, Brest, France}
\address[3]{Symbiosis Center for Medical Image Analysis, Symbiosis International University, Pune, India}
\address[4]{Centre Hospitalier Régional et Universitaire (CHRU) de Brest, Brest, France}

\cortext[cor1]{Corresponding author: \href{mailto:arnaud.boutillon@imt-atlantique.fr}{\nolinkurl{arnaud.boutillon@imt-atlantique.fr}} (A. Boutillon)}

\begin{abstract}
Morphological and diagnostic evaluation of pediatric musculoskeletal system is crucial in clinical practice. However, most segmentation models do not perform well on scarce pediatric imaging data. We propose a new pre-trained regularized convolutional encoder-decoder network for the challenging task of segmenting heterogeneous pediatric magnetic resonance (MR) images. To this end, we have conceived a novel optimization scheme for the segmentation network which comprises additional regularization terms to the loss function. In order to obtain globally consistent predictions, we incorporate a shape priors based regularization, derived from a non-linear shape representation learnt by an auto-encoder. Additionally, an adversarial regularization computed by a discriminator is integrated to encourage precise delineations. The proposed method is evaluated for the task of multi-bone segmentation on two scarce pediatric imaging datasets from ankle and shoulder joints, comprising pathological as well as healthy examinations. The proposed method performed either better or at par with previously proposed approaches for Dice, sensitivity, specificity, maximum symmetric surface distance, average symmetric surface distance, and relative absolute volume difference metrics. We illustrate that the proposed approach can be easily integrated into various bone segmentation strategies and can improve the prediction accuracy of models pre-trained on large non-medical images databases. The obtained results bring new perspectives for the management of pediatric musculoskeletal disorders.
\end{abstract}


\maketitle
\section{Introduction}

Semantic segmentation is a crucial step in many medical imaging workflows which aims at identifying and localizing  meaningful anatomical structures by extracting their boundaries. For musculoskeletal system analysis, image segmentation is mainly employed to generate three dimensional (3D) models of bones and muscles, which in turn help clinicians to diagnose pathologies, plan patient-specific therapeutic interventions or study how the morphology evolves in time \cite{hirschmann_artificial_2019}. Knowledge of anatomy in the pediatric pathological population is all the more crucial since the management of pediatric musculoskeletal disorders involves accurate understanding of morphological deformity and associated joint function \cite{balassy_role_2008}. However, segmentation in pediatric MR images is typically performed manually, which is tedious, time-consuming, and suffers from intra- and inter-observer variability. Indeed, the pediatric musculoskeletal system may be composed of thin structures which are more challenging to delineate than their adult counterparts \cite{jaramillo_pediatric_2008}. Moreover, due to the ongoing bone ossification process, non-ossified areas have to be managed along with completely ossified regions \cite{balassy_role_2008, jaramillo_pediatric_2008}. Hence, to reduce the time and increase the reliability of morphological assessment, employing robust and fully-automated segmentation techniques becomes a necessity. 

The development of automatic segmentation strategies faces numerous challenges including the scarcity of medical imaging datasets whose conception is a slow and onerous process \cite{kohli_medical_2017}. Due to limited imaging resources, it is challenging to develop generalizable tools which could be integrated into clinical practice while achieving reliable delineations on unseen images \cite{hirschmann_artificial_2019}. In recent years, deep learning has achieved promising results for natural image processing compared to traditional variational, model-based, or graph-partitioning learning schemes \cite{krizhevsky_imagenet_2012}. Consequently, the medical imaging community has adopted this technique as a way to enhance performance and generalization capabilities, without relying on hand-crafted features \cite{litjens_survey_2017}.

Recent works aim at incorporating regularization into deep learning-based segmentation models to further avoid over-fitting and improve generalizability \cite{goodfellow_deep_2016, ioffe_batch_2015, kukacka_regularization_2017, srivastava_dropout_2014}. Regularization schemes can arise from different prior information such as boundaries \cite{chen_dcan_2017}, shape models \cite{josephson_segmentation_2005}, atlas models \cite{gauriau_multi-organ_2015} or topology. Exploiting prior knowledge is found to be effective in achieving more precise and consistent results for traditional medical segmentation applications \cite{nosrati_incorporating_2016}. Regularization techniques can alleviate the presence of image artefacts that are inherently embedded in an image during acquisition \cite{nosrati_incorporating_2016}. For their part, pediatric pathological imaging examinations also exhibit irregular and complex pathological structures which are difficult to delineate due to alterations in shape and appearance \cite{balassy_role_2008, jaramillo_pediatric_2008}. In this context, regularization appears as a key strategy to enhance segmentation outcomes and model's generalization abilities when targeting scarce and heterogeneous pediatric imaging datasets.

\subsection{Clinical motivations}

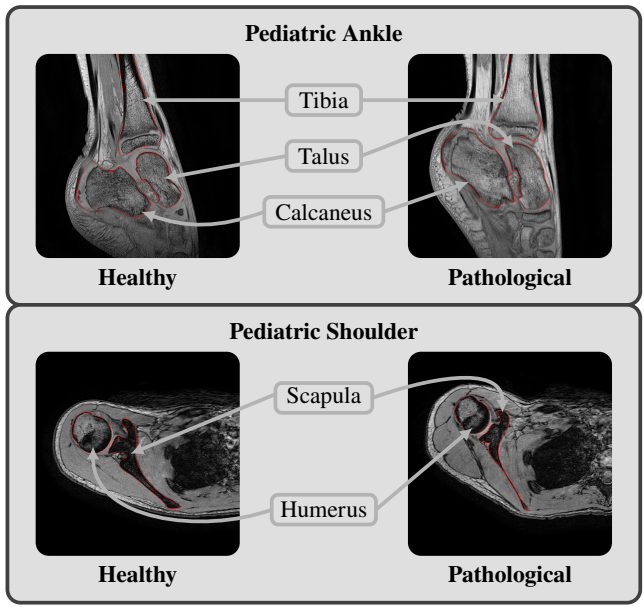
\begin {figure}[t]
\vspace{-0.25cm} 
\centering
\begin{adjustbox}{width=\textwidth}
\begin{tikzpicture}

\draw[line width=0.1mm, color=darkgray, rounded corners=1, fill=white!50!lightgray] (-.35,-.4) -- (1.35,-.4) -- (1.35,.4) -- (-.35, .4) -- cycle;
\draw[line width=0.1mm, color=darkgray, rounded corners=1, fill=white!50!lightgray] (-.35, -.4) -- (1.35,-.4) -- (1.35,-1.2) -- (-.35,-1.2)   -- cycle;

\node[inner sep=0pt] (mri_a_h) at (0,0)
    {\includegraphics[width=.068\textwidth]{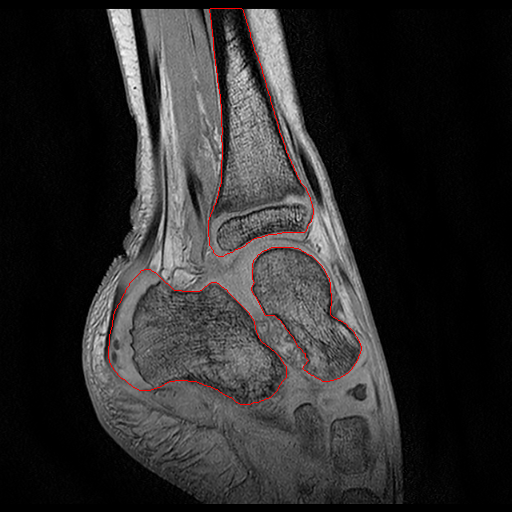}};
\draw[white!50!lightgray, rounded corners=1.5, line width=1] (mri_a_h.north west) -- (mri_a_h.north east) -- (mri_a_h.south east) -- (mri_a_h.south west) -- cycle;
\node[inner sep=0pt] (mri_a_p)  at (1,0)
    {\includegraphics[width=.068\textwidth]{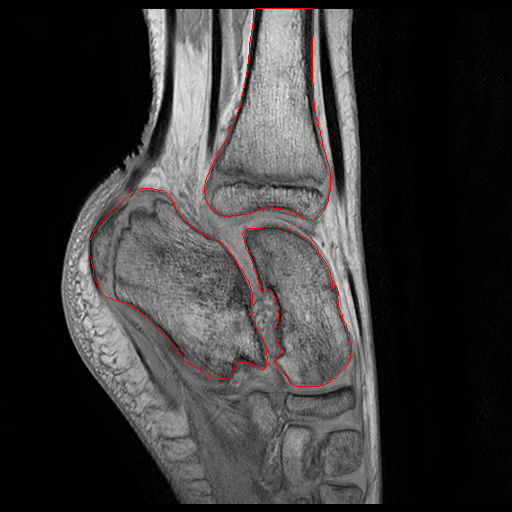}};
\draw[white!50!lightgray, rounded corners=1.5, line width=1] (mri_a_p.north west) -- (mri_a_p.north east) -- (mri_a_p.south east) -- (mri_a_p.south west) -- cycle;

\node[inner sep=0pt] (mri_s_h) at (0,-.8)
    {\includegraphics[width=.068\textwidth]{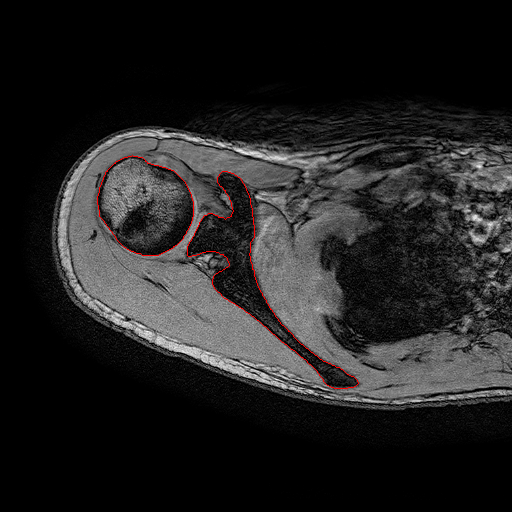}};
\draw[white!50!lightgray, rounded corners=1.5, line width=1] (mri_s_h.north west) -- (mri_s_h.north east) -- (mri_s_h.south east) -- (mri_s_h.south west) -- cycle;
\node[inner sep=0pt] (mri_s_p) at (1,-.8) {\includegraphics[width=.068\textwidth]{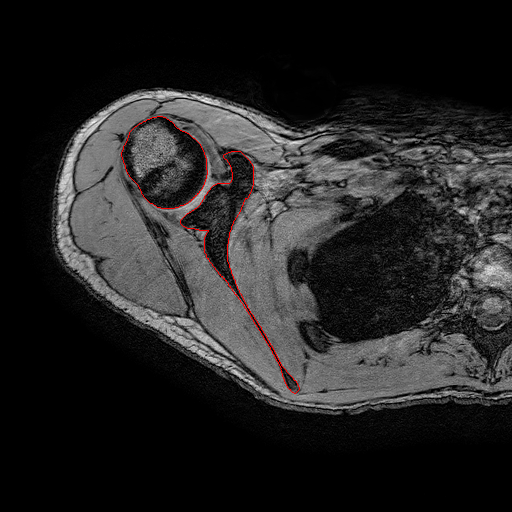}};
\draw[white!50!lightgray, rounded corners=1.5, line width=1] (mri_s_p.north west) -- (mri_s_p.north east) -- (mri_s_p.south east) -- (mri_s_p.south west) -- cycle;

\node at (.5, .33)  {\fontsize{1.8}{1.8}\selectfont \textbf{Pediatric Ankle}};
\node at (.5, .15) {\fontsize{1.8}{1.8}\selectfont Tibia};
\node at (.5, 0) {\fontsize{1.8}{1.8}\selectfont Talus};
\node at (.5, -.15) {\fontsize{1.8}{1.8}\selectfont Calcaneus};

\node at (.5, -.47) {\fontsize{1.8}{1.8}\selectfont \textbf{Pediatric Shoulder}};
\node at (.5, -.65) {\fontsize{1.8}{1.8}\selectfont Scapula};
\node at (.5, -.95) {\fontsize{1.8}{1.8}\selectfont Humerus};

\node[scale=0.3] at (0, -1.13) {\fontsize{6}{6}\selectfont \textbf{Healthy}};
\node[scale=0.3] at (1, -1.13) {\fontsize{6}{6}\selectfont \textbf{Pathological}};

\node[scale=0.3] at (0, -.33) {\fontsize{6}{6}\selectfont \textbf{Healthy}};
\node[scale=0.3] at (1, -.33) {\fontsize{6}{6}\selectfont \textbf{Pathological}};

\draw[line width=0.1mm, color=white!40!gray, -{Latex[length=1pt, width=1pt]}] (.4,.15) -- (.01,.15);
\draw[line width=0.1mm, color=white!40!gray, -{Latex[length=1pt, width=1pt]}] (.6,.15) -- (0.99,.15);

\draw[line width=0.1mm, color=white!40!gray, -{Latex[length=1pt, width=1pt]}] (.4,0) -- (.07,-.06);
\draw[line width=0.1mm, color=white!40!gray, -{Latex[length=1pt, width=1pt]}] (.6,0) to[out=20,in=125] (1.01,0.02);

\draw[line width=0.1mm, color=white!40!gray, -{Latex[length=1pt, width=1pt]}] (.34,-.15) to[out=200,in=340] (.01,-.15);
\draw[line width=0.1mm, color=white!40!gray, -{Latex[length=1pt, width=1pt]}] (.66,-.15) -- (.89,-.07);

\draw[line width=0.1mm, color=white!40!gray, -{Latex[length=1pt, width=1pt]}] (.37,-.64) -- (-.02,-.8);
\draw[line width=0.1mm, color=white!40!gray, -{Latex[length=1pt, width=1pt]}] (.63,-.65) to[out=20,in=90] (.98,-0.69);

\draw[line width=0.1mm, color=white!40!gray, -{Latex[length=1pt, width=1pt]}] (.36,-.95) to[out=200,in=270] (-.12,-.78);
\draw[line width=0.1mm, color=white!40!gray, -{Latex[length=1pt, width=1pt]}] (.64,-.95) -- (.90,-.73);

\draw[line width=0.1mm, color=white!40!gray, rounded corners=.5] (.4,.11) -- (.6,.11) -- (.6,.19) -- (.4,.19) -- cycle;
\draw[line width=0.1mm, color=white!40!gray, rounded corners=.5] (.4,-.04) -- (.6,-.04) -- (.6,.04) -- (.4,.04) -- cycle;
\draw[line width=0.1mm, color=white!40!gray, rounded corners=.5] (.34,-.19) -- (.66,-.19) -- (.66,.-.11) -- (.34,-.11) -- cycle;

\draw[line width=0.1mm, color=white!40!gray, rounded corners=.5] (.37,-.69) -- (.63,-.69) -- (.63,.-.61) -- (.37,-.61) -- cycle;
\draw[line width=0.1mm, color=white!40!gray, rounded corners=.5] (.36,-.99) -- (.64,-.99) -- (.64,-.91) -- (.36,-.91) -- cycle;

\end{tikzpicture}
\end{adjustbox}

\caption{MR image samples from pediatric ankle and shoulder datasets comprising healthy and pathological examinations. Anatomical structures of interest consist of calcaneus, talus and tibia in ankle joint and humerus and scapula in shoulder joint. Ground truth delineations are in red (\textcolor{red}{\raisebox{1.8pt}{\rule{5pt}{1pt}}}).}
\label{fig:introduction}
\end{figure}

The musculoskeletal system consists of different tissue types (bone, cartilage, muscle, tendon, etc.) and multiple anatomical structures that form several articulating joints (ankle, shoulder, knee, elbow, etc.). As shown in Figure \ref{fig:introduction}, our work focuses on the segmentation of multiple bones in two pediatric musculoskeletal joints: ankle and shoulder. The structures of interest include the tibia, talus, and calcaneus bones for ankle joint and humerus and scapular bones for shoulder joint. Both imaging datasets manifest a high level of heterogeneity due to the presence of different age groups and a mixture of healthy and pathological examinations. In particular, we investigate two pediatric musculoskeletal disorders: ankle equinus condition in children with cerebral palsy and obstetrical brachial plexus palsy (OBPP) nerve injury leading to shoulder deformities.

Equinus is a clinical condition that affects the ankle joint's function by restricting its range of motion \cite{deheer_equinus_2017}. However, the etiology of this condition is poorly understood and numerous causes have been discussed such as muscle spasticity due to cerebral palsy, bone block in ankle joint, tendon or calf muscle stiffness \cite{charles_static_2010}. Equinus deformity is characterized by reduced dorsiflexion of the joint and can also be associated with increased forefoot loading, aggravated risk of ankle sprain, and more importantly bone deformity due to abnormal bone growth \cite{charles_static_2010}. Understanding morphological modifications occurring in the joint is imperative to reduce complications associated with equinus.

Obstetrical brachial plexus palsy is a common birth injury associated with difficult or assisted delivery during which the peripheral nervous system is disrupted \cite{zafeiriou_obstetrical_2008}. This nerve injury, which occurs around 1.4 per 1000 births \cite{chauhan_neonatal_2014}, results in shoulder muscle atrophy, impedes bone growth, and leads to osseous deformity. More precisely, OBPP is associated with delayed ossification and malformed bones including hypo-plastic humeral head, non-spherical humeral head, hypoplastic scapula, elevated scapula, and abnormal scapula glenoid \cite{hoeksma_shoulder_2003}. Modifications in muscle and bone morphology lead to shoulder strength imbalance and reduction in joint range of motion \cite{pons_shoulder_2017}. Therefore, patient-specific information related to the degree of bone deformity is key for treatment planning and follow-up.

For both these pathologies, fully-automated and reliable bone segmentation of pediatric examinations could provide a rapid evaluation of the patient's level of impairment, guide surgery, and help optimize rehabilitation programs. Furthermore, patient-specific 3D bone models could also assist clinicians to analyse strength imbalance as well as the kinematics and dynamics of pathological joints.

\subsection{Technical challenges and related works}

Recent advances in medical image processing are linked to the development of deep learning based techniques which, contrary to traditional methods, are aimed at learning hierarchical feature representations in a purely data-driven manner. More specifically, convolutional neural networks (CNN) have proven to outperform other state-of-the-art methods in numerous medical imaging applications such as classification, detection, registration, and segmentation \cite{litjens_survey_2017}. Many neural network architectures have been proposed for medical image semantic segmentation \cite{kamnitsas_deepmedic_2016, long_fully_2015}. The most commonly used methods are based on UNet \cite{ronneberger_u-net_2015} and its 3D counterpart VNet \cite{milletari_v-net_2016}, with impressive performances compared to other CNN architectures. Recently, numerous UNet extensions have been proposed, a typical example being the Attention UNet (Att-UNet) \cite{oktay_attention_2018} which embeds attention gates to focus on salient features. One can also mention more complex architectures incorporating dense, Inception, residual or more recently Transformer \cite{zhang_dense-inception_2020, valanarasu_medical_2021, cheng_resganet_2022} modules to provide more efficient optimization and enhanced performance. Additionally, networks integrating encoders (e.g. \texttt{VGG19} \cite{conze_abdominal_2021}, \texttt{ResNet34} \cite{kalinin_medical_2020}) pre-trained on ImageNet \cite{russakovsky_imagenet_2015} leverage low-level features typically shared between different image types. More specifically, transfer learning and fine tuning from large non-medical datasets has become a widespread method in medical image analysis and has revealed improved performance compared to models with randomly initialized weights \cite{conze_abdominal_2021, kalinin_medical_2020, conze_healthy_2020, raghu_transfusion_2019}. However, the results of transfer learning depend on the task and dataset characteristics, with larger impact in very small data regimes \cite{raghu_transfusion_2019}. Hence, employing pre-trained models appears essential to address the data scarcity issue encountered in pediatric imaging.

UNet and VNet architectures have already been applied for segmentation of musculoskeletal structures in MR images, including adult knee bones, muscles and cartilage \cite{ambellan_automated_2019, zhou_deep_2018}, adult shoulder bones \cite{he_effective_2019} as well as pediatric shoulder muscles \cite{conze_healthy_2020}, pediatric ankle and shoulder bones \cite{boutillon_combining_2020, boutillon_multi-structure_2021, boutillon_multi-task_2021}. However, works addressing pediatric bone segmentation remain scarce in the literature. Furthermore, two bone segmentation strategies emerge in the literature: in the first one, a single network predicts all segmentation classes \cite{ambellan_automated_2019, zhou_deep_2018, he_effective_2019} whereas, in the second one, specific networks are trained for each object of interest \cite{conze_healthy_2020}. Finally, post-processing based on conditional random field \cite{dou_3d_2017}, deformable models \cite{zhou_deep_2018} or statistical shape models \cite{ambellan_automated_2019} have been developed to constrain the predicted shapes. Nevertheless, these methods fail to regularize and incorporate shape information directly into the segmentation network.

In deep learning, the regularization concept covers techniques that can affect the network architecture, the training data or the loss function \cite{kukacka_regularization_2017}. The UNet architecture already contains regularization in the form of convolutional layers which enforce local and translation-equivariant hidden units, pooling-layers which impose translation invariant feature extraction, and skip-connections which assume a correlation between low-level and high-level features \cite{goodfellow_deep_2016}. Moreover, data augmentation and batch normalization are two data-based regularization techniques which are commonly incorporated into UNet models. Data augmentation incites the network to learn invariance, covariance and robustness properties \cite{goodfellow_deep_2016} while the randomization inherent in batch normalization enforces robust data representations \cite{ioffe_batch_2015}. Even though studies dedicated to the loss function-based regularization are rare, defining a suitable loss function for training deep learning based models can lead to improved performance. In this context, two loss function-based regularization methodologies have shown promising results: shape priors based regularization \cite{ dalca_anatomical_2018, myronenko_3d_2019, oktay_anatomically_2018, pham_deep_2019, ravishankar_learning_2017} and adversarial regularization \cite{conze_abdominal_2021, nie_adversarial_2020, singh_breast_2020, xue_segan_2018}. 

Incorporating shape information into medical imaging segmentation algorithms has already proven to be useful in reducing the effect of noise, low contrast, and artefacts \cite{oktay_anatomically_2018}. Recent contributions have proposed to learn a representation of the anatomy directly from ground truth annotations using a deep auto-encoder \cite{dalca_anatomical_2018, myronenko_3d_2019, oktay_anatomically_2018, pham_deep_2019, ravishankar_learning_2017}. Due to the constrained nature of anatomical structures, data-driven models are suitable for learning shape prior information. The learnt non-linear shape representation can then be integrated in the segmentation network during optimization, thanks to a specifically designed regularization term which enforces the predicted segmentation to be close to the ground truth in shape space \cite{oktay_anatomically_2018, ravishankar_learning_2017}. Consequently, such regularization encourages globally consistent shape predictions.

Inspired by image-to-image translation approaches \cite{isola_image--image_2017}, medical imaging researchers have also employed adversarial networks to refine segmentation outputs. In these frameworks, a segmentation network and a discriminator are concurrently trained in a two-player game fashion in which the former learns to produce valid segmentation while the latter learns to discriminate between synthetic and real data \cite{conze_abdominal_2021, nie_adversarial_2020, singh_breast_2020, xue_segan_2018}. An adversarial term computed by the discriminator is added during the segmentation network optimization, which in turn, encourages UNet to fool the discriminator, and produces more plausible segmentation masks. 

\subsection{Contributions}

\begin{figure}[t]
\centering
\begin{adjustbox}{width=\textwidth}
\tikzstyle{dashed}=[dash pattern=on .9pt off .9pt]
\begin{tikzpicture}

\node[inner sep=0pt] (mri) at (0,0)
    {\includegraphics[width=.0525\textwidth]{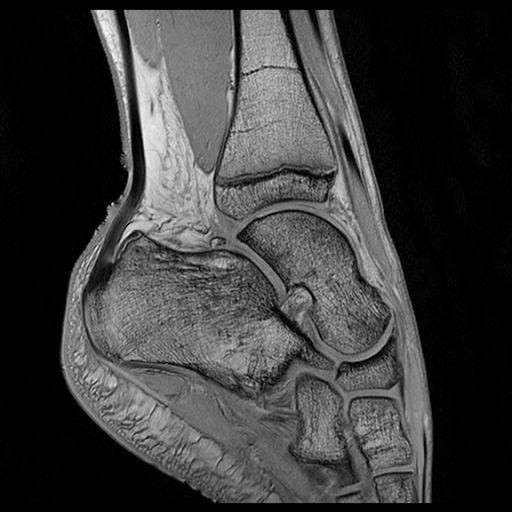}};
\node[inner sep=0pt] (pred) at (1.3,0)
    {\includegraphics[width=.0525\textwidth]{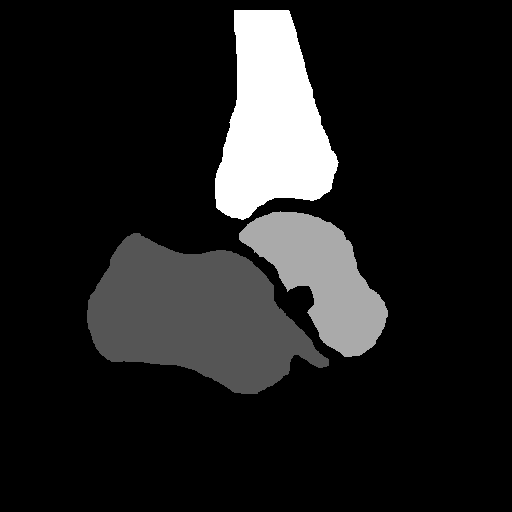}};
\node[inner sep=0pt] (gt) at (1.3,-.9)
    {\includegraphics[width=.0525\textwidth]{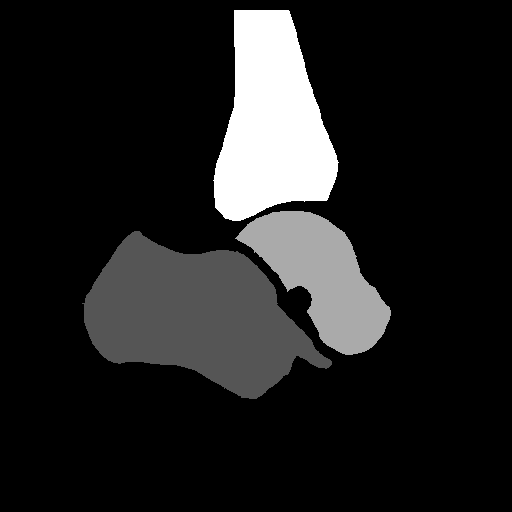}};

\draw[line width=0.01mm, fill=cyan!30] (.35,.25) -- (.65,.125) -- (.95,.25) -- (.95,-.25) -- (.65,-.125) -- (.35,-.25) -- cycle;
\node at (.65,.04) {\fontsize{2.3}{2.3}\selectfont Res-UNet};
\node at (.65,-0.04) {\fontsize{2.3}{2.3}\selectfont $\bm{S}$};

\draw[line width=0.01mm, fill=orange!40] (1.65,.25) -- (1.95,.125) -- (1.95,-.125) -- (1.65,-.25) -- cycle;
\draw[line width=0.01mm, fill=orange!40] (1.65,-.65) -- (1.95,-.775) -- (1.95,-1.025) -- (1.65,-1.15) -- cycle;
\node at (1.8,.08) {\scalebox{.21}{Shape}};
\node at (1.8,0) {\scalebox{.21}{encoder}};
\node at (1.8,-.08) {\scalebox{.21}{$\bm{F}$}};
\node at (1.8, -.82) {\scalebox{.21}{Shape}};
\node at (1.8,-.9) {\scalebox{.21}{encoder}};
\node at (1.8,-.98) {\scalebox{.21}{$\bm{F}$}};
\draw[line width=0.01mm, fill=red!40] (1.05,.38) -- (1.55,.38) -- (1.425,.68) -- (1.175,.68) -- cycle;
\node at (1.3, .53) {\scalebox{.17}{Discriminator}};
\node at (1.3, .60) {\scalebox{.17}{$\bm{D}$}};

\draw[line width=0.01mm] (1.975,-.4) rectangle (2.225,-.5) node[pos=.5] {\scalebox{.24}{{$\ell_{shape}$}}};
\draw[line width=0.01mm] (1.425,-.4) rectangle (1.175,-.5)  node[pos=.5] {\scalebox{.24}{{$\ell_{CE}$}}};
\draw[line width=0.01mm] (1.425,.68) rectangle (1.175,.78) node[pos=.5] {\scalebox{.24}{{$\ell_{adv}$}}};

\draw[line width=0.01mm, -{Latex[length=1.5pt, width=1.5pt]}] (.22, 0) -- (.35,0);
\draw[line width=0.01mm, -{Latex[length=1.5pt, width=1.5pt]}] (.95,0) -- (1.08,0);
\draw[line width=0.01mm, -{Latex[length=1.5pt, width=1.5pt]}] (1.52,0) -- (1.65,0);
\draw[line width=0.01mm] (1.3, -.22) -- (1.3,-.24);
\draw[line width=0.01mm, -{Latex[length=1.5pt, width=1.5pt]}] (1.3,-.32) -- (1.3,-.4);
\draw[line width=0.01mm, -{Latex[length=1.5pt, width=1.5pt]}] (1.19,-.68) -- (1.19,-.62) -- (1.3,-.62) -- (1.3,-.5);
\draw[line width=0.01mm, -{Latex[length=1.5pt, width=1.5pt]}] (1.52, -.9) -- (1.65,-.9);

\draw[line width=0.01mm] (1.95,0) -- (2.1,0);
\draw[line width=0.01mm, -{Latex[length=1.5pt, width=1.5pt]}] (2.1,0) -- (2.1, -.4);

\draw[line width=0.01mm] (1.95,-.9) -- (2.1, -.9);
\draw[line width=0.01mm, -{Latex[length=1.5pt, width=1.5pt]}] (2.1, -.9) -- (2.1,-.5);

\draw[line width=0.01mm, dashed, -{Latex[length=1.5pt, width=1.5pt]}] (2.225, -0.45) -- (2.275, -0.45) -- (2.275,-1.25) -- (.65,-1.25) -- (.65,-.525);

\draw[line width=0.01mm, dashed, -{Latex[length=1.5pt, width=1.5pt]}] (1.3,.78) -- (1.3,.83) -- (-.27,.83) -- (-.27,-.45) -- (.575,-.45);

\draw[line width=0.01mm, dashed, -{Latex[length=1.5pt, width=1.5pt]}] (1.175, -0.45) -- (.725, -0.45);

\draw[line width=0.01mm,dashed, -{Latex[length=1.5pt, width=1.5pt]}] (.65, -.375) -- (.65, -.125);

\draw[line width=0.01mm] (.65,-.45) circle (.075);
\draw[line width=0.01mm] (.65,-.4) -- (.65,-.5);
\draw[line width=0.01mm] (.6,-.45) -- (.7,-.45);

\draw[line width=0.01mm, -{Latex[length=1.5pt, width=1.5pt]}] (1.3,.22) -- (1.3,.38);
\draw[line width=0.01mm, -{Latex[length=1.5pt, width=1.5pt]}] (0,.22) -- (0,.28) -- (1.175,.28) -- (1.175,.38);
\draw[line width=0.01mm] (1.41,-.68) -- (1.41,-.62) -- (1.585,-.62) -- (1.585,-.04);
\draw[line width=0.01mm, -{Latex[length=1.5pt, width=1.5pt]}] (1.585,.04) -- (1.585,.28) -- (1.425,.28) -- (1.425,.38);
\draw[line width=0.01mm] (1.585,.04) arc (90:270:.04);

\node at (0,-.28) {\fontsize{2.5}{2.5}\selectfont MRI};
\node at (1.3,-.28) {\fontsize{2.5}{2.5}\selectfont Prediction};
\node at (1.3,-1.18) {\fontsize{2.5}{2.5}\selectfont Ground truth};

\draw[line width=0.01mm, -{Latex[length=1.5pt, width=1.5pt]}] (-.2,-.8) -- (0, -.8);
\draw[line width=0.01mm, dashed, -{Latex[length=1.5pt, width=1.5pt]}] (-.2,-1.05) -- (0, -1.05);

\node[anchor=west] at (-.1, -.75) {\fontsize{2.5}{2.5}\selectfont Forward };
\node[anchor=west] at (-.1, -.85) {\fontsize{2.5}{2.5}\selectfont propagation};
\node[anchor=west] at (-.1, -1) {\fontsize{2.5}{2.5}\selectfont Backward };
\node[anchor=west] at (-.1, -1.1) {\fontsize{2.5}{2.5}\selectfont propagation};

\node at (.72, -.29) {\scalebox{.24}{$\ell_{S}$}};
\node at (.72, -.65) {\scalebox{.24}{$\lambda_1$}};
\node at (.45, -.39) {\scalebox{.24}{$\lambda_2$}};

\end{tikzpicture}
\end{adjustbox}
  \caption{Proposed regularized segmentation network $\bm{S}$ based on Res-UNet exploiting cross-entropy loss $\ell_{CE}$, shape priors based $\ell_{shape}$ and adversarial $\ell_{adv}$ regularizations respectively computed by a shape encoder $\bm{F}$ with fixed weights and a discriminator $\bm{D}$ trained in competition with Res-UNet. The shape encoder corresponds to the encoder component of an auto-encoder previously optimized on ground truth segmentation masks, while the discriminator learns the plausibility of segmentation masks conditioned by their corresponding intensity image. $\lambda_1$ and $\lambda_2$ are two empirical weighting hyper-parameters.}
  \label{fig:framework}
\end{figure}
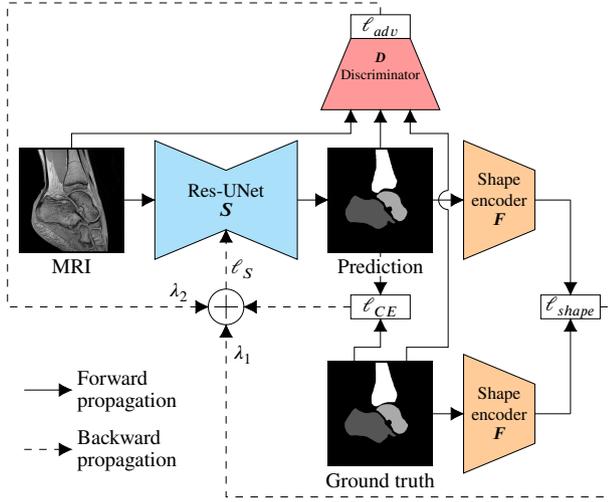

In this study, we propose a multi-structure bone segmentation framework based on a partially pre-trained deep learning architecture combining shape priors with adversarial regularizations (Figure \ref{fig:framework}). Unlike previous methods \cite{conze_abdominal_2021, myronenko_3d_2019, oktay_anatomically_2018, pham_deep_2019, ravishankar_learning_2017, nie_adversarial_2020, singh_breast_2020, xue_segan_2018}, our framework simultaneously leverages both the regularizations to guide the segmentation network to make anatomically consistent predictions and produce precise delineations. Furthermore, we demonstrate the usefulness of employing pre-trained models along with combining different regularization schemes for deep learning based medical image segmentation in order to alleviate data scarcity limitations. Thus, the contributions of this paper are three-fold:
\begin{enumerate}
    \item A deep learning multi-structure segmentation framework exploiting a combination of shape priors and an adversarial regularizer to reduce the data scarcity issue while improving model generalizability,
    \item An automatic and multi-bone pediatric segmentation method for scarce and heterogeneous MR images,
    \item An in-depth evaluation of the proposed method's performance extending \cite{boutillon_combining_2020} on two heterogeneous pediatric imaging datasets of the musculoskeletal system.
\end{enumerate}

\section{Method}

In this section, we explain the proposed segmentation network built upon Res-UNet and additional regularization terms incorporated into the loss function. We first briefly recall the partially pre-trained Res-UNet architecture and standard optimization procedure (Section \ref{sec:segmentaton_network_with_pre-trained_encoder}). We then incorporate shape priors based regularization into our model (Section \ref{sec:incorporating_shape_priors_based_regularization}). Finally, we combine regularization from shape priors with conditional adversarial network (Section \ref{sec:combining_shape_priors_based_regularization_with_adversarial_regularization}). 

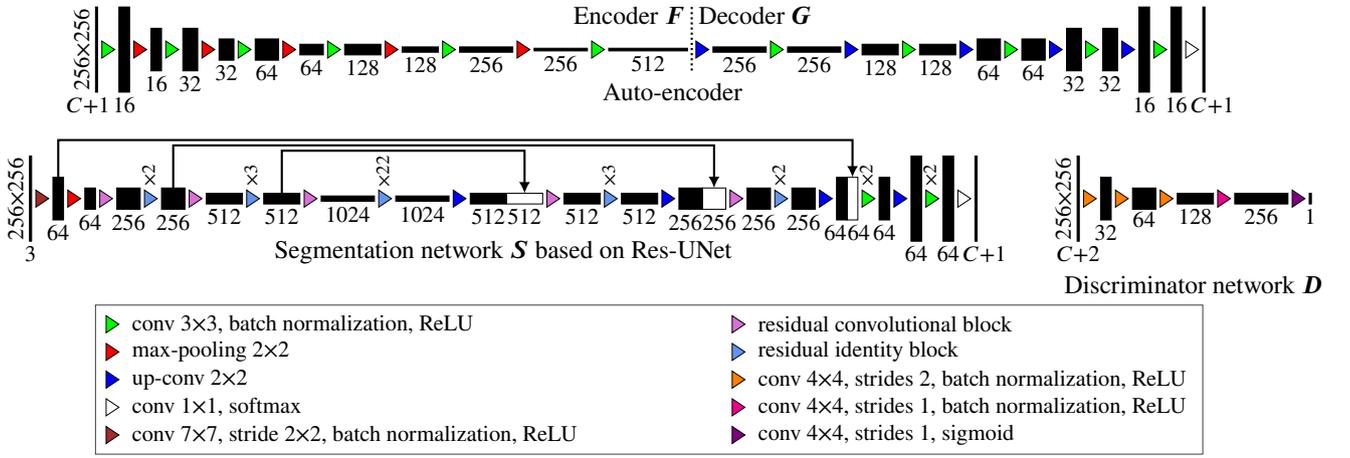
\begin{figure*}
\centering
\begin{adjustbox}{width=\textwidth}

\begin{tikzpicture}

\draw[line width=0.01mm, fill=black] (-.1,-2) rectangle (0,2);
\draw[line width=0.01mm, fill=black] (1,-1) rectangle (1.5,1);
\draw[line width=0.01mm, fill=black] (2.5,-.5) rectangle (3,.5);

\draw[line width=0.01mm, fill=black] (4,-.5) rectangle (5.1,.5);
\draw[line width=0.01mm, fill=black] (6.1,-.5) rectangle (7.2,.5);

\draw[line width=0.01mm, fill=black] (8.2,-.25) rectangle (9.9,.25);
\draw[line width=0.01mm, fill=black] (10.9,-.25) rectangle (12.6,.25);

\draw[line width=0.01mm, fill=black] (13.6,-.125) rectangle (16.1,.125);
\draw[line width=0.01mm, fill=black] (17.1,-.125) rectangle (19.6,.125);

\draw[line width=0.01mm, fill=black] (20.6,-.25) rectangle (22.3,.25);
\draw[line width=0.01mm] (22.3,-.25) rectangle (24,.25);
\draw[line width=0.01mm, fill=black] (25,-.25) rectangle (26.7,.25);
\draw[line width=0.01mm, fill=black] (27.7,-.25) rectangle (29.4,.25);

\draw[line width=0.01mm, fill=black] (30.4,-.5) rectangle (31.5,.5);
\draw[line width=0.01mm] (31.5,-.5) rectangle (32.6,.5);
\draw[line width=0.01mm, fill=black] (33.6,-.5) rectangle (34.7,.5);
\draw[line width=0.01mm, fill=black] (35.7,-.5) rectangle (36.8,.5);

\draw[line width=0.01mm, fill=black] (37.8,-1) rectangle (38.3,1);
\draw[line width=0.01mm] (38.3,-1) rectangle (38.8,1);
\draw[line width=0.01mm, fill=black] (39.8,-1) rectangle (40.3,1);

\draw[line width=0.01mm, fill=black] (41.3,-2) rectangle (41.8,2);
\draw[line width=0.01mm, fill=black] (42.8,-2) rectangle (43.3,2);
\draw[line width=0.01mm, fill=black] (44.3,-2) rectangle (44.4,2);

\draw[line width=1mm, -{Latex[length=15pt, width=15pt]}] (1.25,1) -- (1.25,2.75) -- (38.55,2.75) -- (38.55,1);
\draw[line width=1mm, -{Latex[length=15pt, width=15pt]}] (6.65,.5) -- (6.65,2.5) -- (32.05,2.5) -- (32.05,.5);
\draw[line width=1mm, -{Latex[length=15pt, width=15pt]}] (11.75,.25) -- (11.75,2.25) -- (23.15,2.25) -- (23.15,.25);

\draw[line width=0.01mm, fill=Brown] (.2,.4) -- (.2,-.4) -- (.8,0) -- cycle;
\draw[line width=0.01mm, fill=red] (1.7,.4) -- (1.7,-.4) -- (2.3,0) -- cycle;
\draw[line width=0.01mm, fill=Orchid] (3.2,.4) -- (3.2,-.4) -- (3.8,0) -- cycle;

\draw[line width=0.01mm, fill=CornflowerBlue] (5.3,.4) -- (5.3,-.4) -- (5.9,0) -- cycle;
\node[rotate=90] at (5.5,1.1) {\fontsize{25}{25}\selectfont $\times$2};
\draw[line width=0.01mm, fill=Orchid] (7.4,.4) -- (7.4,-.4) -- (8,0) -- cycle;

\draw[line width=0.01mm, fill=CornflowerBlue] (10.1,.4) -- (10.1,-.4) -- (10.7,0) -- cycle;
\node[rotate=90] at (10.3,1.1) {\fontsize{25}{25}\selectfont $\times$3};
\draw[line width=0.01mm, fill=Orchid] (12.8,.4) -- (12.8,-.4) -- (13.4,0) -- cycle;

\draw[line width=0.01mm, fill=CornflowerBlue] (16.3,.4) -- (16.3,-.4) -- (16.9,0) -- cycle;
\node[rotate=90] at (16.5,1.3) {\fontsize{25}{25}\selectfont $\times$22};
\draw[line width=0.01mm, fill=blue] (19.8,.4) -- (19.8,-.4) -- (20.4,0) -- cycle;

\draw[line width=0.01mm, fill=Orchid] (24.2,.4) -- (24.2,-.4) -- (24.8,0) -- cycle;
\draw[line width=0.01mm, fill=CornflowerBlue] (26.9,.4) -- (26.9,-.4) -- (27.5,0) -- cycle;
\node[rotate=90] at (27.1,1.1) {\fontsize{25}{25}\selectfont $\times$3};
\draw[line width=0.01mm, fill=blue] (29.6,.4) -- (29.6,-.4) -- (30.2,0) -- cycle;

\draw[line width=0.01mm, fill=Orchid] (32.8,.4) -- (32.8,-.4) -- (33.4,0) -- cycle;
\draw[line width=0.01mm, fill=CornflowerBlue] (34.9,.4) -- (34.9,-.4) -- (35.5,0) -- cycle;
\node[rotate=90] at (35.1,1.1) {\fontsize{25}{25}\selectfont $\times$2};
\draw[line width=0.01mm, fill=blue] (37,.4) -- (37,-.4) -- (37.6,0) -- cycle;

\draw[line width=0.01mm, fill=green] (39,.4) -- (39,-.4) -- (39.6,0) -- cycle;
\node[rotate=90] at (39.2,1.1) {\fontsize{25}{25}\selectfont $\times$2};
\draw[line width=0.01mm, fill=blue] (40.5,.4) -- (40.5,-.4) -- (41.1,0) -- cycle;

\draw[line width=0.01mm, fill=green] (42,.4) -- (42,-.4) -- (42.6,0) -- cycle;
\node[rotate=90] at (42.2,1.1) {\fontsize{25}{25}\selectfont $\times$2};
\draw[line width=0.01mm] (43.5,.4) -- (43.5,-.4) -- (44.1,0) -- cycle;

\node[rotate=90, anchor=south] at (-0.25,0) {\fontsize{30}{30}\selectfont 256$\times$256};
\node at (-0.05,-2.6) {\fontsize{30}{30}\selectfont 3};
\node at (1.25,-1.6) {\fontsize{30}{30}\selectfont 64};
\node at (2.75,-1.1) {\fontsize{30}{30}\selectfont 64};

\node at (4.55,-1.1) {\fontsize{30}{30}\selectfont 256};
\node at (6.65,-1.1) {\fontsize{30}{30}\selectfont 256};

\node at (9.05,-.85) {\fontsize{30}{30}\selectfont 512};
\node at (11.75,-.85) {\fontsize{30}{30}\selectfont 512};

\node at (14.85,-.725) {\fontsize{30}{30}\selectfont 1024};
\node at (18.35,-.725) {\fontsize{30}{30}\selectfont 1024};

\node at (21.45,-.85) {\fontsize{30}{30}\selectfont 512};
\node at (23.15,-.85) {\fontsize{30}{30}\selectfont 512};
\node at (25.85,-.85) {\fontsize{30}{30}\selectfont 512};
\node at (28.55,-.85) {\fontsize{30}{30}\selectfont 512};

\node at (30.7,-1.1) {\fontsize{30}{30}\selectfont 256};
\node at (32.3,-1.1) {\fontsize{30}{30}\selectfont 256};
\node at (34.15,-1.1) {\fontsize{30}{30}\selectfont 256};
\node at (36.25,-1.1) {\fontsize{30}{30}\selectfont 256};

\node at (37.75,-1.6) {\fontsize{30}{30}\selectfont 64};
\node at (38.85,-1.6) {\fontsize{30}{30}\selectfont 64};
\node at (40.05,-1.6) {\fontsize{30}{30}\selectfont 64};

\node at (41.55,-2.6) {\fontsize{30}{30}\selectfont 64};
\node at (43.05,-2.6) {\fontsize{30}{30}\selectfont 64};
\node at (44.75,-2.6) {\fontsize{30}{30}\selectfont $C$+1};

\draw[line width=0.01mm, fill=black] (3,5) rectangle (3.1,9);
\draw[line width=0.01mm, fill=black] (4.1,5) rectangle (4.6,9);

\draw[line width=0.01mm, fill=black] (5.6,6) rectangle (6.1,8);
\draw[line width=0.01mm, fill=black] (7.1,6) rectangle (7.8,8);

\draw[line width=0.01mm, fill=black] (8.8,6.5) rectangle (9.5,7.5);
\draw[line width=0.01mm, fill=black] (10.5,6.5) rectangle (11.6,7.5);

\draw[line width=0.01mm, fill=black] (12.6,6.75) rectangle (13.7,7.25);
\draw[line width=0.01mm, fill=black] (14.7,6.75) rectangle (16.4,7.25);

\draw[line width=0.01mm, fill=black] (17.4,6.875) rectangle (19.1,7.125);
\draw[line width=0.01mm, fill=black] (20.1,6.875) rectangle (22.6,7.125);

\draw[line width=0.01mm, fill=black] (23.6,6.9375) rectangle (26.1,7.0625);
\draw[line width=0.01mm, fill=black] (27.1,6.9375) rectangle (30.8,7.0625);

\draw[line width=0.01mm, fill=black] (32,6.875) rectangle (34.5,7.125);
\draw[line width=0.01mm, fill=black] (35.5,6.875) rectangle (38,7.125);

\draw[line width=0.01mm, fill=black] (39,6.75) rectangle (40.7,7.25);
\draw[line width=0.01mm, fill=black] (41.7,6.75) rectangle (43.4,7.25);

\draw[line width=0.01mm, fill=black] (44.4,6.5) rectangle (45.5,7.5);
\draw[line width=0.01mm, fill=black] (46.5,6.5) rectangle (47.6,7.5);

\draw[line width=0.01mm, fill=black] (48.6,6) rectangle (49.3,8);
\draw[line width=0.01mm, fill=black] (50.3,6) rectangle (51,8);

\draw[line width=0.01mm, fill=black] (52,5) rectangle (52.5,9);
\draw[line width=0.01mm, fill=black] (53.5,5) rectangle (54,9);
\draw[line width=0.01mm, fill=black] (55,5) rectangle (55.1,9);

\draw[line width=1mm, loosely dotted ] (31,9) -- (31,6);
\node[anchor=west] at (31.2,8.6) {\scalebox{3.4}{Decoder $\bm{G}$}};
\node[anchor=east] at (30.8,8.6) {\scalebox{3.4}{Encoder $\bm{F}$}};

\draw[line width=0.01mm, fill=green] (3.3,6.6) -- (3.3,7.4) -- (3.9,7) -- cycle;
\draw[line width=0.01mm, fill=red] (4.8,6.6) -- (4.8,7.4) -- (5.4,7) -- cycle;

\draw[line width=0.01mm, fill=green] (6.3,6.6) -- (6.3,7.4) -- (6.9,7) -- cycle;
\draw[line width=0.01mm, fill=red] (8,6.6) -- (8,7.4) -- (8.6,7) -- cycle;

\draw[line width=0.01mm, fill=green] (9.7,6.6) -- (9.7,7.4) -- (10.3,7) -- cycle;
\draw[line width=0.01mm, fill=red] (11.8,6.6) -- (11.8,7.4) -- (12.4,7) -- cycle;

\draw[line width=0.01mm, fill=green] (13.9,6.6) -- (13.9,7.4) -- (14.5,7) -- cycle;
\draw[line width=0.01mm, fill=red] (16.6,6.6) -- (16.6,7.4) -- (17.2,7) -- cycle;

\draw[line width=0.01mm, fill=green] (19.3,6.6) -- (19.3,7.4) -- (19.9,7) -- cycle;
\draw[line width=0.01mm, fill=red] (22.8,6.6) -- (22.8,7.4) -- (23.4,7) -- cycle;

\draw[line width=0.01mm, fill=green] (26.3,6.6) -- (26.3,7.4) -- (26.9,7) -- cycle;
\draw[line width=0.01mm, fill=blue] (31.2,6.6) -- (31.2,7.4) -- (31.8,7) -- cycle;

\draw[line width=0.01mm, fill=green] (34.7,6.6) -- (34.7,7.4) -- (35.3,7) -- cycle;
\draw[line width=0.01mm, fill=blue] (38.2,6.6) -- (38.2,7.4) -- (38.8,7) -- cycle;

\draw[line width=0.01mm, fill=green] (40.9,6.6) -- (40.9,7.4) -- (41.5,7) -- cycle;
\draw[line width=0.01mm, fill=blue] (43.6,6.6) -- (43.6,7.4) -- (44.2,7) -- cycle;

\draw[line width=0.01mm, fill=green] (45.7,6.6) -- (45.7,7.4) -- (46.3,7) -- cycle;
\draw[line width=0.01mm, fill=blue] (47.8,6.6) -- (47.8,7.4) -- (48.4,7) -- cycle;

\draw[line width=0.01mm, fill=green] (49.5,6.6) -- (49.5,7.4) -- (50.1,7) -- cycle;
\draw[line width=0.01mm, fill=blue] (51.2,6.6) -- (51.2,7.4) -- (51.8,7) -- cycle;

\draw[line width=0.01mm, fill=green] (52.7,6.6) -- (52.7,7.4) -- (53.3,7) -- cycle;
\draw[line width=0.01mm] (54.2,6.6) -- (54.2,7.4) -- (54.8,7) -- cycle;

\node[rotate=90, anchor=south] at (2.85,7) {\fontsize{30}{30}\selectfont 256$\times$256};
\node at (2.65,4.4) {\fontsize{30}{30}\selectfont $C$+1};
\node at (4.35,4.4) {\fontsize{30}{30}\selectfont 16};

\node at (5.85,5.4) {\fontsize{30}{30}\selectfont 16};
\node at (7.45,5.4) {\fontsize{30}{30}\selectfont 32};

\node at (9.15,5.9) {\fontsize{30}{30}\selectfont 32};
\node at (11.05,5.9) {\fontsize{30}{30}\selectfont 64};

\node at (13.15,6.15) {\fontsize{30}{30}\selectfont 64};
\node at (15.55,6.15) {\fontsize{30}{30}\selectfont 128};

\node at (18.25,6.275) {\fontsize{30}{30}\selectfont 128};
\node at (21.35,6.275) {\fontsize{30}{30}\selectfont 256};

\node at (24.85,6.3375) {\fontsize{30}{30}\selectfont 256};
\node at (28.95,6.3375) {\fontsize{30}{30}\selectfont 512};

\node at (33.25,6.275) {\fontsize{30}{30}\selectfont 256};
\node at (36.75,6.275) {\fontsize{30}{30}\selectfont 256};

\node at (39.85,6.15) {\fontsize{30}{30}\selectfont 128};
\node at (42.45,6.15) {\fontsize{30}{30}\selectfont 128};

\node at (44.95,5.9) {\fontsize{30}{30}\selectfont 64};
\node at (47.05,5.9) {\fontsize{30}{30}\selectfont 64};

\node at (48.95,5.4) {\fontsize{30}{30}\selectfont 32};
\node at (50.65,5.4) {\fontsize{30}{30}\selectfont 32};

\node at (52.25,4.4) {\fontsize{30}{30}\selectfont 16};
\node at (53.75,4.4) {\fontsize{30}{30}\selectfont 16};
\node at (55.45,4.4) {\fontsize{30}{30}\selectfont $C$+1};

\draw[line width=0.01mm, fill=black] (49.1,2) rectangle (49.2,-2);
\draw[line width=0.01mm, fill=black] (50.2,1) rectangle (50.7,-1);
\draw[line width=0.01mm, fill=black] (51.7,.5) rectangle (52.8,-.5);
\draw[line width=0.01mm, fill=black] (53.8,.25) rectangle (55.5,-.25);
\draw[line width=0.01mm, fill=black] (56.5,.25) rectangle (59,-.25);
\draw[line width=0.01mm, fill=black] (60,.25) rectangle (60.1,-.25);

\draw[line width=0.01mm, fill=orange] (49.4,.4) -- (49.4,-.4) -- (50,0) -- cycle;
\draw[line width=0.01mm, fill=orange] (50.9,.4) -- (50.9,-.4) -- (51.5,0) -- cycle;
\draw[line width=0.01mm, fill=orange] (53,.4) -- (53,-.4) -- (53.6,0) -- cycle;
\draw[line width=0.01mm, fill=magenta] (55.7,.4) -- (55.7,-.4) -- (56.3,0) -- cycle;
\draw[line width=0.01mm, fill=violet] (59.2,.4) -- (59.2,-.4) -- (59.8,0) -- cycle;

\node[rotate=90, anchor=south] at (48.95,0) {\fontsize{30}{30}\selectfont 256$\times$256};
\node at (49.15,-2.6) {\fontsize{30}{30}\selectfont $C$+2};
\node at (50.45,-1.6) {\fontsize{30}{30}\selectfont 32};
\node at (52.25,-1) {\fontsize{30}{30}\selectfont 64};
\node at (54.65,-.85) {\fontsize{30}{30}\selectfont 128};
\node at (57.75,-.85) {\fontsize{30}{30}\selectfont 256};
\node at (60.05,-.85) {\fontsize{30}{30}\selectfont 1};

\node at (30.1,5) {\scalebox{3.4}{Auto-encoder}};
\node at (22.15,-2.5) {\scalebox{3.4}{Segmentation network $\bm{S}$ based on Res-UNet}};
\node at (54.6,-4.05) {\scalebox{3.4}{Discriminator network $\bm{D}$}};

\draw[line width=0.1mm, color=darkgray] (3,-5) rectangle (55,-12);

\draw[line width=0.01mm, fill=green] (3.5,-5.5) -- (3.5,-6.3) -- (4.1,-5.9) -- cycle;
\draw[line width=0.01mm, fill=red] (3.5,-6.8) -- (3.5,-7.6) -- (4.1,-7.2) -- cycle;
\draw[line width=0.01mm, fill=blue] (3.5,-8.1) -- (3.5,-8.9) -- (4.1,-8.5) -- cycle;
\draw[line width=0.01mm] (3.5,-9.4) -- (3.5,-10.2) -- (4.1,-9.8) -- cycle;
\draw[line width=0.01mm, fill=Brown] (3.5,-10.7) -- (3.5,-11.5) -- (4.1,-11.1) -- cycle;

\node[anchor=west] at (4.6,-5.9) {\fontsize{30}{30}\selectfont conv 3$\times$3, batch normalization, ReLU};
\node[anchor=west] at (4.6,-7.2) {\fontsize{30}{30}\selectfont max-pooling 2$\times$2};
\node[anchor=west] at (4.6,-8.5) {\fontsize{30}{30}\selectfont up-conv 2$\times$2};
\node[anchor=west] at (4.6,-9.8) {\fontsize{30}{30}\selectfont conv 1$\times$1, softmax};
\node[anchor=west] at (4.6,-11.1) {\fontsize{30}{30}\selectfont conv 7$\times$7, stride 2$\times$2, batch normalization, ReLU};

\draw[line width=0.01mm, fill=Orchid] (32.9,-5.5) -- (32.9,-6.3) -- (33.5,-5.9) -- cycle; 
\draw[line width=0.01mm, fill=CornflowerBlue] (32.9,-6.8) -- (32.9,-7.6) -- (33.5,-7.2) -- cycle; 
\draw[line width=0.01mm, fill=orange] (32.9,-8.1) -- (32.9,-8.9) -- (33.5,-8.5) -- cycle; 
\draw[line width=0.01mm, fill=magenta] (32.9,-9.4) -- (32.9,-10.2) -- (33.5,-9.8) -- cycle;
\draw[line width=0.01mm, fill=violet] (32.9,-10.7) -- (32.9,-11.5) -- (33.5,-11.1) -- cycle;

\node[anchor=west] at (34,-5.9) {\fontsize{30}{30}\selectfont residual convolutional block};
\node[anchor=west] at (34,-7.2) {\fontsize{30}{30}\selectfont residual identity block};
\node[anchor=west] at (34,-8.5) {\fontsize{30}{30}\selectfont conv 4$\times$4, strides 2, batch normalization, ReLU};
\node[anchor=west] at (34,-9.8) {\fontsize{30}{30}\selectfont conv 4$\times$4, strides 1, batch normalization, ReLU};
\node[anchor=west] at (34,-11.1) {\fontsize{30}{30}\selectfont conv 4$\times$4, strides 1, sigmoid};

\end{tikzpicture}
\end{adjustbox}
\caption{Proposed multi-structure deep architectures with $C$ structures of interest: auto-encoder comprising encoder $\bm{F}$ and decoder $\bm{G}$ (top), segmentation network $\bm{S}$ based on Res-UNet (bottom left) as well as discriminator $\bm{D}$ (bottom right). The auto-encoder allows learning a non-linear shape representation from ground truth segmentations, while the discriminator outputs a one-channel likelihood map consisting of values ranging from 0 (fake) to 1 (real). During $\bm{S}$ training, the shape encoder $\bm{F}$ and discriminator $\bm{D}$ respectively compute the shape priors based and adversarial regularizations to constrain the segmentation network (Figure \ref{fig:framework}). Finally, $\bm{S}$ integrates \texttt{ResNet50} as a pre-trained encoder and to fit the image dimensions, we extended input MR images from single grayscale channel to 3 channels.}
\label{fig:architecture}
\end{figure*}

\subsection{Residual segmentation network with pre-trained encoder}
\label{sec:segmentaton_network_with_pre-trained_encoder}
Let $\bm{x} = \{x_u \in \mathbb{R}, u \in \Omega\}$ be a grayscale image with $\Omega \subset \mathbb{N}\times\mathbb{N}$ the image grid. The corresponding image class labels $\bm{y} = \{y_{c,u} \in \{0,1\}, c \in \mathscr{C}, u \in \Omega\}$ represent the different anatomical objects of interest (plus background) $\mathscr{C} = \{0,...,C\}$. In CNN-based segmentation approaches, the aim is to learn a mapping $\bm{S}: \bm{x} \mapsto \bm{S}(\bm{x}; \bm{\Theta_S})$ between intensity $\bm{x}$ and class labels $\bm{y}$ images. The function $\bm{S}$ is a segmentation network composed of a succession of layers whose parameters $\bm{\Theta_{S}}$ must be optimized during training. Let $\{\bm{x_n}, \bm{y_n}\}_{1\leq n \leq N}$ be a training set of $N$ couples of images and corresponding segmentation maps. In the following, we note $\bm{\hat{y}_n} = \bm{S}(\bm{x_n}; \bm{\Theta_S})$ as the estimate of $\bm{y_n}$ having observed $\bm{x_n}$. During training, we optimized the loss function $\mathcal{L}_S$ using stochastic gradient descent to estimate the optimal weights $\bm{\Theta_{S}^{*}}$ as follows:
\begin{align}
    & \bm{\Theta_{S}^{*}} = \argmin_{\bm{\Theta_S}} \mathcal{L}_S(\bm{\Theta_S}) \\
    & \mathcal{L}_S(\bm{\Theta_S}) = \dfrac{1}{N} \sum_{n=1}^{N} \ell_S (\bm{\hat{y}_n},\bm{y_n})
\end{align}
\noindent where $\ell_S$ was the employed per-image loss function.

The neural network $\bm{S}$ is based on the UNet architecture \cite{ronneberger_u-net_2015} whose encoder components is replaced by a classification network with weights previously trained on an image classification task. More precisely, we employed \texttt{ResNet50} \cite{he_deep_2016} which incorporates residual blocks to allow faster convergence, increase network depth and improve predictive performance. First, to fit the \texttt{ResNet50} image dimensions, we concatenated 3 copies of each MR slice to extend them from single grayscale channel to 3 channels. The encoder branch built on residual convolutional and identity blocks \cite{he_deep_2016} generated a 1024 dimensional feature map, which corresponds to the central part between the contracting and expanding paths (Figure \ref{fig:architecture}). We then constructed a symmetrical decoder branch with additional convolutional layers, features channels and residual blocks (Figure \ref{fig:architecture}). Contrary to encoder weights that are pre-trained on ImageNet, the decoder weights were randomly initialized. A final 1×1 convolutional layer with softmax activation function achieved pixel-wise segmentation.

We employed a loss function based on cross-entropy defined as follows:
\begin{align}
    & \ell_{CE}(\bm{\hat{y}_n}, \bm{y_n}) = \dfrac{1}{\vert\mathscr{C}\vert \vert\Omega\vert} \sum_{c \in \mathscr{C}}  \sum_{u \in \Omega} -y_{n,c,u}\log(\hat{y}_{n,c,u})
\end{align}
\noindent The segmentation model is trained through a loss function which operates on individual pixel-level class predictions. However, this pixel-wise loss function fails to exploit contextual inter-structure relationships arising from segmentation masks. Indeed, this loss integrates regional context thr\-ugh the receptive field of the network but fails to include global context \cite{oktay_anatomically_2018, pham_deep_2019, ravishankar_learning_2017}. Hence, we propose to incorporate additional regularization terms which assess the global similarity between predicted and ground truth masks.

\subsection{Incorporating shape priors based regularization}
\label{sec:incorporating_shape_priors_based_regularization}
In the context of medical image segmentation, one can assume that the ground truth segmentation masks lie in a manifold of true shapes, due to the constrained nature of anatomical structures. However, the output prediction of a segmentation network may not lie on the  true shapes manifold, and it is hence needed to perform a projection onto the correct manifold \cite{oktay_anatomically_2018, ravishankar_learning_2017}. While many choices exist for linear and non-linear representations of segmentation shape priors, a convolutional auto-encoder allows us to efficiently learn such low-dimensional shape representation from ground truth segmentation masks, and to easily compute the projection of segmentation masks using its encoder component \cite{ravishankar_learning_2017}. 

Specifically, an auto-encoder is a neural network composed of an encoder $\bm{F}:\bm{y} \mapsto \bm{F}(\bm{y}; \bm{\Theta_F})$ and a decoder $\bm{G}:\bm{F}(\bm{y}; \bm{\Theta_F}) \mapsto \bm{G}(\bm{F}(\bm{y}; \bm{\Theta_F}); \bm{\Theta_G})$. The encoder $\bm{F}$ maps the input to a low-dimensional feature space and the decoder $\bm{G}$ reconstructs the original input from the compact representation. After optimizing the auto-encoder, its encoder component is able to produce a feature map $\bm{F}(\bm{y}; \bm{\Theta_F})$ which compactly encodes the most salient characteristics of the input mask and each value represents a global feature of a crop of the input binary mask. 

The auto-encoder training procedure minimizes a loss function $\mathcal{L}_{AE}$ which penalizes the reconstruction $(\bm{G}\circ\bm{F})(\bm{y_n})$ $= \bm{G}(\bm{F}(\bm{y_n}; \bm{\Theta_F}); \bm{\Theta_G})$ to be dissimilar from the original input $\bm{y_n}$. Usual training schemes are based on mean-squared error, Dice or cross-entropy loss to enforce the auto-encoder to learn the global shape features arising from ground truth annotations \cite{oktay_anatomically_2018, ravishankar_learning_2017}. The cross-entropy loss function to optimize both encoder and decoder weights $\bm{\Theta_F}$ and $\bm{\Theta_G}$ is given as follows:
\begin{align}
    & \bm{\Theta_F^{*}}, \bm{\Theta_G^{*}} = \argmin_{\bm{\Theta_F}, \bm{\Theta_G}} \mathcal{L}_{AE}(\bm{\Theta_F}, \bm{\Theta_G})\\
    &\mathcal{L}_{AE}(\bm{\Theta_F}, \bm{\Theta_G}) = \dfrac{1}{N} \sum_{n=1}^{N} \ell_{CE}((\bm{G}\circ\bm{F})(\bm{y_n}), \bm{y_n})
\end{align}

As a first step, the auto-encoder was trained on ground truth annotations using cross-entropy to learn a shape space in the form of a non-linear low-dimensional manifold which is simply represented by the latent space at the output of its encoder component. While some approaches leverage contour information (e.g. Hausdorff distance) to enforce shape constraints \cite{karimi_reducing_2020}, our shape representation is based on complete segmentation masks. In practice, an auto-encoder would have difficulty learning a shape representation based on contours due to the high imbalance between contours and background pixels. Hence, we employed a mask-based shape regularization and the architecture of the convolutional auto-encoder incorporated traditional convolutional and up-conv\-olutional layers (Figure \ref{fig:architecture}).

After training the auto-encoder, we integrated its encoder component into the baseline segmentation network by computing a shape regularization term $\ell_{shape}$. To this end, both predictions $\bm{\hat{y}_n}$ and ground truth labels $\bm{y_n}$ were projected onto the latent shape space by the shape encoder with learnt weights $\bm{\Theta_F}$ (Figure \ref{fig:framework}). The shape regularization term computed the Euclidean distance between both latent shape representations \cite{oktay_anatomically_2018}, as follows: 
\begin{align}
    & \ell_{shape}(\bm{\hat{y}_n},\bm{y_n}; \bm{\Theta_F}) = \norm{\bm{F}(\bm{\hat{y}_n}; \bm{\Theta_F}) - \bm{F}(\bm{y_n}; \bm{\Theta_F})}_2^2 
\end{align}

The shape regularization loss enforced the predicted segmentation $\bm{\hat{y}_n}$ to be in the same low-dimensional manifold as the ground truth mask $\bm{y_n}$ (i.e. true shapes manifold) and thus encouraged anatomically consistent class label prediction \cite{oktay_anatomically_2018}. More precisely, minimizing the Euclidean distance led to similar feature maps at the output of the shape encoder (i.e. shape codes) for both segmentation masks. It should be emphasized that because the weights of the shape encoder were fixed, the two feature maps were in correspondence, with each value encoding the same global shape feature for both ground truth and predicted segmentation masks. However, due to the black-box nature of deep learning models, the interpretability of each shape feature remained limited in practice. We combined both cross-entropy and shape regularization losses during training, and the updated optimization problem was defined as follows:
\begin{align}
    \bm{\Theta_S^{*}} &= \argmin_{\bm{\Theta_S}} \mathcal{L}_S(\bm{\Theta_S}; \bm{\Theta_F}) \\
    \mathcal{L}_S(\bm{\Theta_S}; \bm{\Theta_F}) &= \dfrac{1}{N} \sum_{n=1}^{N} \ell_S(\bm{\hat{y}_n}, \bm{y_n};  \bm{\Theta_F}) \\
    \ell_S(\bm{\hat{y}_n}, \bm{y_n}; \bm{\Theta_F}) &= \ell_{CE}(\bm{\hat{y}_n}, \bm{y_n}) \nonumber\\
    & \hspace{1.05em} + \lambda_1 \ell_{shape}(\bm{\hat{y}_n},\bm{y_n}; \bm{\Theta_F}) 
\end{align}
\noindent where $\lambda_1$ was an empirically set weighting factor.

Even though the shape regularization enforces global sh\-ape consistency in model predictions, it fails to assess the global accuracy of generated masks given intensity images. Indeed, this regularization only exploits mask-based information and does not allow to evaluate the accuracy of the segmentation with respect to the input intensity image. To mitigate this issue, we propose to incorporate into the pipeline a conditional discriminator to reinforce the global realistic aspect of predicted delineations.

\subsection{Combining shape priors based regularization with adversarial regularization}
\label{sec:combining_shape_priors_based_regularization_with_adversarial_regularization}
For semantic segmentation, a conditional discriminator $\bm{D}:\bm{y}, \bm{x} \mapsto \bm{D}(\bm{y}, \bm{x}; \bm{\Theta_D})$ can assess if a binary mask is fake or not, given the corresponding grayscale image which is provided as a condition. The discriminator $\bm{D}$ is a neural network which returns a one-channel likelihood map $\bm{D}(\bm{y}, \bm{x}; \bm{\Theta_D})$. Each value of the likelihood map (Figure \ref{fig:architecture}) represents the degree of likelihood of correct segmentation of a crop of the input image, ranging from 0 (fake) to 1 (plausible or real). The likelihood is learnt from the ground truth and generated data and the discriminator architecture consisted of $4\times4$ convolutional layers to obtain a large receptive field.

Although traditional GAN approaches aim at generating new images with the same characteristics (i.e. statistics) as the training set, in a segmentation context, the discriminator instead enables to constrain the segmentation network through an adversarial regularization. Specifically, instead of generating new images, the segmentation model predicts synthetic (i.e. fake) masks from intensity images, which should be indistinguishable from ground truth (i.e real) segmentation. Following typical adversarial learning schemes, the discriminator and the segmentation networks are trained alternatively and competitively, with the role of $\bm{S}$ being similar to that of the generator. More precisely, the optimization of weights $\bm{\Theta_S}$ (respectively $\bm{\Theta_D}$) is done using the loss function $\mathcal{L}_S$ (respectively $\mathcal{L}_D$) while parameters $\bm{\Theta_D}$ (respectively $\bm{\Theta_S}$) are fixed. These losses are defined in such a way that the discriminator learns to differentiate real from synthetic segmentation masks while the segmentation network learns to generate increasingly plausible masks.

The binary cross entropy loss $\ell_{BCE}$ is typically used to train the discriminator, with real and fake labels for the likelihood maps of ground truth and generated masks respectively. $\ell_{BCE}$ maximizes the loss value associated with the likelihood map of ground truth masks $\bm{y_n}$ and minimizes the loss corresponding to the likelihood map of predicted masks $\bm{\hat{y}_n}$, given the intensity image $\bm{x_n}$. Therefore, the discriminator learns to discriminate ground truth (i.e real) from generated (i.e. fake) segmentations during the optimization of $\bm{\Theta_D}$ \cite{conze_abdominal_2021, nie_adversarial_2020, singh_breast_2020, xue_segan_2018}. More precisely, the discriminator is optimized to yield a likelihood map with values equal to 1 (respectively 0) for ground truth (respectively predicted) masks. In the following, we adopt a compact notation $\bm{D}(\bm{y_n}, \bm{x_n}) = \bm{D}(\bm{y_n}, \bm{x_n}; \bm{\Theta_D})$:
\begin{align}
    & \hspace{3.78em}\bm{\Theta_D^{*}} = \argmin_{\bm{\Theta_D}} \mathcal{L}_D(\bm{\Theta_D};\bm{\Theta_S})\\
    &\mathcal{L}_D(\bm{\Theta_D};\bm{\Theta_S}) = \dfrac{1}{N} \sum_{n=1}^{N} \ell_{BCE}(\bm{D}(\bm{\hat{y}_n}, \bm{x_n}), \bm{D}(\bm{y_n}, \bm{x_n}))\\
    &\ell_{BCE}(\bm{D}(\bm{\hat{y}_n}, \bm{x_n}), \bm{D}(\bm{y_n}, \bm{x_n})) = - \log(1 - \bm{D}(\bm{\hat{y}_n}, \bm{x_n})) \nonumber\\ 
    &\hspace{12.55em} - \log(\bm{D}(\bm{y_n}, \bm{x_n}))
\end{align}

The discriminator was integrated into our segmentation framework by computing an adversarial term $\ell_{adv}$ derived from the probability that the network considered the generated mask to be the ground truth segmentation for a given grayscale image (Figure \ref{fig:framework}) \cite{conze_abdominal_2021, nie_adversarial_2020, singh_breast_2020, xue_segan_2018}. The loss computed from the discriminator likelihood map given $\bm{\hat{y}_n}$, $\bm{x_n}$ and with fixed weights $\bm{\Theta_D}$ was defined as follows:
\begin{align}
    & \ell_{adv}(\bm{\hat{y}_n}, \bm{x_n}; \bm{\Theta_D}) = - \log(\bm{D}(\bm{\hat{y}_n}, \bm{x_n}))
\end{align}

We modified the segmentation training strategy to combine shape priors based and conditional adversarial regularizations. The optimization of the adversarial term encouraged the segmentation network to fool the discriminator (i.e. discriminator predicting a likelihood map equals to one for a synthetic mask), resulting in more plausible segmentation mask with respect to the conditional intensity image. At first, the segmentation network will provide a rough prediction of the mask shape, and as training progresses, the discriminator will foster an increasingly accurate mask outline, resulting in more precise delineations of the targeted structures \cite{conze_abdominal_2021, nie_adversarial_2020, singh_breast_2020, xue_segan_2018}. The proposed loss function was a linear combination of cross-entropy, shape priors and adversarial regularizations. The novel optimization procedure was defined as follows:
\begin{align}
    &  \bm{\Theta_S^{*}} = \arg \min_{\bm{\Theta_S}} \max_{\bm{\Theta_D}} \mathcal{L}_S(\bm{\Theta_S}; \bm{\Theta_F}, \bm{\Theta_D})\\
    &\mathcal{L}_S(\bm{\Theta_S}; \bm{\Theta_F}, \bm{\Theta_D}) = \dfrac{1}{N} \sum_{n=1}^{N} \ell_S(\bm{\hat{y}_n}, \bm{y_n}, \bm{x_n}; \bm{\Theta_F}, \bm{\Theta_D}) \\
    &\ell_S(\bm{\hat{y}_n}, \bm{y_n}, \bm{x_n}; \bm{\Theta_F}, \bm{\Theta_D}) = \ell_{CE}(\bm{\hat{y}_n}, \bm{y_n}) \nonumber \\
    &\hspace{10.65em} + \lambda_1 \ell_{shape}(\bm{\hat{y}_n},\bm{y_n}; \bm{\Theta_F}) \nonumber\\
    &\hspace{10.65em} + \lambda_2 \ell_{adv}(\bm{\hat{y}_n}, \bm{x_n}; \bm{\Theta_D})
\end{align}
\noindent where $\lambda_1$ and $\lambda_2$ were empirically determined.

\section{Experiments}

In this section, we describe the experiments conducted with the proposed multi-structure automatic segmentation network on scarce pediatric imaging datasets. We first describe the imaging datasets (Section \ref{sec:imaging_datasets}) followed by the compared segmentation methods (Section \ref{sec:segmentation_performance}). We then explain the implementation details (Section \ref{sec:implementation_details}) while describing and determining the metrics used for the method evaluation (Section \ref{sec:assessment_of_predicted_segmentation}) and our proposed ranking system to measure the performance of each evaluated strategy (Section \ref{sec:ranking_system}).

\subsection{Imaging datasets}
\label{sec:imaging_datasets}

Experiments were conducted on two pediatric MR image datasets acquired at Centre Hospitalier Régional Universitaire (CHRU), La Cavale Blanche, Brest, France, using a 3.0T Achieva scanner (Philips Healthcare, Best, Netherlands). The ankle and shoulder datasets were retrospectively and independently acquired from the radiology department of CHRU and each contained healthy and impaired pediatric individuals. MRI data acquisition was performed in line with the principles of the Declaration of Helsinki. Ethical approval was granted by the Ethics Committee (Comit\'e Protection de Personnes Ouest VI) of CHRU Brest (2015-A01409-40).\\

\noindent \textbf{Ankle joint dataset.} MR images of $17$ ankles were obtained from pediatric individuals aged from 7 to 13 years (average age: $10 \pm 2$ years) which included 7 pathological ($A_{P,1},...,A_{P,7}$) and 10 healthy ($A_{H,1},...,A_{H,10}$) cases. Images were acquired using a T1-weighted gradient echo sequence (TR: 7.9 ms, TE: 2.8 ms, FOV: $140\times161$ mm$^{2}$), with resolutions varying from $0.25\times0.25\times0.50$ mm$^3$ to $0.28\times0.28\times0.80$ mm$^3$. All images were annotated by a medically trained expert (15 years of experience) to get ground truth delineations of calcaneus, talus and tibia bones, with specific label for each bone using ITK-SNAP software (\url{http://www.itksnap.org/}).\\

\noindent \textbf{Shoulder joint dataset.} MR images of $15$ shoulder joints were acquired from pediatric subjects consisting of 7 pathological ($S_{P,1},...,S_{P,7}$) and 8 healthy ($S_{H,1},...,S_{H,8}$) examinations with age group ranging from 5 to 17 years old (average age: $12 \pm 4$ years). An eTHRIVE (enhanced T1-weighted High-Resolution Isotropic Volume Examination) sequence was employed during image acquisition (TR: 8.4 ms, TE: 4.2 ms, FOV: $260\times210$ mm$^{2}$). Image resolution varied across subjects from $0.24\times0.24\times0.60$ mm$^3$ to $0.37\times0.37\times1.00$ mm$^3$. Segmentation masks of the humerus and scapula bones were manually obtained following the same protocol as for ankle joints using ITK-SNAP software. \\

For each dataset, all 2D slices were downsampled to $256 \times 256$ pixels and intensities were normalized to have zero-mean and unit variance.

\subsection{Segmentation performance}
\label{sec:segmentation_performance}

\subsubsection{Experimental setups}
\label{sec:experimental_setups}

\begin{figure}
\centering
\begin{adjustbox}{width=\textwidth}
\begin{tikzpicture}

\node[inner sep=0pt] (mri_s_i) at (0,0)
    {\includegraphics[width=.0525\textwidth]{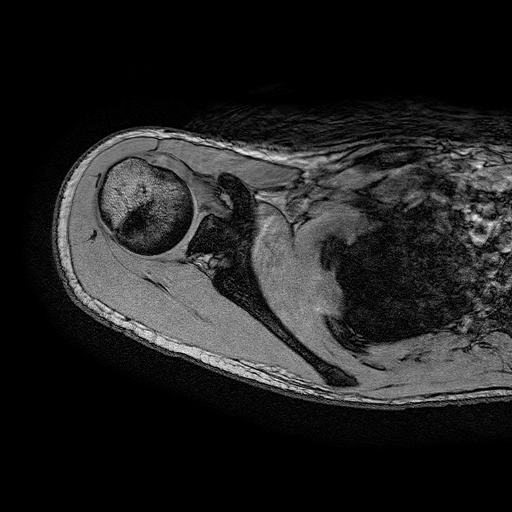}};
\node[inner sep=0pt] (pred_i_h) at (-.4,-1.2)
    {\includegraphics[width=.0525\textwidth]{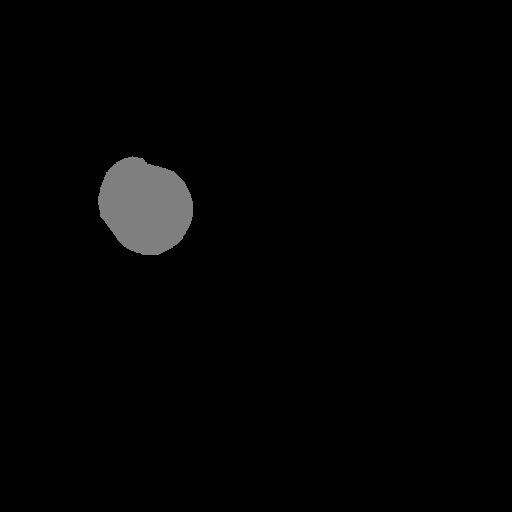}};
\node[inner sep=0pt] (pred_i_s) at (.4,-1.2)
    {\includegraphics[width=.0525\textwidth]{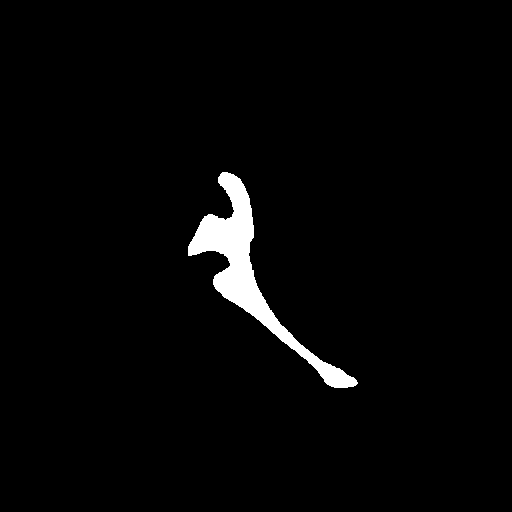}};

\node[inner sep=0pt] (mri_s_g) at (1,0)
    {\includegraphics[width=.0525\textwidth]{experiments_mri_s.png}};
\node[inner sep=0pt] (pred_g) at (1,-1.2)
    {\includegraphics[width=.0525\textwidth]{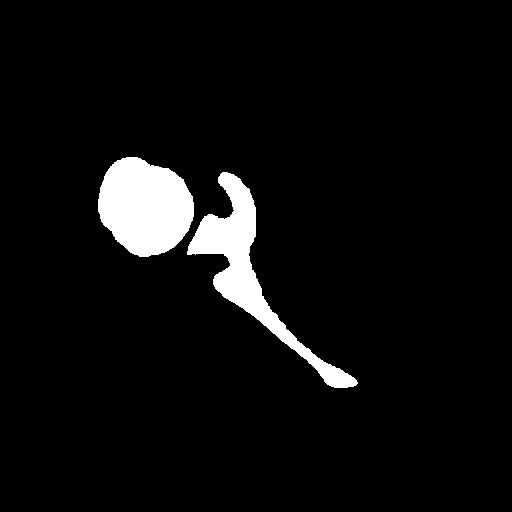}};

\node[inner sep=0pt] (mri_s_m) at (1.6,0)
    {\includegraphics[width=.0525\textwidth]{experiments_mri_s.png}};
\node[inner sep=0pt] (pred_m) at (1.6,-1.2)
    {\includegraphics[width=.0525\textwidth]{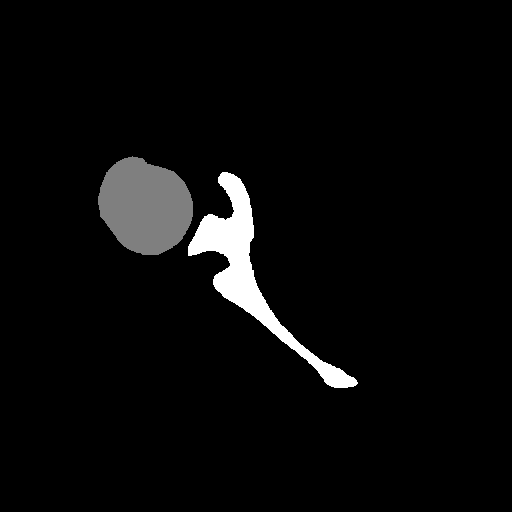}};

\draw[line width=0.01mm, fill=cyan!30] (-.65,-.3) -- (-.15,-.3) -- (-.275,-.6) -- (-.15,-.9) -- (-.65,-.9) -- (-.525,-.6) -- cycle;
\node[rotate=90, scale=0.3] at (-.45,-.6) {\fontsize{9}{9}\selectfont Humerus};
\node[rotate=90, scale=0.3] at (-.35,-.6) {\fontsize{9}{9}\selectfont Network};

\draw[line width=0.01mm, fill=cyan!30] (.65,-.3) -- (.15,-.3) -- (.275,-.6) -- (.15,-.9) -- (.65,-.9) -- (.525,-.6) -- cycle;
\node[rotate=90, scale=0.3] at (.35,-.6) {\fontsize{9}{9}\selectfont Scapula};
\node[rotate=90, scale=0.3] at (.45,-.6) {\fontsize{9}{9}\selectfont Network};

\draw[line width=0.01mm, fill=cyan!30] (1.25,-.3) -- (.75,-.3) -- (.875,-.6) -- (.75,-.9) -- (1.25,-.9) -- (1.125,-.6) -- cycle;
\node[rotate=90, scale=0.3] at (.95,-.6) {\fontsize{9}{9}\selectfont Global-class};
\node[rotate=90, scale=0.3] at (1.05,-.6) {\fontsize{9}{9}\selectfont Network};

\draw[line width=0.01mm, fill=cyan!30] (1.85,-.3) -- (1.35,-.3) -- (1.475,-.6) -- (1.35,-.9) -- (1.85,-.9) -- (1.725,-.6) -- cycle;
\node[rotate=90, scale=0.3] at (1.55,-.6) {\fontsize{9}{9}\selectfont Multi-class};
\node[rotate=90, scale=0.3] at (1.65,-.6) {\fontsize{9}{9}\selectfont Network};

\draw[line width=0.01mm, -{Latex[length=1pt, width=1pt]}] (.4,0) -- (.4,-.3);
\draw[line width=0.01mm, -{Latex[length=1pt, width=1pt]}] (-.4,0) -- (-.4,-.3);
\draw[line width=0.01mm] (mri_s_i.east) -- (.4,0);
\draw[line width=0.01mm] (mri_s_i.west) -- (-.4,0);

\draw[line width=0.01mm, -{Latex[length=1pt, width=1pt]}] (-.4,-.9) -- (pred_i_h.north);
\draw[line width=0.01mm, -{Latex[length=1pt, width=1pt]}] (.4,-.9) -- (pred_i_s.north);

\draw[line width=0.01mm, -{Latex[length=1pt, width=1pt]}] (mri_s_g.south) -- (1,-.3);
\draw[line width=0.01mm, -{Latex[length=1pt, width=1pt]}] (1,-.9) -- (pred_g.north);

\draw[line width=0.01mm, -{Latex[length=1pt, width=1pt]}] (mri_s_m.south) -- (1.6,-.3);
\draw[line width=0.01mm, -{Latex[length=1pt, width=1pt]}] (1.6,-.9) -- (pred_m.north);

\draw[line width=0.1mm, color=darkgray, rounded corners=1] (.7, .3) -- (.7,-1.65) -- (-.7,-1.65) -- (-.7,.3) -- cycle;
\draw[line width=0.1mm, color=darkgray, rounded corners=1] (.7, .3) -- (.7,-1.65) -- (1.3,-1.65) -- (1.3,.3) -- cycle;
\draw[line width=0.1mm, color=darkgray, rounded corners=1] (1.9, .3) -- (1.9,-1.65) -- (1.3,-1.65) -- (1.3,.3) -- cycle;

\node[scale=0.3] at (0,-1.55) {\fontsize{9}{9} \selectfont(a) Individual - $\bm{y^{i}}$};
\node[scale=0.3] at (1,-1.55) {\fontsize{9}{9}\selectfont (b) Global - $\bm{y^{g}}$};
\node[scale=0.3] at (1.6,-1.55) {\fontsize{9}{9}\selectfont (c) Multi - $\bm{y^{m}}$};

\end{tikzpicture}
\end{adjustbox}
\caption{Proposed bone segmentation strategies: (a) individual strategy comprising a specific network for each bone of interest $\bm{y^{i}}$, (b) global strategy constituted of a unique bone class $\bm{y^{g}}$ and (c) multi strategy based on segmentation maps $\bm{y^{m}}$ containing multiple classes.}
  \label{fig:bone_segmentation_strategies}
\end{figure}

In the following experiments, we employed the Att-UNet as backbone architecture to investigate three bone segmentation strategies and to assess which approach would enforce better segmentation outcomes. The segmentation strategies were based on individual-class $\bm{y^{i}}$, global-class $\bm{y^{g}}$ and multi-class $\bm{y^{m}}$ labels (Figure \ref{fig:bone_segmentation_strategies}), which were defined as follows:
\begin{align}
      & \bm{y^{i}} = \{\{y_{c,u} \in \{0,1\}, u \in \Omega \}, c \in \{1,...,C\}\}\\
      & \bm{y^{g}} = \{y_{u} \in \{0,1\}, u \in \Omega \} \\
      & \bm{y^{m}} = \{y_{c,u} \in \{0,1\}, c \in \mathscr{C}, u \in \Omega\}
\end{align}

In the individual-class scheme, for each anatomical class of interest $c$, we trained an Att-UNet, an auto-encoder and a discriminator on the individual-class binary masks. The individual-class networks were optimized on each class of interest, and the learnt weights were thus specific to a single bone. For the global-class approach, we concatenated the different bone classes into a unique bone-class, and the learnt weights were specific to the global bone class. Finally, in the multi-class strategy, the networks were trained on ground truth segmentation maps containing multiple classes, and the learnt weights were shared across all anatomical structures. We evaluated whether the multi-class strategy promoted more accurate bone segmentation than individual and specific networks, while the global-class approach correspond\-ed to an intermediate strategy between multi-class and indivi\-dual-class schemes. 

Due to the presence of individual-class (respectively glo\-bal-class) binary masks in the individual (respectively global) scheme, we modified the last activation function of the individual (respectively global) Att-UNet and auto-encoder from a softmax function to a sigmoid activation resulting in a binary one-channel prediction map. Consequently, we used a binary cross-entropy loss instead of a multi-class cross-entropy loss function during optimization. The input of the auto-encoder (i.e. segmentation mask) was also characterized by one channel, while the input of the discriminator (i.e. concatenation of the intensity image and segmentation mask) presented two channels. Moreover, in the individual-class scheme, as the predictions produced by the different networks were independent, a pixel could be predicted as belonging to several classes simultaneously. In this eventuality, we selected the class with highest probability (i.e prediction with highest confidence). Finally, to perform a fair comparison between bone segmentation strategies, predicted individual and multi segmentation masks were transformed into global segmentation masks.

Furthermore, to evaluate the contributions of each regularization term, we performed an ablation study for each bone segmentation strategies and compared the baseline Att-UNet \cite{oktay_attention_2018}, Att-UNet with shape priors based regularization \cite{oktay_anatomically_2018}, Att-UNet with adversarial regularization \cite{singh_breast_2020} and Att-UNet with proposed combined regularization scheme. Both hyper-parameters $\lambda_1$ and $\lambda_2$ were fixed to $0$ to train baseline Att-UNet. We set $\lambda_1$ (respectively $\lambda_2$) to $0$ to train Att-UNet with adversarial (respectively shape priors based) regularization. All training hyper-parameters (except $\lambda_1$ and $\lambda_2$) remained fixed across all methods and all networks were trained from scratch, without relying on any transfer learning and fine tuning scheme.

\subsubsection{Pre-trained architectures performance}
\label{sec:pre-trained_architectures_performance}

In the following experiments, we assessed the performance of our method based on a pre-trained Res-UNet with multi-class strategy and proposed combined regularization (CombReg$^{\text{Multi}}_{\text{Res-UNet}}$) against other backbone architectures pre-trained on a large natural images database. Specifically, we employed two backbone architectures (VGG-UNet \cite{simonyan_very_2015} and Dense-UNet \cite{huang_densely_2017}) and compared pre-trained models with and without combined regularization. Moreover, as the work of Raghu et al. provided an in-depth study of the benefits of employing transferred weights compared to randomly initialized ones for medical image analysis, especially in low data regime \cite{raghu_transfusion_2019}, we omitted such evaluation in our experiments.

The VGG-UNet (respectively Dense-UNet) architecture referred to a UNet model whose encoder was replaced by a \texttt{VGG19} \cite{simonyan_very_2015} (respectively \texttt{DenseNet121} \cite{huang_densely_2017}) classifier network pre-trained on ImageNet. Similarly to Res-UNet, the decoder components of VGG-UNet and Dense-UNet were extended by adding convolutional filters and more features (as well as dense blocks \cite{huang_densely_2017} for Dense-UNet) to get symmetrical networks. Finally, we employed only the multi-class strategy with combined regularization in these transfer learning experiments as this scheme reached the best performance in the previous comparisons (Section \ref{sec:rankings}).

\subsection{Implementation details}
\label{sec:implementation_details}

Our training method consisted of two steps. The auto-encoder was first trained using cross-entropy loss. We explored different hyper-parameters: Adam optimizer with initial learning rate $1\text{e-}2$, batch size set $32$ and $10$ epochs were found to be optimal. As a second step, the segmentation network and the discriminator were trained alternatively, one optimization step for both networks at each batch. We used Adam optimizer with initial learning rate set to $1\text{e-}3$ for Att-UNet and to $1\text{e-}4$ for VGG-UNet, Dense-UNet and Res-UNet. The batch size and number of epochs were set to $32$ and $20$ for Att-UNet, $32$ and $10$ for VGG-UNet, $16$ and $10$ for Dense-UNet and Res-UNet. Additionally, each architecture was characterized by distinct model complexity (i.e. number of trainable parameters): auto-encoder (3.1 million), discriminator (0.7 million), Att-UNet (7.9 million), VGG-UNet (34.7 million), Dense-UNet (18.5 million), and Res-UNet (13.6 million). Finally, since the individual scheme involved individual-class networks, this scheme thus involved $C$ times more networks and parameters than the global- and multi-class strategies.

We explored different regularization weighting parameters values and observed $\lambda_1 = 1\text{e-}1$ and $\lambda_2 = 1\text{e-}2$ to be the best combination across all backbone models. All architectures were trained on 2D slices with extensive on-the-fly data augmentation due to limited available training data. Data augmentation comprised random scaling ($\pm20\%$), rotation ($\pm20^{\circ}$), shifting ($\pm20\%$), and flipping in both directions to teach the networks the desired invariance, covariance and robustness properties. Deep learning architectures were implemented using Keras and optimized using a Nvidia RTX 2080 Ti GPU with 12 GB of RAM. 

As a post-processing step, the obtained 2D segmentation masks were stacked together to form a 3D volume. In the individual-class and multi-class schemes, for each anatomical structure, we selected the largest connected set as final 3D predicted mask. In the global-class scheme, we retained the $C$ largest connected sets with $C$ corresponding to the number of bones of interest (3 in ankle and 2 in shoulders). Finally, we applied morphological closing (5$\times$5$\times$5 spherical kernel) to smooth the resulting boundaries.

\subsection{Assessment of predicted segmentation}
\label{sec:assessment_of_predicted_segmentation}
To assess the performance of the different methods, the accuracy of the generated 3D segmentation masks were evaluated against manually annotated ground truths. We computed the Dice coefficient, sensitivity, specificity, maximum symmetric surface distance (MSSD), average symmetric surface distance (ASSD) and relative absolute volume difference (RAVD), as described in the supplementary materials. Dice coefficient measures the similarity between the two 3D sets while sensitivity and specificity calculates the true positive and true negative rates. MSSD and ASSD assesses the models' ability to generate the same contours as those produced manually. Finally, RAVD determines the volumetric difference between volumes.

To evaluate the generalization abilities of each method, experiments were performed in a leave-one-out fashion such that one examination was retained for validation, one for test and the remaining data were used to train the model. The procedure was repeated over all the samples in the dataset to compute the mean and standard deviation for each metric. The hyper-parameters values ($\lambda_1$, $\lambda_2$, batch size, learning rate, etc.) were selected based on the performance of the model on the validation set. Moreover, an expert visually validated the global shape consistency and plausibility of each predicted segmentation. Finally, due to the scarce amount of 3D examinations, we performed the statistical analysis between methods on 2D MR images. We employed the Wilcoxon signed-rank non-parametric test using Dice, sensitivity and specificity scores obtained from the 1446 ankle (respectively 3357 shoulder) 2D slices containing at least one of the bone of interest and which corresponded  to the 17 ankle (respectively 15 shoulder) 3D MR images. The statistical analysis is reported in the supplementary materials.

\subsection{Ranking system}
\label{sec:ranking_system}
\begin{table}[t]
\centering
\caption{Metrics wise threshold values employed in the ranking system. Metrics included Dice, sensitivity, MSSD, ASSD and RAVD. $\delta$ is the longest possible distance in 3D examinations.}
\begin{tabular}{ | c | c | c | c | }

\hline
Metric & Best & Worst & Threshold \\
\hline
\hline
Dice (\%) & 100 & 0 & $>80$ \\
Sensitivity (\%) & 100 & 0 & $>80$ \\
MSSD (mm) & 0 & $\delta$ & $<30$ \\
ASSD (mm) & 0 & $\delta$ & $<4$ \\
RAVD (\%) & 0 & 100 & $<10$ \\
\hline

\end{tabular}
\label{tab:ranking_system}
\end{table}

Although it is essential to employ complementary metrics to assess the performance of each segmentation model, simultaneously comparing the performance of each segmentation strategy across multiple metrics can be challenging, hence we propose to employ a metric-based ranking system. More specifically, we converted the metrics outputs to normalized scores and used the average scores from all the datasets as a ranking system \cite{kavur_chaos_2021}. The proposed ranking system was created based on Dice, sensitivity, MSSD, ASSD and RAVD. Specificity was disregarded as excellent results ($>99.3\%$ in Tables \ref{tab:quantitative_results_ablation_study} and \ref{tab:quantitative_results}) were obtained for all the methods. Furthermore, for each metric, a threshold was defined based on expert knowledge to remove non-satisfactory results. Then, we mapped the metric value between the corresponding best value and the threshold (Table \ref{tab:ranking_system}) to the normalized interval $[0,100]$. Metric values outside this acceptable range were assigned zero scores. The score of the predicted 3D segmentation corresponded to the average over all metric scores, and methods were ranked according to the obtained scalar scores. Separate ranking was performed on each of the shoulder and ankle datasets.

Ranking results via multiple metrics is an arduous task as the selection of thresholds may have an impact on the final ranking \cite{maier-hein_why_2018}. Hence, to assess the robustness of the ranking system, we analysed the effect of the modification of threshold values (each resulting in a different ranking system), as explained in the supplementary materials.

\subsection{Qualitative assessment of predicted segmentation}
\label{sec:qualitative_assessment_of_predicted_segmentation}

We performed visual comparison of predicted segmentation masks at three levels. First, we compared the results of the bone segmentation strategies (individual, global and multi) using Att-UNet with combined regularization. Second, we compared the regularization methods (baseline, shape priors only, adversarial only and proposed combined) by employing a multi-class Att-UNet. Third, we compared the pre-trained backbone architectures including VGG-UNet, Dense-UNet and Res-UNet, employed with multi-class segmentation strategy and combined regularization.

\section{Results}
The proposed CombReg$^{\text{Multi}}_{\text{Res-UNet}}$ method based on pre-trained multi-class Res-UNet with combined regularization was evaluated on two pediatric datasets. In this section we report quantitative results (Section \ref{sec:quantitative_assessment}), ranking scores (Section \ref{sec:rankings}), and qualitative comparisons (Section \ref{sec:qualitative_assessment}) for each dataset.

\subsection{Quantitative assessment}
\label{sec:quantitative_assessment}

\begin{table*}[t]
\centering
\caption{Leave-one-out quantitative assessment of Att-UNet \cite{oktay_attention_2018} on ankle and shoulder datasets. Regularization methods include: baseline, shape priors \cite{oktay_anatomically_2018}, adversarial \cite{singh_breast_2020} and proposed combined; while bone segmentation strategies comprise: individual, global and multi. Metrics encompass Dice (\%), sensitivity (\%), specificity (\%), MSSD (mm), ASSD (mm) and RAVD (\%). First and second best results for each dataset and for each metric are in bold and underlined italic respectively.}
\begin{tabular}{||P{.3cm}||P{.3cm}|P{.95cm}|P{1.7cm}||P{1.5cm}|P{1.5cm}|P{1.5cm}|P{1.5cm}|P{1.5cm}|P{1.5cm}||}
    \hline
    & \multicolumn{3}{c||}{Method} & Dice $\uparrow$ & Sens. $\uparrow$ & Spec. $\uparrow$ & MSSD $\downarrow$ & ASSD $\downarrow$ & RAVD $\downarrow$ \\
    \hline
    \hline
    \multirow{12}{*}{\rotatebox[origin=c]{90}{Ankle Dataset}} & \multirow{12}{*}{\rotatebox[origin=c]{90}{Att-UNet}} & \multirow{4}{*}{Indiv} & Base & 84.0$\pm$6.4 & 82.7$\pm$9.5 & 99.3$\pm$0.8 & 20.4$\pm$12.2 & 2.0$\pm$1.4 & 17.9$\pm$16.0 \\\cline{4-10}
    & & & ShapeReg & 87.1$\pm$3.9 & 85.8$\pm$7.3 & 99.5$\pm$0.4 & 18.4$\pm$11.2 & 1.5$\pm$0.6 & 11.5$\pm$7.5 \\\cline{4-10}
    & & & AdvReg & 86.4$\pm$3.8 & 83.8$\pm$8.4 & 99.5$\pm$0.4 & 16.5$\pm$9.4 & 1.5$\pm$0.7 & 15.1$\pm$8.8 \\\cline{4-10}
    & & & CombReg & 88.0$\pm$5.4 & 86.9$\pm$8.0 & 99.5$\pm$0.3 & 18.0$\pm$13.3 & 1.4$\pm$0.7 & 9.1$\pm$5.8 \\\cline{3-10}
    & & \multirow{4}{*}{Global} & Base & 87.6$\pm$7.8 & \textit{\underline{\smash{92.6$\pm$5.4}}} & 99.0$\pm$1.2 & 19.9$\pm$14.7 & 1.8$\pm$1.6 & 17.9$\pm$27.5 \\\cline{4-10}
    & & & ShapeReg & 87.8$\pm$6.2 & 90.6$\pm$7.7 & 99.1$\pm$1.0 & 18.0$\pm$12.3 & 1.7$\pm$1.4 & 13.7$\pm$14.3 \\\cline{4-10}
    & & & AdvReg & 87.8$\pm$5.0 & \textbf{93.0$\pm$3.8} & 99.1$\pm$0.7 & 18.2$\pm$11.3 & 1.7$\pm$1.2 & 14.4$\pm$11.4 \\\cline{4-10}
    & & & CombReg & 89.8$\pm$2.4 & 90.9$\pm$4.9 & 99.4$\pm$0.3 & 18.3$\pm$12.4 & 1.3$\pm$0.7 & \textit{\underline{\smash{7.3$\pm$4.8}}} \\\cline{3-10}
    & & \multirow{4}{*}{Multi} & Base & 88.4$\pm$6.2 & 86.6$\pm$10.1 & \textit{\underline{\smash{99.6$\pm$0.4}}} & 17.0$\pm$12.4 & 1.3$\pm$0.9 & 12.5$\pm$10.8 \\\cline{4-10}
    & & & ShapeReg & \textit{\underline{\smash{89.9$\pm$6.2}}} & 89.1$\pm$9.1 & \textit{\underline{\smash{99.6$\pm$0.3}}} & \textbf{11.1$\pm$4.3} & \textit{\underline{\smash{1.0$\pm$0.6}}} & 10.1$\pm$7.0 \\\cline{4-10}
    & & & AdvReg & \textit{\underline{\smash{89.9$\pm$3.5}}} & 88.9$\pm$6.8 & \textit{\underline{\smash{99.6$\pm$0.3}}} & \textit{\underline{\smash{13.6$\pm$7.5}}} & 1.1$\pm$0.5 & 9.5$\pm$5.3 \\\cline{4-10}
    & & & CombReg & \textbf{90.7$\pm$3.2} & 88.8$\pm$6.3 & \textbf{99.7$\pm$0.2} & \textbf{11.1$\pm$3.4} & \textbf{0.9$\pm$0.3} & \textbf{7.1$\pm$5.7} \\
    \hline
    \hline
   
   \multirow{12}{*}{\rotatebox[origin=c]{90}{Shoulder Dataset}} & \multirow{12}{*}{\rotatebox[origin=c]{90}{Att-UNet}} & \multirow{4}{*}{Indiv} & Base & 82.6$\pm$8.9 & 82.7$\pm$10.9 & \textit{\underline{\smash{99.8$\pm$0.2}}} & 59.9$\pm$31.1 & 4.6$\pm$3.9 & 12.3$\pm$12.3 \\\cline{4-10}
    & & & ShapeReg & 84.5$\pm$7.3 & 81.4$\pm$11.2 & \textbf{99.9$\pm$0.1} & 38.1$\pm$27.9 & 2.3$\pm$1.7 & 11.1$\pm$11.6 \\\cline{4-10}
    & & & AdvReg & 84.3$\pm$6.4 & 82.0$\pm$10.0 & \textit{\underline{\smash{99.8$\pm$0.1}}} & 28.6$\pm$16.0 & 1.7$\pm$1.0 & 10.8$\pm$8.0 \\\cline{4-10}
    & & & CombReg & 85.7$\pm$5.6 & 83.4$\pm$9.7 & \textit{\underline{\smash{99.8$\pm$0.1}}} & 30.5$\pm$19.7 & 1.8$\pm$1.3 & 9.4$\pm$10.0 \\\cline{3-10}
    & & \multirow{4}{*}{Global} & Base & 82.6$\pm$9.1 & 80.3$\pm$8.7 & \textit{\underline{\smash{99.8$\pm$0.3}}} & 30.2$\pm$17.1 & 2.0$\pm$1.5 & 11.0$\pm$7.5 \\\cline{4-10}
    & & & ShapeReg & 84.5$\pm$9.1 & 83.5$\pm$13.7 & \textit{\underline{\smash{99.8$\pm$0.2}}} & 21.6$\pm$9.9 & 1.5$\pm$1.3 & 14.9$\pm$12.1 \\\cline{4-10}
    & & & AdvReg & 84.3$\pm$9.3 & 84.5$\pm$13.0 & \textit{\underline{\smash{99.8$\pm$0.3}}} & 26.7$\pm$11.3 & 1.7$\pm$1.3 & 13.6$\pm$15.0 \\\cline{4-10}
    & & & CombReg & 86.1$\pm$5.2 & 85.5$\pm$7.1 & \textit{\underline{\smash{99.8$\pm$0.2}}} & 25.8$\pm$8.9 & \textit{\underline{\smash{1.4$\pm$0.7}}} & \textit{\underline{\smash{8.2$\pm$11.1}}} \\\cline{3-10}
    & & \multirow{4}{*}{Multi} & Base & 84.0$\pm$12.3 & 82.8$\pm$16.5 & \textit{\underline{\smash{99.8$\pm$0.2}}} & 24.7$\pm$16.6 & 2.0$\pm$3.3 & 14.8$\pm$17.4 \\\cline{4-10}
    & & & ShapeReg & \textit{\underline{\smash{86.9$\pm$5.9}}} & 84.8$\pm$9.1 & \textbf{99.9$\pm$0.1} & \textit{\underline{\smash{21.7$\pm$10.5}}} & \textbf{1.2$\pm$0.9} & 8.8$\pm$9.2 \\\cline{4-10}
    & & & AdvReg & 85.7$\pm$7.1 & \textit{\underline{\smash{86.4$\pm$8.2}}} & \textit{\underline{\smash{99.8$\pm$0.3}}} & 23.7$\pm$18.5 & 1.6$\pm$1.5 & 10.4$\pm$11.5 \\\cline{4-10}
    & & & CombReg & \textbf{87.8$\pm$5.2} & \textbf{87.1$\pm$5.9} & \textbf{99.9$\pm$0.1} & \textbf{21.2$\pm$13.3} & \textbf{1.2$\pm$1.1} & \textbf{4.8$\pm$4.7} \\
    \hline
  
\end{tabular}
\label{tab:quantitative_results_ablation_study}
\end{table*}

\begin{table*}[t]
\centering
\caption{Leave-one-out quantitative assessment of the three pre-trained architectures: VGG-UNet \cite{simonyan_very_2015}, Dense-UNet \cite{huang_densely_2017} and Res-UNet \cite{he_deep_2016} on ankle and shoulder datasets. Baseline and combined regularization methods are employed along with the multi-structure strategy. Metrics encompass Dice (\%), sensitivity (\%), specificity (\%), MSSD (mm), ASSD (mm) and RAVD (\%). First and second best results for each dataset and for each metric are in bold and underlined italic respectively.}
\begin{tabular}{||P{.3cm}||P{.3cm}|P{.95cm}|P{1.7cm}||P{1.5cm}|P{1.5cm}|P{1.5cm}|P{1.5cm}|P{1.5cm}|P{1.5cm}||}
    \hline
    & \multicolumn{3}{c||}{Method} & Dice $\uparrow$ & Sens. $\uparrow$ & Spec. $\uparrow$ & MSSD $\downarrow$ & ASSD $\downarrow$ & RAVD $\downarrow$ \\
    \hline
    \hline
    \multirow{6}{*}{\rotatebox[origin=c]{90}{Ankle Dataset}}
    & \multirow{2}{*}{\rotatebox[origin=c]{90}{VGG}} & \multirow{2}{*}{Multi} & Base & 92.9$\pm$1.4 & \textbf{93.6$\pm$3.7} & \textit{\underline{\smash{99.7$\pm$0.2}}} & 9.1$\pm$4.2 & \textit{\underline{\smash{0.7$\pm$0.1}}} & 7.3$\pm$3.1 \\\cline{4-10}
    & & & CombReg & 93.1$\pm$1.6 & \textbf{93.6$\pm$3.6} & \textit{\underline{\smash{99.7$\pm$0.2}}} & 8.1$\pm$2.6 & \textit{\underline{\smash{0.7$\pm$0.1}}} & 5.9$\pm$4.0 \\\cline{2-10}
    & \multirow{2}{*}{\rotatebox[origin=c]{90}{\small{Dense}}} & \multirow{2}{*}{Multi} & Base & 93.6$\pm$2.0 & 91.7$\pm$4.7 & \textbf{99.8$\pm$0.1} & 7.9$\pm$3.9 & \textit{\underline{\smash{0.7$\pm$0.2}}} & 6.8$\pm$4.9 \\\cline{4-10}
    & & & CombReg & \textit{\underline{\smash{93.9$\pm$1.8}}} & 92.2$\pm$4.5 & \textbf{99.8$\pm$0.1} & \textit{\underline{\smash{6.8$\pm$3.4}}} & \textbf{0.6$\pm$0.2} & 6.1$\pm$4.4 \\\cline{2-10}
    & \multirow{2}{*}{\rotatebox[origin=c]{90}{Res}} & \multirow{2}{*}{Multi} & Base & 93.8$\pm$1.8 & \textit{\underline{\smash{92.4$\pm$4.3}}} & \textbf{99.8$\pm$0.1} & 7.3$\pm$3.2 & \textbf{0.6$\pm$0.2} & \textit{\underline{\smash{5.3$\pm$4.4}}} \\\cline{4-10}
    & & & CombReg & \textbf{94.3$\pm$1.1} & \textbf{93.6$\pm$3.1} & \textbf{99.8$\pm$0.1} & \textbf{6.1$\pm$2.8} & \textbf{0.6$\pm$0.1} & \textbf{4.7$\pm$2.9} \\
    \hline
    \hline
   
   \multirow{6}{*}{\rotatebox[origin=c]{90}{Shoulder Dataset}}
    & \multirow{2}{*}{\rotatebox[origin=c]{90}{VGG}} & \multirow{2}{*}{Multi} & Base & 88.7$\pm$4.5 & \textit{\underline{\smash{91.5$\pm$5.0}}} & \textit{\underline{\smash{99.8$\pm$0.2}}} & 24.6$\pm$26.9 & 1.3$\pm$1.6 & 8.9$\pm$11.6 \\\cline{4-10}
    & & & CombReg & 89.2$\pm$3.7 & \textbf{92.1$\pm$3.5} & \textit{\underline{\smash{99.8$\pm$0.1}}} & \textit{\underline{\smash{21.7$\pm$22.1}}} & \textit{\underline{\smash{1.0$\pm$0.8}}} & 6.8$\pm$6.1 \\\cline{2-10}
    & \multirow{2}{*}{\rotatebox[origin=c]{90}{\small{Dense}}} & \multirow{2}{*}{Multi} & Base & \textit{\underline{\smash{90.5$\pm$3.2}}} & 91.1$\pm$3.0 & \textbf{99.9$\pm$0.1} & 29.8$\pm$26.4 & 1.1$\pm$0.9 & 4.5$\pm$4.0 \\\cline{4-10}
    & & & CombReg & \textbf{90.7$\pm$3.0} & 90.2$\pm$3.6 & \textbf{99.9$\pm$0.1} & 21.8$\pm$21.0 & \textbf{0.8$\pm$0.6} & \textit{\underline{\smash{4.4$\pm$2.2}}} \\\cline{2-10}
    & \multirow{2}{*}{\rotatebox[origin=c]{90}{Res}} & \multirow{2}{*}{Multi} & Base & 90.1$\pm$3.5 & 90.4$\pm$3.1 & \textbf{99.9$\pm$0.1} & 23.7$\pm$22.6 & \textit{\underline{\smash{1.0$\pm$1.2}}} & \textbf{3.5$\pm$3.7} \\\cline{4-10}
    & & & CombReg & \textbf{90.7$\pm$3.0} & 90.7$\pm$3.6 & \textbf{99.9$\pm$0.1} & \textbf{19.3$\pm$14.2} & \textbf{0.8$\pm$0.5} & \textbf{3.5$\pm$3.4} \\
    \hline
  
\end{tabular}
\label{tab:quantitative_results}
\end{table*}

Results obtained using the Att-UNet architecture demonstrated that the segmentation method based on a multi-class model with proposed combined regularization achieved the best results on all metrics, except for sensitivity on ankle dataset (Table \ref{tab:quantitative_results_ablation_study}). For the ankle dataset, the method improved Dice ($+0.8\%$), specificity ($+0.1\%$), MSSD ($-2.5$ mm), ASSD ($-0.1$ mm) and RAVD ($-0.2\%$) metrics while remaining $4.2\%$ lower than the best in sensitivity metric. For shoulder examinations, the method outperformed other approaches in Dice ($+0.9\%$), sensitivity ($+0.7\%$), specificity ($+0.1\%$), MSSD ($-0.5$ mm), ASSD ($-0.2$ mm) and RAVD ($-3.4\%$).

Furthermore, the proposed CombReg$^{\text{Multi}}_{\text{Res-UNet}}$ method outperformed state-of-the-art pre-trained methods on all metrics, except for sensitivity on shoulder dataset (Table \ref{tab:quantitative_results}). All methods reached excellent specificity scores ($>99.7\%$). For ankle examinations, the proposed CombReg$^{\text{Multi}}_{\text{Res-UNet}}$ method ranked best in Dice ($94.3\%$), sensitivity ($93.6\%$), MSSD ($6.1$ mm), ASSD ($0.6$ mm) and RAVD ($5.1\%$) metrics. For the shoulder dataset, the proposed CombReg$^{\text{Multi}}_{\text{Res-UNet}}$ method achi\-eved the best results in Dice ($90.7\%$), MSSD ($19.3$ mm), ASSD ($0.8$ mm) and RAVD ($3.5\%$) metrics while remaining marginally lower in sensitivity ($0.4\%$ lower than the best). It is also worth mentioning that, while the performance improvements are lower than in Att-UNet experiments (Table \ref{tab:quantitative_results_ablation_study}), the proposed combined regularization consistently improved performance across all architectures and metrics except for VGG-UNet and Dense-UNet shoulder sensitivity, and the proposed  CombReg$^{\text{Multi}}_{\text{Res-UNet}}$ method was associated with the lowest variance in all metrics except for ankle MSSD and shoulder sensitivity and RAVD. Finally, the statistical analysis performed on 2D slices using Dice, sensitivity and specificity metrics (Table \ref{tab:statistical analysis}) indicated that the proposed CombReg$^{\text{Multi}}_{\text{Res-UNet}}$ model produced significant improvements ($p$-values $< 0.05$).

\subsection{Rankings}
\label{sec:rankings}

\begin{table*}[t]
\centering
\caption{Scores of the four backbone architectures: Att-UNet \cite{oktay_attention_2018}, VGG-UNet \cite{simonyan_very_2015}, Dense-UNet \cite{huang_densely_2017} and Res-UNet \cite{he_deep_2016} on ankle and shoulder datasets. Regularization methods include: baseline, shape priors \cite{oktay_anatomically_2018}, adversarial \cite{singh_breast_2020} and proposed combined; while bone segmentation strategies comprise: individual, global and multi. Results encompass mean, standard deviation (STD) and associated rank. Methods were ranked according to their mean score. Best results are in bold.}
\begin{tabular}{||P{.3cm}|P{.95cm}|P{1.7cm}||P{2cm}|P{1cm}||P{2cm}|P{1cm}||}
    \hline
    \multicolumn{3}{||c||}{\multirow{2}{*}{Method}} & \multicolumn{2}{c||}{Ankle Dataset} & \multicolumn{2}{c||}{Shoulder Dataset} \\\cline{4-7}
    \multicolumn{3}{||c||}{} & Mean $\pm$ STD & Rank & Mean $\pm$ STD & Rank  \\
    \hline
    \hline
    \multirow{12}{*}{\rotatebox[origin=c]{90}{Att-UNet}} & \multirow{4}{*}{Indiv} & Base & 33.3$\pm$16.2 & 18 & 25.2$\pm$18.8 & 18 \\\cline{3-7}
    & & ShapeReg & 40.4$\pm$10.9 & 16 & 33.0$\pm$23.9 & 15 \\\cline{3-7}
    & & AdvReg & 37.7$\pm$13.2 & 17 & 31.9$\pm$16.3 & 16 \\\cline{3-7}
    & & CombReg & 44.2$\pm$16.3 & 15 & 36.8$\pm$21.3 & 14 \\\cline{2-7}
    & \multirow{4}{*}{Global} & Base & 46.2$\pm$18.3 & 14 & 28.7$\pm$20.4 & 17 \\\cline{3-7}
    & & ShapeReg & 47.4$\pm$20.7 & 13 & 37.2$\pm$22.6 & 12 \\\cline{3-7}
    & & AdvReg & 47.6$\pm$17.2 & 12 & 37.2$\pm$18.3 & 13 \\\cline{3-7}
    & & CombReg & 50.9$\pm$12.7 & 9 & 40.9$\pm$18.6 & 11 \\\cline{2-7}
    & \multirow{4}{*}{Multi} & Base & 47.9$\pm$21.9 & 11 & 41.6$\pm$23.3 & 10 \\\cline{3-7}
    & & ShapeReg & 54.4$\pm$18.5 & 8 & 42.4$\pm$20.3 & 9 \\\cline{3-7}
    & & AdvReg & 49.3$\pm$15.7 & 10 & 42.5$\pm$23.4 & 8 \\\cline{3-7}
    & & CombReg & 56.7$\pm$16.2 & 7 & 48.7$\pm$21.3 & 7 \\\hline
    \multirow{2}{*}{\rotatebox[origin=c]{90}{VGG}} & \multirow{2}{*}{Multi} & Base & 63.0$\pm$7.8 & 6 & 53.7$\pm$18.4 & 6 \\\cline{3-7}
    & & CombReg & 66.5$\pm$9.2 & 4 & 54.6$\pm$16.0 & 5 \\\hline
    \multirow{2}{*}{\rotatebox[origin=c]{90}{\small{Dense}}} & \multirow{2}{*}{Multi} & Base & 65.2$\pm$13.8 & 5 & 55.2$\pm$12.0 & 4 \\\cline{3-7}
    & & CombReg & 67.7$\pm$13.3 & 3 & 56.4$\pm$10.6 & 3 \\\hline
    \multirow{2}{*}{\rotatebox[origin=c]{90}{Res}} & \multirow{2}{*}{Multi} & Base & 68.5$\pm$14.2 & 2 & 56.7$\pm$12.9 & 2 \\\cline{3-7}
    & & CombReg & \textbf{71.4$\pm$10.0} & \textbf{1} & \textbf{59.2$\pm$13.9} & \textbf{1} \\
    \hline
  
\end{tabular}
\label{tab:rankings}
\end{table*}

CombReg$^{\text{Multi}}_{\text{Res-UNet}}$ ranked first in performance (Table \ref{tab:rankings}) for both datasets with mean scores of 71.4 on ankle dataset and 59.2 on shoulder dataset. Baseline Res-UNet ranked second on both datasets, while individual-class baseline Att-UNet ranked last on ankle (mean score of 33.3) and shoulder (mean score of 25.2) datasets. It was observed in the experiments based on Att-UNet architecture that for a fixed regularization scheme, the multi-class strategy outperformed the global-class strategy, which in turn outranked the individual-class scheme. Furthermore, for a fixed bone segmentation strategy, shape priors based and adversarial regularization improved the baseline performance, while a combined regularization resulted in the best overall performance. Additionally, the ranks achieved by the pre-trained architectures (VGG-UNet, Dense-UNet and Res-UNet) further demonstr\-ated that the proposed combined regularization promoted better performance as compared to baseline training. Hence, the combined regularization consistently outperformed compared the methods covering various segmentation strategies (individual, global, and multi) and distinct architectures (Att-UNet, VGG-UNet, Dense-UNet, and Res-UNet) demonstrating the effectiveness of the proposed approach. Finally, to assess the robustness of our ranking system and these observations, several threshold values were tested as reported in the supplementary materials and our conclusions remained unchanged on every transformed rankings.

\subsection{Qualitative assessment}
\label{sec:qualitative_assessment}

\begin {figure*}[ht!]
\centering
\begin{adjustbox}{width=\textwidth}
\begin{tikzpicture}
\begin{scope}[spy using outlines=
      {circle, magnification=3, size=.4cm, connect spies, rounded corners}]

\node[inner sep=0pt]  at (0,0)
    {\includegraphics[width=.08125\textwidth]{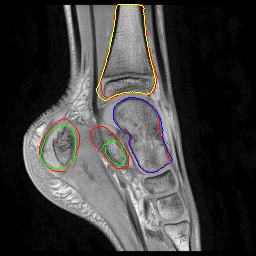}};
\node[inner sep=0pt]  at (3.25,0)
    {\includegraphics[width=.08125\textwidth]{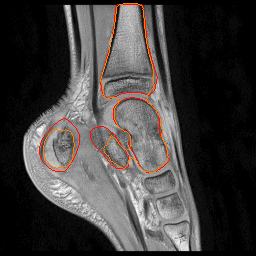}};
\node[inner sep=0pt]  at (6.5,0)
    {\includegraphics[width=.08125\textwidth]{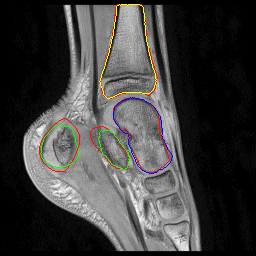}};

\node[inner sep=0pt] at (0,-1.7)
    {\includegraphics[width=.08125\textwidth]{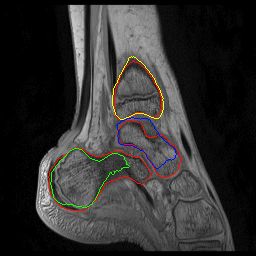}};
\node[inner sep=0pt]  at (3.25,-1.7)
    {\includegraphics[width=.08125\textwidth]{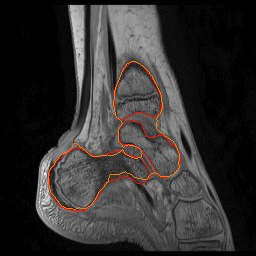}};
\node[inner sep=0pt]  at (6.5,-1.7)
    {\includegraphics[width=.08125\textwidth]{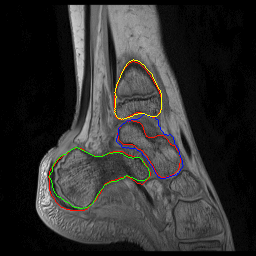}};

\node[inner sep=0pt] at (1.5,0)
    {\includegraphics[width=.08125\textwidth]{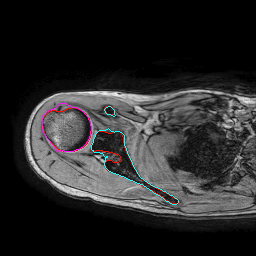}};
\node[inner sep=0pt]  at (4.75,0)
    {\includegraphics[width=.08125\textwidth]{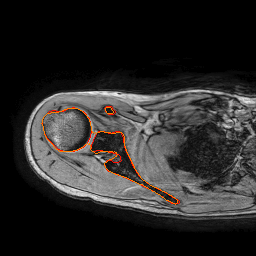}};
\node[inner sep=0pt]  at (8,0)
    {\includegraphics[width=.08125\textwidth]{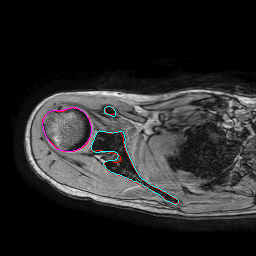}};
    
\node[inner sep=0pt] at (1.5,-1.7)
    {\includegraphics[width=.08125\textwidth]{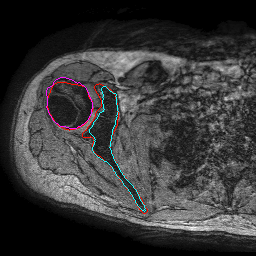}};
\node[inner sep=0pt]  at (4.75,-1.7)
    {\includegraphics[width=.08125\textwidth]{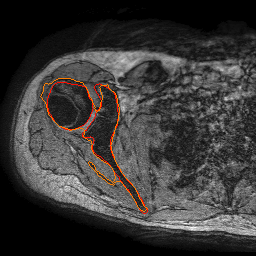}};
\node[inner sep=0pt]  at (8,-1.7)
    {\includegraphics[width=.08125\textwidth]{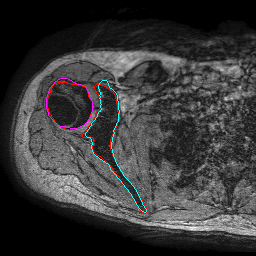}};

\spy [Dandelion] on (-.14, .20) in node [left] at (-.27,.47);
\spy [Dandelion] on (3.11,.20) in node [left] at (2.98,.47);
\spy [Dandelion] on (6.36,.20) in node [left] at (6.23,.47);

\spy [Dandelion] on (.01,-.14) in node [left] at (.67,-.47);
\spy [Dandelion] on (3.26,-.14) in node [left] at (3.92,-.47);
\spy [Dandelion] on (6.51,-.14) in node [left] at (7.17,-.47);

\spy [Dandelion] on (-.37,-1.89) in node [left] at (-.27,-1.23);
\spy [Dandelion] on (2.88,-1.89) in node [left] at (2.98,-1.23);
\spy [Dandelion] on (6.13,-1.89) in node [left] at (6.23,-1.23);

\spy [Dandelion] on (0.17,-1.62) in node [left] at (.67,-1.23);
\spy [Dandelion] on (3.42,-1.62) in node [left] at (3.92,-1.23);
\spy [Dandelion] on (6.67,-1.62) in node [left] at (7.17,-1.23);

\spy [Dandelion] on (.15,-1.94) in node [left] at (.67,-2.17);
\spy [Dandelion] on (3.4,-1.94) in node [left] at (3.92,-2.17);
\spy [Dandelion] on (6.65,-1.94) in node [left] at (7.17,-2.17);

\spy [Dandelion] on (1.14,.11) in node [left] at (1.23,.47);
\spy [Dandelion] on (4.39,.11) in node [left] at (4.48,.47);
\spy [Dandelion] on (7.64,.11) in node [left] at (7.73,.47);

\spy [Dandelion] on (1.42,-.16) in node [left] at (1.23,-.47);
\spy [Dandelion] on (4.67,-.16) in node [left] at (4.48,-.47);
\spy [Dandelion] on (7.92,-.16) in node [left] at (7.73,-.47);

\spy [Dandelion] on (1.4,.1) in node [left] at (2.17,.47);
\spy [Dandelion] on (4.65,.1) in node [left] at (5.42,.47);
\spy [Dandelion] on (7.9,.1) in node [left] at (8.67,.47);

\spy [Dandelion] on (1.15,-1.45) in node [left] at (2.17,-1.23);
\spy [Dandelion] on (4.4,-1.45) in node [left] at (5.42,-1.23);
\spy [Dandelion] on (7.65,-1.45) in node [left] at (8.67,-1.23);

\spy [Dandelion] on (1.26,-1.72) in node [left] at (1.23,-2.17);
\spy [Dandelion] on (4.51,-1.72) in node [left] at (4.48,-2.17);
\spy [Dandelion] on (7.76,-1.72) in node [left] at (7.73,-2.17);

\end{scope}


\node[inner sep=0pt] at (.75, .85) {\scalebox{.5}{CombReg$_{\text{Att-UNet}}^{\text{Indiv}}$}};
\node[inner sep=0pt] at (4, .85) {\scalebox{.5}{CombReg$_{\text{Att-UNet}}^{\text{Global}}$}};
\node[inner sep=0pt] at (7.25, .85) {\scalebox{.5}{CombReg$_{\text{Att-UNet}}^{\text{Multi}}$}};

\node[inner sep=0pt] at (0, -.85) {\scalebox{.5}{$A_{H,1}$}};
\node[inner sep=0pt] at (3.25, -.85) {\scalebox{.5}{$A_{H,1}$}};
\node[inner sep=0pt] at (6.5, -.85) {\scalebox{.5}{$A_{H,1}$}};

\node[inner sep=0pt] at (0, -2.55) {\scalebox{.5}{$A_{P,3}$}};
\node[inner sep=0pt] at (3.25, -2.55) {\scalebox{.5}{$A_{P,3}$}};
\node[inner sep=0pt] at (6.5, -2.55) {\scalebox{.5}{$A_{P,3}$}};

\node[inner sep=0pt] at (1.5, -.85) {\scalebox{.5}{$S_{H,3}$}};
\node[inner sep=0pt] at (4.75, -.85) {\scalebox{.5}{$S_{H,3}$}};
\node[inner sep=0pt] at (8, -.85) {\scalebox{.5}{$S_{H,3}$}};

\node[inner sep=0pt] at (1.5, -2.55) {\scalebox{.5}{$S_{P,6}$}};
\node[inner sep=0pt] at (4.75, -2.55) {\scalebox{.5}{$S_{P,6}$}};
\node[inner sep=0pt] at (8, -2.55) {\scalebox{.5}{$S_{P,6}$}};

\end{tikzpicture}
\end{adjustbox}
\caption{\textbf{Visual comparison of bone segmentation strategies using Att-UNet with combined regularization.} Automatic segmentation of ankle and shoulder bones based on Att-UNet with combined regularization using individual-class, global-class and multi-class strategies. Ground truth delineations are in red (\textcolor{red}{\raisebox{1.8pt}{\rule{5pt}{1pt}}}) while predicted bones comprising calcaneus, talus, tibia, humerus and scapula appear in green (\textcolor{green}{\raisebox{1.8pt}{\rule{5pt}{1pt}}}), blue (\textcolor{blue}{\raisebox{1.8pt}{\rule{5pt}{1pt}}}), yellow (\textcolor{yellow}{\raisebox{1.8pt}{\rule{5pt}{1pt}}}), magenta (\textcolor{magenta}{\raisebox{1.8pt}{\rule{5pt}{1pt}}}) and cyan (\textcolor{cyan}{\raisebox{1.8pt}{\rule{5pt}{1pt}}}) respectively. Predicted global bone class is in orange (\textcolor{orange}{\raisebox{1.8pt}{\rule{5pt}{1pt}}}).}
\label{fig:comparison_bone_structures}
\end{figure*}

We first visually compared the combined regularization method using individual-class, global-class, and multi-class Att-UNet models to assess the anatomical validity of the segmentation prediction (Figure \ref{fig:comparison_bone_structures}). The individual-class Att-UNet models produced masks based on weights specific to each bone, the global-class Att-UNet models exploited shared features between bones, and multi-class Att-UNet models utilized both specific and shared bone features. It was observed that the global-class mask predictions included fused bone errors in both ankle and shoulder datasets ($A_{H,1}$, $A_{P,3}$, $S_{H,3}$ and $S_{P,6}$ examinations). Thus, exploiting specific bone annotations was necessary to prevent fused-bone errors in predicted delineations. Moreover, shared feature learning in the global-class strategy enforced more accurate delineations ($A_{H,1}$ and $S_{H,3}$). Hence, multi-class Att-UNet leveraged the benefits of simultaneously learning specific and sha\-red bone features and avoided fused-bones in estimated segmentation while producing precise delineations.

\begin {figure*}[ht!]
\centering
\begin{adjustbox}{width=\textwidth}
\begin{tikzpicture}
\begin{scope}[spy using outlines=
      {circle, magnification=3, size=.4cm, connect spies, rounded corners}]

\node[inner sep=0pt]  at (0,0)
    {\includegraphics[width=.08125\textwidth]{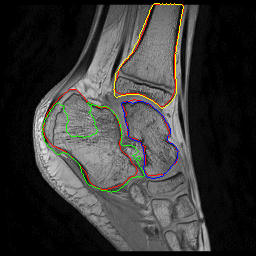}};
\node[inner sep=0pt]  at (0,-1.7)
    {\includegraphics[width=.08125\textwidth]{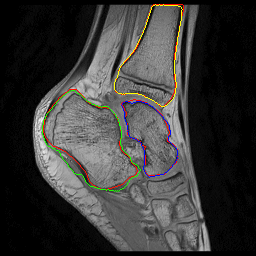}};
\node[inner sep=0pt]  at (4.75,0)
    {\includegraphics[width=.08125\textwidth]{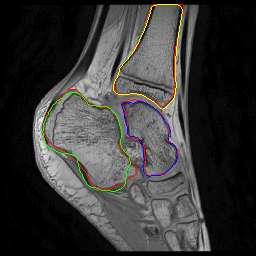}};
\node[inner sep=0pt] at (4.75,-1.7)
    {\includegraphics[width=.08125\textwidth]{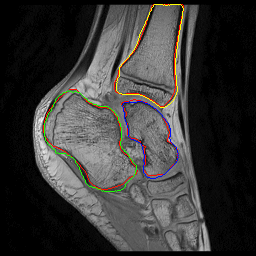}};

\node[inner sep=0pt] at (1.5,0)
    {\includegraphics[width=.08125\textwidth]{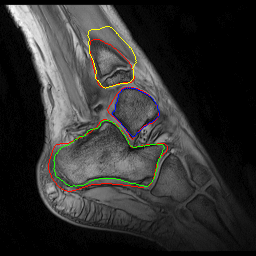}};
\node[inner sep=0pt]  at (1.5,-1.7)
    {\includegraphics[width=.08125\textwidth]{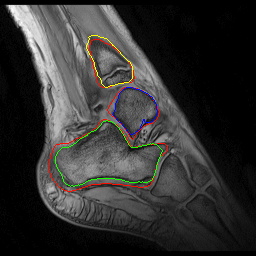}};
\node[inner sep=0pt]  at (6.25,0)
    {\includegraphics[width=.08125\textwidth]{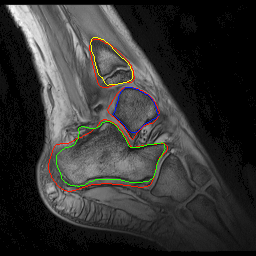}};
\node[inner sep=0pt] at (6.25,-1.7)
    {\includegraphics[width=.08125\textwidth]{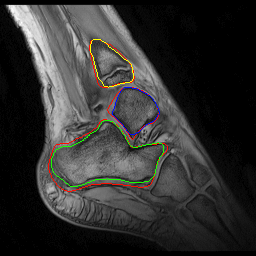}};
    
\node[inner sep=0pt] at (3,0)
    {\includegraphics[width=.08125\textwidth]{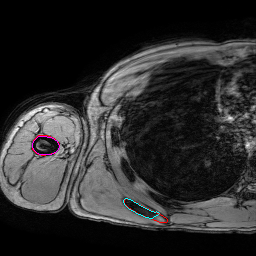}};
\node[inner sep=0pt]  at (3,-1.7)
    {\includegraphics[width=.08125\textwidth]{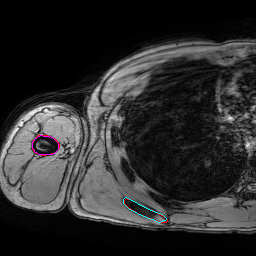}};
\node[inner sep=0pt]  at (7.75,0)
    {\includegraphics[width=.08125\textwidth]{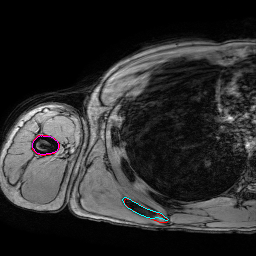}};
\node[inner sep=0pt] at (7.75,-1.7)
    {\includegraphics[width=.08125\textwidth]{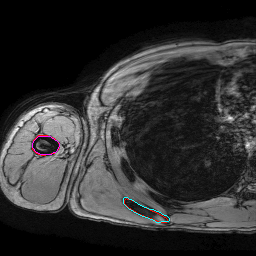}};

\spy [Dandelion] on (-0.31,.19) in node [left] at (-.27,.47);
\spy [Dandelion] on (-0.31,-1.51) in node [left] at (-.27,-1.23);
\spy [Dandelion] on (4.44,.19) in node [left] at (4.48,.47);
\spy [Dandelion] on (4.44,-1.51) in node [left] at (4.48,-1.23);

\spy [Dandelion] on (-.05,.2) in node [left] at (.67,.47);
\spy [Dandelion] on (-.05,-1.5) in node [left] at (.67,-1.23);
\spy [Dandelion] on (4.7,.2) in node [left] at (5.42,.47);
\spy [Dandelion] on (4.7,-1.5) in node [left] at (5.45,-1.23);

\spy [Dandelion] on (.08,-.24) in node [left] at (.67,-.47);
\spy [Dandelion] on (.08,-1.94) in node [left] at (.67,-2.17);
\spy [Dandelion] on (4.83,-.24) in node [left] at (5.42,-.47);
\spy [Dandelion] on (4.83,-1.94) in node [left] at (5.42,-2.17);

\spy [Dandelion] on (1.32,.48) in node [left] at (2.17,.47);
\spy [Dandelion] on (1.32,-1.22) in node [left] at (2.17,-1.23);
\spy [Dandelion] on (6.07,.48) in node [left] at (6.92,.47);
\spy [Dandelion] on (6.07,-1.22) in node [left] at (6.92,-1.23);

\spy [Dandelion] on (1.69,-0.12) in node [left] at (2.17,-.47);
\spy [Dandelion] on (1.69,-1.82) in node [left] at (2.17,-2.17);
\spy [Dandelion] on (6.44,-0.12) in node [left] at (6.92,-.47);
\spy [Dandelion] on (6.44,-1.82) in node [left] at (6.92,-2.17);

\spy [Dandelion] on (3.18,-.49) in node [left] at (3.67,.47);
\spy [Dandelion] on (3.18,-2.19) in node [left] at (3.67,-1.23);
\spy [Dandelion] on (7.93,-.49) in node [left] at (8.42,.47);
\spy [Dandelion] on (7.93,-2.19) in node [left] at (8.42,-1.23);

\end{scope}

\node[inner sep=0pt,rotate=90] at (-.85, 0) {\scalebox{.5}{Base$_{\text{Att-UNet}}^{\text{Multi}}$}};
\node[inner sep=0pt,rotate=90] at (-.85, -1.7) {\scalebox{.5}{ShapeReg$_{\text{Att-UNet}}^{\text{Multi}}$}};
\node[inner sep=0pt,rotate=90] at (3.9, 0) {\scalebox{.5}{AdvReg$_{\text{Att-UNet}}^{\text{Multi}}$}};
\node[inner sep=0pt,rotate=90] at (3.9, -1.7) {\scalebox{.5}{CombReg$_{\text{Att-UNet}}^{\text{Multi}}$}};

\node[inner sep=0pt] at (0, -.85) {\scalebox{.5}{$A_{P,4}$}};
\node[inner sep=0pt] at (0, -2.55) {\scalebox{.5}{$A_{P,4}$}};
\node[inner sep=0pt] at (4.75, -.85) {\scalebox{.5}{$A_{P,4}$}};
\node[inner sep=0pt] at (4.75, -2.55) {\scalebox{.5}{$A_{P,4}$}};

\node[inner sep=0pt] at (1.5, -.85) {\scalebox{.5}{$A_{H,5}$}};
\node[inner sep=0pt] at (1.5, -2.55) {\scalebox{.5}{$A_{H,5}$}};
\node[inner sep=0pt] at (6.25, -.85) {\scalebox{.5}{$A_{H,5}$}};
\node[inner sep=0pt] at (6.25, -2.55) {\scalebox{.5}{$A_{H,5}$}};

\node[inner sep=0pt] at (3, -.85) {\scalebox{.5}{$S_{H,1}$}};
\node[inner sep=0pt] at (3, -2.55) {\scalebox{.5}{$S_{H,1}$}};
\node[inner sep=0pt] at (7.75, -.85) {\scalebox{.5}{$S_{H,1}$}};
\node[inner sep=0pt] at (7.75, -2.55) {\scalebox{.5}{$S_{H,1}$}};

\end{tikzpicture}
\end{adjustbox}
\caption{\textbf{Visual comparison of regularizations methods using Att-UNet with multi-structure strategy.} Automatic segmentation of ankle and shoulder bones based on Att-UNet \cite{oktay_attention_2018} with multi-structure strategy using baseline, shape priors \cite{oktay_anatomically_2018}, adversarial \cite{singh_breast_2020} and combined regularizations. Ground truth delineations are in red (\textcolor{red}{\raisebox{1.8pt}{\rule{5pt}{1pt}}}) while predicted bones comprising calcaneus, talus, tibia, humerus and scapula appear in green (\textcolor{green}{\raisebox{1.8pt}{\rule{5pt}{1pt}}}), blue (\textcolor{blue}{\raisebox{1.8pt}{\rule{5pt}{1pt}}}), yellow (\textcolor{yellow}{\raisebox{1.8pt}{\rule{5pt}{1pt}}}), magenta (\textcolor{magenta}{\raisebox{1.8pt}{\rule{5pt}{1pt}}}) and cyan (\textcolor{cyan}{\raisebox{1.8pt}{\rule{5pt}{1pt}}}) respectively.}
\label{fig:comparison_regularization}
\end{figure*}
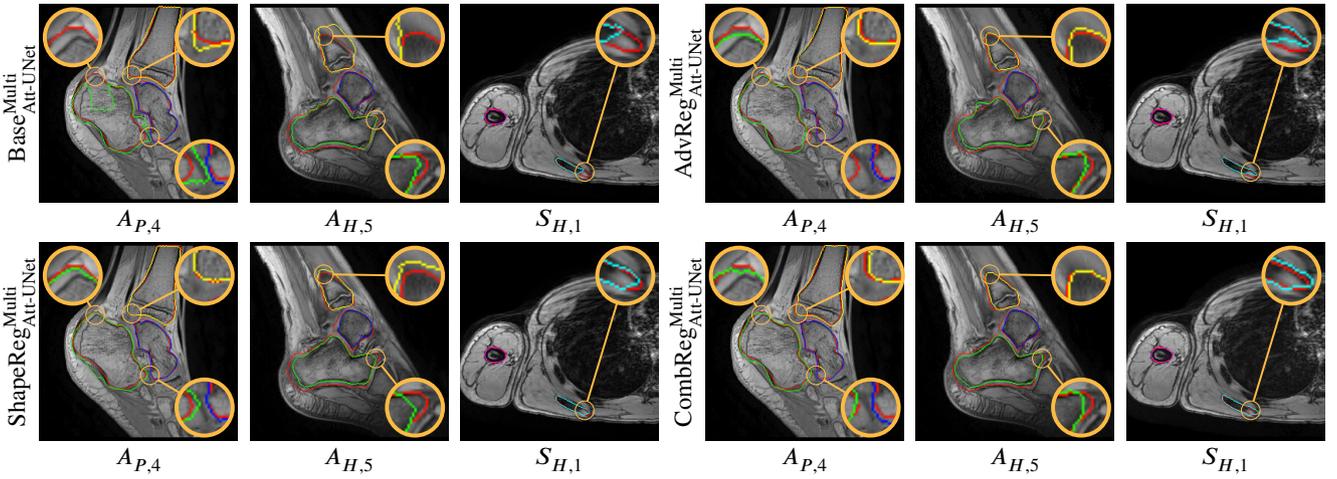

Visual comparison of the four regularization approaches (baseline, shape priors, adversarial and the proposed combined regularizations) provided the visual evidence on stepwise improvements in segmentation quality from baseline to combined regularization (Figure \ref{fig:comparison_regularization}). It was clearly observed that each additional regularization improved the segmentation prediction over baseline Att-UNet. Furthermore, baseline Att-UNet did not segment the complete non-ossified area of the scapula, contrary to the compared regularized methods which incorporated prior knowledge ($S_{H,1}$). More specifically, the shape regularization enforced the model to follow the learnt shape representation and promoted smoother bone delineations ($A_{P,4}$), while the adversarial regularization encouraged the model to generate more realistic mask and incited more precise bone delineations ($A_{H,5}$). Meanwhile, the proposed combined regularization fostered the advantages of both former regularizations and provided smoother and more realistic bone extraction ($A_{P,4}$, $A_{H,5}$ and $S_{H,1}$).

\begin {figure*}[ht!]
\centering
\begin{adjustbox}{width=\textwidth}
\begin{tikzpicture}
\begin{scope}[spy using outlines=
      {circle, magnification=3, size=.4cm, connect spies, rounded corners}]

\node[inner sep=0pt]  at (0,0)
    {\includegraphics[width=.08125\textwidth]{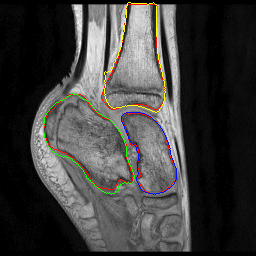}};
\node[inner sep=0pt]  at (3.25,0)
    {\includegraphics[width=.08125\textwidth]{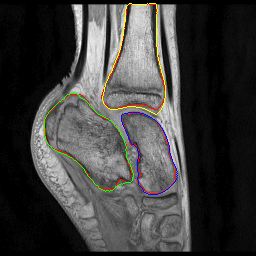}};
\node[inner sep=0pt]  at (6.5,0)
    {\includegraphics[width=.08125\textwidth]{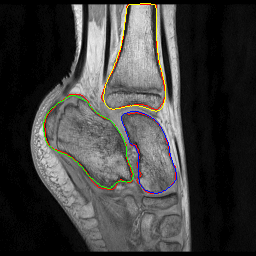}};

\node[inner sep=0pt] at (0,-1.7)
    {\includegraphics[width=.08125\textwidth]{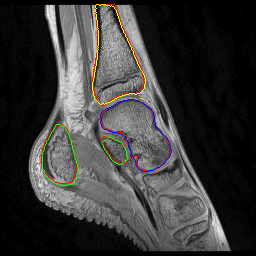}};
\node[inner sep=0pt]  at (3.25,-1.7)
    {\includegraphics[width=.08125\textwidth]{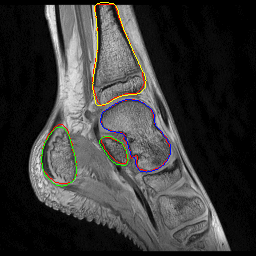}};
\node[inner sep=0pt]  at (6.5,-1.7)
    {\includegraphics[width=.08125\textwidth]{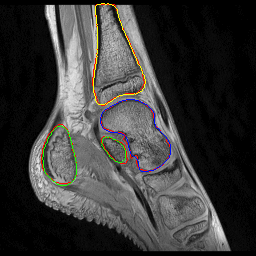}};

\node[inner sep=0pt] at (1.5,0)
    {\includegraphics[width=.08125\textwidth]{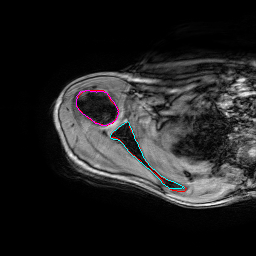}};
\node[inner sep=0pt]  at (4.75,0)
    {\includegraphics[width=.08125\textwidth]{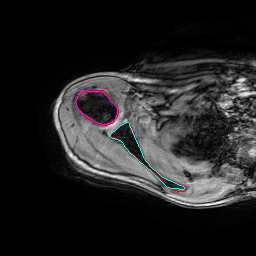}};
\node[inner sep=0pt]  at (8,0)
    {\includegraphics[width=.08125\textwidth]{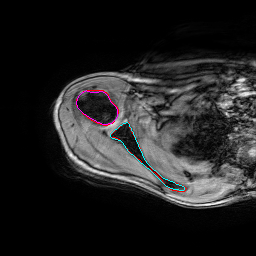}};
    
\node[inner sep=0pt] at (1.5,-1.7)
    {\includegraphics[width=.08125\textwidth]{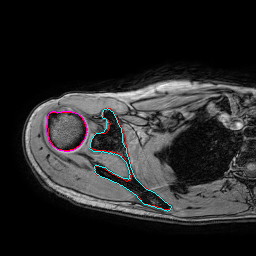}};
\node[inner sep=0pt]  at (4.75,-1.7)
    {\includegraphics[width=.08125\textwidth]{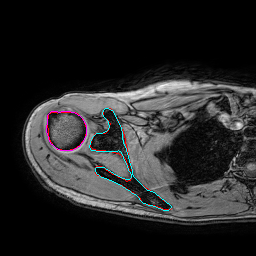}};
\node[inner sep=0pt]  at (8,-1.7)
    {\includegraphics[width=.08125\textwidth]{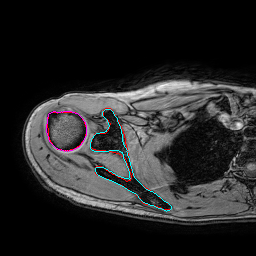}};

\spy [Dandelion] on (-.11,.18) in node [left] at (.67,.47);
\spy [Dandelion] on (3.14,.18) in node [left] at (3.92,.47);
\spy [Dandelion] on (6.39,.18) in node [left] at (7.17,.47);

\spy [Dandelion] on (-.4,-.11) in node [left] at (-.27,-.47);
\spy [Dandelion] on (2.85,-.11) in node [left] at (2.98,-.47);
\spy [Dandelion] on (6.1,-.11) in node [left] at (6.23,-.47);

\spy [Dandelion] on (-0.12,-1.78) in node [left] at (.67,-1.23);
\spy [Dandelion] on (3.13,-1.78) in node [left] at (3.92,-1.23);
\spy [Dandelion] on (6.38,-1.78) in node [left] at (7.17,-1.23);

\spy [Dandelion] on (0,-1.88) in node [left] at (.67,-2.17);
\spy [Dandelion] on (3.25,-1.88) in node [left] at (3.92,-2.17);
\spy [Dandelion] on (6.5,-1.88) in node [left] at (7.17,-2.17);

\spy [Dandelion] on (1.38,-1.69) in node [left] at (2.17,-1.23);
\spy [Dandelion] on (4.63,-1.69) in node [left] at (5.42,-1.23);
\spy [Dandelion] on (7.88,-1.69) in node [left] at (8.67,-1.23);

\spy [Dandelion] on (1.51,-1.98) in node [left] at (1.23,-2.17);
\spy [Dandelion] on (4.75,-1.98) in node [left] at (4.48,-2.17);
\spy [Dandelion] on (8,-1.98) in node [left] at (7.73,-2.17);

\spy [Dandelion] on (1.25,.18) in node [left] at (2.17,.47);
\spy [Dandelion] on (4.5,.18) in node [left] at (5.42,.47);
\spy [Dandelion] on (7.75,.18) in node [left] at (8.67,.47);

\spy [Dandelion] on (1.79,-.33) in node [left] at (1.23,-.47);
\spy [Dandelion] on (5.04,-.33) in node [left] at (4.48,-.47);
\spy [Dandelion] on (8.29,-.33) in node [left] at (7.73,-.47);

\end{scope}


\node[inner sep=0pt] at (.75, .85) {\scalebox{.5}{CombReg$_{\text{VGG-UNet}}^{\text{Multi}}$}};
\node[inner sep=0pt] at (4, .85) {\scalebox{.5}{CombReg$_{\text{Dense-UNet}}^{\text{Multi}}$}};
\node[inner sep=0pt] at (7.25, .85) {\scalebox{.5}{CombReg$_{\text{Res-UNet}}^{\text{Multi}}$}};

\node[inner sep=0pt] at (0, -.85) {\scalebox{.5}{$A_{P,1}$}};
\node[inner sep=0pt] at (3.25, -.85) {\scalebox{.5}{$A_{P,1}$}};
\node[inner sep=0pt] at (6.5, -.85) {\scalebox{.5}{$A_{P,1}$}};

\node[inner sep=0pt] at (0, -2.55) {\scalebox{.5}{$A_{P,2}$}};
\node[inner sep=0pt] at (3.25, -2.55) {\scalebox{.5}{$A_{P,2}$}};
\node[inner sep=0pt] at (6.5, -2.55) {\scalebox{.5}{$A_{P,2}$}};

\node[inner sep=0pt] at (1.5, -.85) {\scalebox{.5}{$S_{H,2}$}};
\node[inner sep=0pt] at (4.75, -.85) {\scalebox{.5}{$S_{H,2}$}};
\node[inner sep=0pt] at (8, -.85) {\scalebox{.5}{$S_{H,2}$}};

\node[inner sep=0pt] at (1.5, -2.55) {\scalebox{.5}{$S_{H,4}$}};
\node[inner sep=0pt] at (4.75, -2.55) {\scalebox{.5}{$S_{H,4}$}};
\node[inner sep=0pt] at (8, -2.55) {\scalebox{.5}{$S_{H,4}$}};

\end{tikzpicture}
\end{adjustbox}
\caption{\textbf{Visual comparison of pre-trained architectures using multi-class strategy with combined regularization.} Automatic segmentation of ankle and shoulder bones based on VGG-UNet \cite{simonyan_very_2015}, Dense-UNet \cite{huang_densely_2017} and Res-UNet \cite{he_deep_2016} using combined regularization with multi-class strategy. Ground truth delineations are in red (\textcolor{red}{\raisebox{1.8pt}{\rule{5pt}{1pt}}}) while predicted bones comprising calcaneus, talus, tibia, humerus and scapula appear in green (\textcolor{green}{\raisebox{1.8pt}{\rule{5pt}{1pt}}}), blue (\textcolor{blue}{\raisebox{1.8pt}{\rule{5pt}{1pt}}}), yellow (\textcolor{yellow}{\raisebox{1.8pt}{\rule{5pt}{1pt}}}), magenta (\textcolor{magenta}{\raisebox{1.8pt}{\rule{5pt}{1pt}}}) and cyan (\textcolor{cyan}{\raisebox{1.8pt}{\rule{5pt}{1pt}}}) respectively. Predicted global bone class is in orange (\textcolor{orange}{\raisebox{1.8pt}{\rule{5pt}{1pt}}}).}
\label{fig:comparison_pre-trained}
\end{figure*}
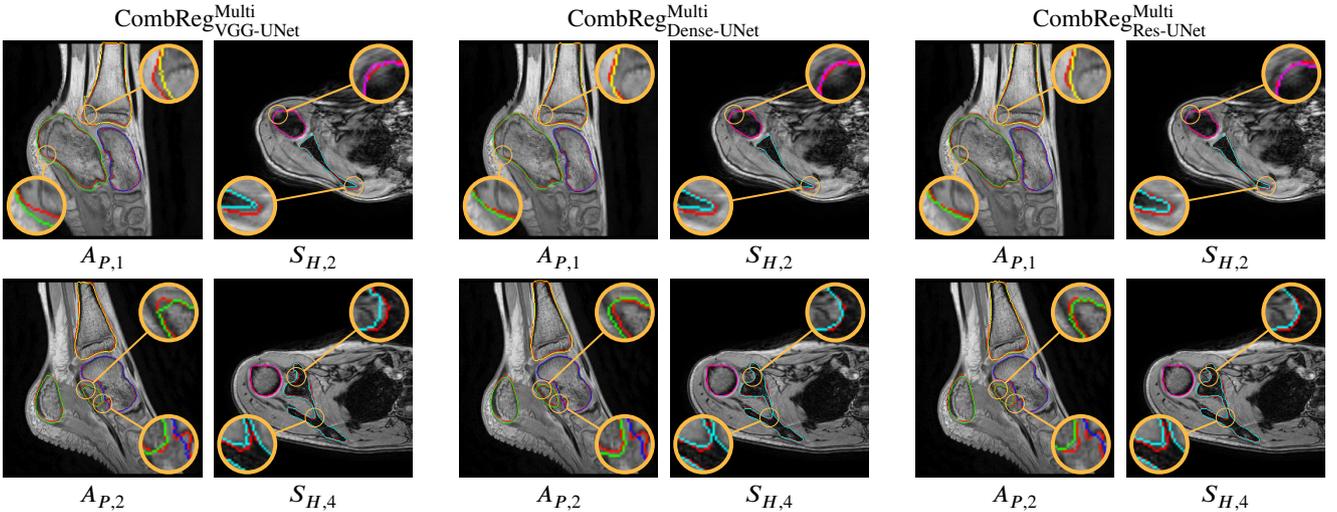

Visual comparison of the pre-trained models (VGG-UNet, Dense-UNet and Res-UNet) demonstrated that networks benefiting from transfer learning produced highly accurate delineations and captured complex bone shapes (Figure \ref{fig:comparison_pre-trained}). The qualitative results further confirmed the advantages of employing networks pre-trained on large non-medical databases along with a combination of shape priors and adversarial regularization to train more generalizable models on scarce pediatric datasets ($A_{P,2}$ and $S_{H,4}$). Most importantly, the proposed pre-trained CombReg$_{\text{Res-UNet}}^{\text{Multi}}$ model together with regularization approaches effectively segmented non ossified areas in addition to ossified bones dealing efficiently with the corresponding intensity variations within a single bone structure ($A_{P,1}$ and $S_{H,2}$). This outcome can be seen as a crucial need for the image analysis of pediatric musculoskeletal systems.

\section{Discussion}
\subsection{Segmentation performance}

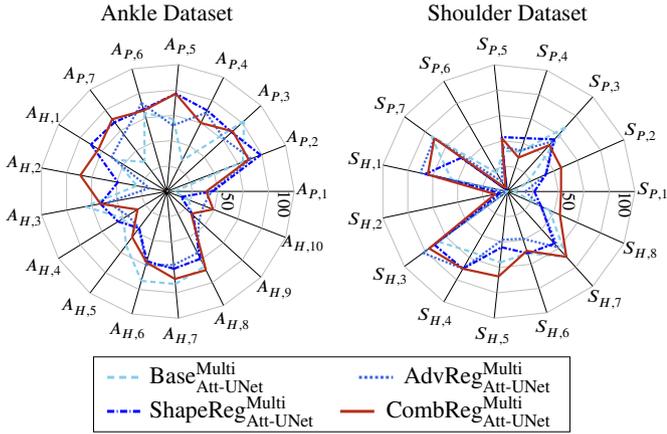
\begin{figure}[t]
\centering
\begin{adjustbox}{width=\textwidth}
\hspace{-1cm}
\begin{minipage}{.8\textwidth}
\begin{tikzpicture}
  \centering

\tkzKiviatDiagram[radial  style/.style ={-},
        label space=.35,
        space = 0,
        gap = .5,  
        step = .5,
        lattice = 5]
        {$A_{P,1}$,$A_{P,2}$,$A_{P,3}$,$A_{P,4}$,$A_{P,5}$,$A_{P,6}$,$A_{P,7}$,$A_{H,1}$,$A_{H,2}$,$A_{H,3}$,$A_{H,4}$,$A_{H,5}$,$A_{H,6}$,$A_{H,7}$,$A_{H,8}$,$A_{H,9}$,$A_{H,10}$}
\tkzKiviatLine[densely dashed,thick,color=SkyBlue,line width=1.25](0.6, 6.4, 8.2, 2.7, 5.7, 6.4, 2.9, 4.7, 2.4, 6.5, 4.0, 5.3, 7.3, 7.3, 6.6, 2.6, 1.9)
\tkzKiviatLine[densely dotted,thick,color=RoyalBlue,line width=1.25](3.5, 7.1, 6.1, 6.9, 5.2, 7.2, 5.6, 5.1, 1.4, 4.7, 3.6, 3.6, 6.0, 5.8, 5.4, 3.4, 3.1) 
\tkzKiviatLine[densely dashdotted,thick,color=Blue,line width=1.25](3.6, 7.9, 6.9, 7.1, 7.7, 6.7, 6.8, 7.0, 3.9, 5.3, 4.5, 3.6, 5.7, 6.1, 5.9, 2.8, 1.2) 
\tkzKiviatLine[thick,color=BrickRed,line width=1.25](3.1, 6.9, 7.0, 6.0, 7.7, 6.6, 7.1, 6.3, 6.9, 5.3, 2.7, 4.5, 5.8, 6.9, 6.9, 2.6, 3.9)
\node at (0,3.5) {\scalebox{1.3}{Ankle Dataset}};
\node[rotate=270] at (1.2,-0.2) {50};
\node[rotate=270] at (2.3,-0.25) {100};
\end{tikzpicture}
\end{minipage}%
\begin{minipage}{.8\textwidth}
\begin{tikzpicture}
\centering

\tkzKiviatDiagram[radial  style/.style ={-},
        label space=.35,
        space = 0,
        gap = .5,  
        step = .5,
        lattice = 5]
        {$S_{P,1}$,$S_{P,2}$,$S_{P,3}$,$S_{P,4}$,$S_{P,5}$,$S_{P,6}$,$S_{P,7}$,$S_{H,1}$,$S_{H,2}$,$S_{H,3}$,$S_{H,4}$,$S_{H,5}$,$S_{H,6}$,$S_{H,7}$,$S_{H,8}$}
\tkzKiviatLine[densely dashed,thick,color=SkyBlue,line width=1.25](0.5, 1.7, 6.8, 3.3, 3.2, 1.7, 7.4, 4.5, 0.0, 6.7, 5.5, 5.6, 5.1, 6.9, 3.6)
\tkzKiviatLine[densely dotted,thick,color=RoyalBlue,line width=1.25](1.4, 3.3, 5.3, 3.4, 3.5, 0.9, 7.1, 7.0, 0.2, 8.2, 7.0, 3.8, 3.9, 5.9, 2.9) 
\tkzKiviatLine[densely dashdotted,thick,color=Blue,line width=1.25](2.1, 3.3, 5.5, 4.4, 4.3, 0.2, 4.5, 6.6, 0.6, 7.1, 7.0, 4.4, 5.1, 5.5, 2.9)
\tkzKiviatLine[thick,color=BrickRed,line width=1.25](4.2, 4.6, 4.9, 2.8, 4.2, 0.3, 7.1, 6.4, 1.0, 7.6, 7.0, 6.7, 4.9, 6.9, 4.5)
\node at (0,3.5) {\scalebox{1.3}{Shoulder Dataset}};
\node[rotate=270] at (1.2,-0.2) {50};
\node[rotate=270] at (2.3,-0.25) {100};

\end{tikzpicture}
\end{minipage}
\end{adjustbox}
\begin{adjustbox}{width=.75\textwidth}
\begin{tikzpicture}

\draw (-3.5,-3.45) -- (3.5,-3.45) -- (3.5,-4.55) -- (-3.5,-4.55) -- cycle;
\draw[densely dashed,line width=1.25,SkyBlue] (-3.3,-3.75) -- (-2.8,-3.75);
\node[anchor=west] at (-2.8,-3.75) {$\text{Base}^{\text{Multi}}_{\text{Att-UNet}}$};
\draw[densely dotted,line width=1.25,RoyalBlue] (.4,-3.75) -- (.9,-3.75);
\node[anchor=east] at (3.35,-3.75) {$\text{AdvReg}^{\text{Multi}}_{\text{Att-UNet}}$};
\draw[densely dashdotted, line width=1.25,Blue] (-3.3,-4.25) -- (-2.8,-4.25);
\node[anchor=west] at (-2.8,-4.25) {$\text{ShapeReg}^{\text{Multi}}_{\text{Att-UNet}}$};
\draw[line width=1.25,BrickRed] (.125,-4.25) -- (.625,-4.25);
\node[anchor=east] at (3.35,-4.25) {$\text{CombReg}^{\text{Multi}}_{\text{Att-UNet}}$};

\end{tikzpicture}
\end{adjustbox}
\caption{Spider graphs showing scores obtained within ankle and shoulder datasets based on multi-class strategy using Att-UNet with baseline \cite{oktay_attention_2018}, shape priors \cite{oktay_anatomically_2018}, adversarial \cite{singh_breast_2020} and proposed combined regularizations. Scores were computed for pathological $A_{P,1},...,A_{P,7}$ and healthy $A_{H,1},...,A_{H,10}$ ankles, as well as for pathological $S_{P,1},...,S_{P,7}$ and healthy $S_{H,1},...,S_{H,8}$ shoulders.}
\label{fig:spider_regularization}
\end{figure}

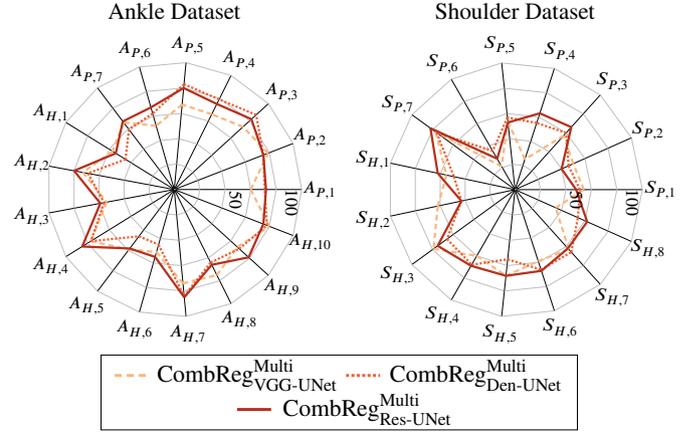
\begin{figure}[t]

\begin{adjustbox}{width=\textwidth}
\hspace{-1cm}
\begin{minipage}{.8\textwidth}
\begin{tikzpicture}
  \centering
\tkzKiviatDiagram[radial  style/.style ={-},
        label space=.35,
        space = 0,
        gap = .5,  
        step = .5,
        lattice = 5]
        {$A_{P,1}$,$A_{P,2}$,$A_{P,3}$,$A_{P,4}$,$A_{P,5}$,$A_{P,6}$,$A_{P,7}$,$A_{H,1}$,$A_{H,2}$,$A_{H,3}$,$A_{H,4}$,$A_{H,5}$,$A_{H,6}$,$A_{H,7}$,$A_{H,8}$,$A_{H,9}$,$A_{H,10}$}
\tkzKiviatLine[densely dashed,color=Apricot,line width=1.25](6.0, 7.9, 7.3, 6.6, 6.7, 5.2, 6.6, 5.7, 7.3, 5.4, 7.7, 5.9, 5.1, 7.4, 7.5, 6.7, 8.0)
\tkzKiviatLine[densely dotted,color=RedOrange,line width=1.25](7.1, 7.5, 8.7, 7.9, 8.3, 6.2, 5.9, 4.5, 6.9, 5.6, 7.7, 4.6, 4.5, 8.4, 6.4, 6.9, 7.8) 
\tkzKiviatLine[thick,color=BrickRed,line width=1.25](7.2, 7.5, 8.2, 7.5, 8.0, 6.7, 6.7, 5.4, 8.0, 5.9, 8.5, 5.8, 5.5, 8.5, 6.6, 7.9, 7.5)
\node at (0,3.5) {\scalebox{1.3}{Ankle Dataset}};
\node[rotate=270] at (1.2,-0.2) {50};
\node[rotate=270] at (2.3,-0.25) {100};
\end{tikzpicture}
\end{minipage}%
\begin{minipage}{.8\textwidth}
\begin{tikzpicture}
\centering

\tkzKiviatDiagram[radial  style/.style ={-},
        label space=.35,
        space = 0,
        gap = .5,
        step = .5,
        lattice = 5]
        {$S_{P,1}$,$S_{P,2}$,$S_{P,3}$,$S_{P,4}$,$S_{P,5}$,$S_{P,6}$,$S_{P,7}$,$S_{H,1}$,$S_{H,2}$,$S_{H,3}$,$S_{H,4}$,$S_{H,5}$,$S_{H,6}$,$S_{H,7}$,$S_{H,8}$}
\tkzKiviatLine[densely dashed,thick,color=Apricot,line width=1.25](5.4, 5.0, 5.7, 2.5, 5.4, 2.1, 7.7, 5.5, 6.2, 7.9, 6.0, 6.8, 5.9, 6.3, 3.5)
\tkzKiviatLine[densely dotted,thick,color=RedOrange,line width=1.25](5.1, 4.6, 6.0, 5.5, 5.7, 3.5, 7.6, 4.9, 4.3, 6.6, 6.8, 5.5, 6.6, 6.6, 5.4) 
\tkzKiviatLine[thick,color=BrickRed,line width=1.25](4.8, 4.0, 6.6, 6.3, 5.3, 2.8, 8.2, 6.2, 4.3, 7.5, 6.9, 6.8, 6.7, 6.2, 6.2)    
\node at (0,3.5) {\scalebox{1.3}{Shoulder Dataset}};
\node[rotate=270] at (1.2,-0.2) {50};
\node[rotate=270] at (2.3,-0.25) {100};
\end{tikzpicture}
\end{minipage}
\end{adjustbox}
\begin{adjustbox}{width=.75\textwidth}
\begin{tikzpicture}

\draw (-3.5,-3.45) -- (3.5,-3.45) -- (3.5,-4.55) -- (-3.5,-4.55) -- cycle;
\draw[densely dashed, line width=1.25,color=Apricot] (-3.3,-3.75) -- (-2.8,-3.75);
\node[anchor=west] at (-2.8,-3.75) {$\text{CombReg}^{\text{Multi}}_{\text{VGG-UNet}}$};
\draw[densely dotted,line width=1.25,RedOrange] (0.1,-3.75) -- (.6,-3.75);
\node[anchor=east] at (3.4,-3.75) {$\text{CombReg}^{\text{Multi}}_{\text{Den-UNet}}$};
\draw[line width=1.25,BrickRed] (-1.5,-4.25) -- (-1,-4.25);
\node[anchor=west] at (-.95,-4.25) {$\text{CombReg}^{\text{Multi}}_{\text{Res-UNet}}$};
\end{tikzpicture}
\end{adjustbox}
\caption{Spider graphs showing scores obtained within ankle and shoulder datasets based on VGG-UNet \cite{simonyan_very_2015}, Dense-UNet \cite{huang_densely_2017} and Res-Net \cite{he_deep_2016} employed with  multi-class strategy and combined regularization. Scores were computed for pathological $A_{P,1},...,A_{P,7}$ and healthy $A_{H,1},...,A_{H,10}$ ankles, as well as for pathological $S_{P,1},...,S_{P,7}$ and healthy $S_{H,1},...,S_{H,8}$ shoulders.}
\label{fig:spider_pre_trained}
\end{figure}

This study explored various bone segmentation strategies, regularization methods and backbone architectures and provided an insight into how combination of regularizations can improve the bone segmentation quality in a pediatric, sparse, and heterogeneous MRI datasets. We analysed the performance of each multi-structure strategy with fixed combined regularization (Figure \ref{fig:spider_multi_bone_structures}), the impact of each regularization scheme with the multi-class scheme (Figure \ref{fig:spider_regularization}) and the performance of pre-trained models with fixed combined regularization (Figure \ref{fig:spider_pre_trained}) for each MRI dataset.

\begin{figure}[t]

\begin{adjustbox}{width=\textwidth}
\hspace{-1cm}
\begin{minipage}{.8\textwidth}
\begin{tikzpicture}
  \centering
\tkzKiviatDiagram[radial  style/.style ={-},
        label space=.35,
        space = 0,
        gap = .5,  
        step = .5,
        lattice = 5]
        {$A_{P,1}$,$A_{P,2}$,$A_{P,3}$,$A_{P,4}$,$A_{P,5}$,$A_{P,6}$,$A_{P,7}$,$A_{H,1}$,$A_{H,2}$,$A_{H,3}$,$A_{H,4}$,$A_{H,5}$,$A_{H,6}$,$A_{H,7}$,$A_{H,8}$,$A_{H,9}$,$A_{H,10}$}
\tkzKiviatLine[densely dashed,color=YellowGreen,line width=1.25](6.9, 5.7, 7.0, 5.6, 6.1, 2.8, 5.1, 4.5, 3.5, 5.9, 2.6, 4.6, 3.4, 1.5, 4.1, 1.7, 3.9)
\tkzKiviatLine[densely dotted,color=ForestGreen,line width=1.25](4.3, 6.5, 7.3, 5.4, 6.0, 5.7, 5.4, 4.1, 6.4, 6.0, 2.4, 4.6, 6.1, 5.5, 3.8, 4.0, 3.2) 
\tkzKiviatLine[thick,color=BrickRed,line width=1.25](3.1, 6.9, 7.0, 6.0, 7.7, 6.6, 7.1, 6.3, 6.9, 5.3, 2.7, 4.5, 5.8, 6.9, 6.9, 2.6, 3.9)
\node at (0,3.5) {\scalebox{1.3}{Ankle Dataset}};
\node[rotate=270] at (1.2,-0.2) {50};
\node[rotate=270] at (2.3,-0.25) {100};
\end{tikzpicture}
\end{minipage}%
\begin{minipage}{.8\textwidth}
\begin{tikzpicture}
\centering

\tkzKiviatDiagram[radial  style/.style ={-},
        label space=.35,
        space = 0,
        gap = .5,
        step = .5,
        lattice = 5]
        {$S_{P,1}$,$S_{P,2}$,$S_{P,3}$,$S_{P,4}$,$S_{P,5}$,$S_{P,6}$,$S_{P,7}$,$S_{H,1}$,$S_{H,2}$,$S_{H,3}$,$S_{H,4}$,$S_{H,5}$,$S_{H,6}$,$S_{H,7}$,$S_{H,8}$}
\tkzKiviatLine[densely dashed,thick,color=YellowGreen,line width=1.25](3.6, 3.0, 4.8, 1.6, 0.0, 1.8, 7.0, 4.6, 0.3, 3.7, 6.2, 6.4, 2.0, 5.4, 4.8)
\tkzKiviatLine[densely dotted,thick,color=ForestGreen,line width=1.25](3.7, 4.4, 4.7, 3.0, 1.6, 1.6, 5.6, 5.1, 0.5, 6.9, 5.8, 4.4, 2.3, 6.2, 5.7) 
\tkzKiviatLine[thick,color=BrickRed,line width=1.25](4.2, 4.6, 4.9, 2.8, 4.2, 0.3, 7.1, 6.4, 1.0, 7.6, 7.0, 6.7, 4.9, 6.9, 4.5)    
\node at (0,3.5) {\scalebox{1.3}{Shoulder Dataset}};
\node[rotate=270] at (1.2,-0.2) {50};
\node[rotate=270] at (2.3,-0.25) {100};
\end{tikzpicture}
\end{minipage}
\end{adjustbox}
\begin{adjustbox}{width=.75\textwidth}
\begin{tikzpicture}

\draw (-3.5,-3.45) -- (3.5,-3.45) -- (3.5,-4.55) -- (-3.5,-4.55) -- cycle;
\draw[densely dashed, line width=1.25,color=YellowGreen] (-3.3,-3.75) -- (-2.8,-3.75);
\node[anchor=west] at (-2.8,-3.75) {$\text{CombReg}^{\text{Indiv}}_{\text{Att-UNet}}$};
\draw[densely dotted,line width=1.25,ForestGreen] (0.1,-3.75) -- (.6,-3.75);
\node[anchor=east] at (3.4,-3.75) {$\text{CombReg}^{\text{Global}}_{\text{Att-UNet}}$};
\draw[line width=1.25,BrickRed] (-1.5,-4.25) -- (-1,-4.25);
\node[anchor=west] at (-.95,-4.25) {$\text{CombReg}^{\text{Multi}}_{\text{Att-UNet}}$};
\end{tikzpicture}
\end{adjustbox}
\caption{Spider graphs showing scores obtained within ankle and shoulder datasets based on Att-UNet with combined regularization using individual, global and multi strategies. Scores were computed for pathological $A_{P,1},...,A_{P,7}$ and healthy $A_{H,1},...,A_{H,10}$ ankles, as well as for pathological $S_{P,1},...,S_{P,7}$ and healthy $S_{H,1},...,S_{H,8}$ shoulders.}
\label{fig:spider_multi_bone_structures}
\end{figure}
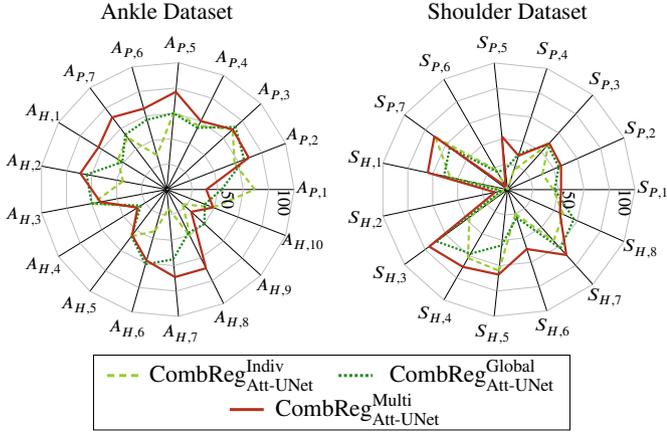{}

{}

From the results obtained on Att-UNet models, we first observed that the multi-class strategy outperformed or at least achieved similar performance compared to individual-class and global-class approaches, on almost all ankle and shoulder examinations (Figure \ref{fig:spider_multi_bone_structures}). However, for two subjects ($A_{P,1}$ and $S_{H,8}$) the multi-class strategy achieved the lowest scores, wherein the extremity of one bone was poorly segmented compared to the other approaches. While, class-wise segmentation provided by the multi and individual strategies yielded bone-specific meshes essential in morphological analysis, global bone tissue masks could also be transformed into class-wise predictions using positional or shape information. However, such post-processing proved difficult to implement in practice due to the fused-bone errors observed in the global scheme (Figure \ref{fig:comparison_bone_structures}). Secondly, our proposed combined regularization outscored or obtained similar scores as the other regularization schemes on almost all ankle and shoulder examinations (Figure \ref{fig:spider_regularization}). However, for $A_{H,4}$ and $S_{P,4}$ subjects, CombReg$^{\text{Multi}}_{\text{Att-UNet}}$ ranked last and produced poor delineations in which the bone extremities were not well segmented either. From these observations, it appeared that bone extremities remained challenging to be managed by Att-UNet models. A possible explanation relies on the fact that compared to 3D or multi-view fusion models for segmentation \cite{milletari_v-net_2016, noori_attention-guided_2019}, our 2D slice by slice approaches do not benefit from 3D spatial information. Although our 2D models do not include 3D contextual information, it is less computationally expensive and requires less GPU memory consumption than 3D approaches.

{}

We reported two outlier examinations $S_{P,6}$ and $S_{H,2}$ for which the Att-UNet models produced poor segmentation results (Figures \ref{fig:spider_multi_bone_structures} and \ref{fig:spider_regularization}). The condition of the patients did not influence the poor segmentation performance, as the two samples were of different types: one pathological $S_{P,6}$ and one healthy $S_{H,2}$. However, both 3D MR images presented a higher level of noise as well as a smaller bone-muscle intensity difference than in the rest of our shoulder dataset. The relatively poor quality of these examinations was due to patient movements during acquisition. Hence, the Att-UNet models did not generalize well on these samples. However, we observed that pre-trained models produced more adequate delineations (mean score of $37.5$) on these outlier examinations (Figure \ref{fig:spider_pre_trained}). More generally, pre-trained models induced better overall performance than Att-UNet models on all 3D MR images, with CombReg$^{\text{Multi}}_{\text{Res-UNet}}$ producing the best results. From these observations, we can assume that pre-training on a large set of non-medical images attenuates the effect of noise on segmentation predictions and imposes more robust and generalizable representations. Specifically, approaches based on transfer learning (i.e. VGG-UNet, Dense-UNet, and Res-UNet) exploit the knowledge (i.e. network's weights) previously gained while solving an image classification problem to provide better initialization for optimization and extract more robust image features that are then used by the decoder to generate segmentation masks. In this study, more complex and deeper architectures with wider convolutional layers (i.e. VGG-UNet), dense modules (i.e. Dense-UNet) and residual blocks (i.e. Res-UNet) allowed for more efficient optimization and enhanced performance compared to standard Att-UNet.

The scores obtained also demonstrated that the performance of the different approaches was not influenced by the pathological or healthy status of patients (Figures \ref{fig:spider_multi_bone_structures}, \ref{fig:spider_regularization} and \ref{fig:spider_pre_trained}). For instance, the proposed CombReg$^{\text{Multi}}_{\text{Res-UNet}}$ achieved mean scores of $74.0$ and $69.6$ on pathological and healthy ankle examinations respectively. As deep models were optimized using a mixture of healthy and impaired joint images, networks were therefore not biased toward any specific population. Moreover, the major differences between pathological and healthy patients was in the shape and relative positioning of the bones rather than in grayscale intensity values. Indeed, both ankle equinus and shoulder OBPP conditions result in osseous deformity and joint malformation \cite{charles_static_2010, hoeksma_shoulder_2003}, while images in each joint dataset were acquired using the same acquisition protocol. For example, the shoulder examination $S_{P,6}$ (Figure \ref{fig:comparison_bone_structures}) exhibited a deformity of the scapular glenoid shape which resulted in an abnormal positioning of the scapula with respect to the humerus bone. It should be emphasized that the difference in bone intensity due to non-ossified areas were present in both healthy and pathological populations (Figure \ref{fig:comparison_regularization}, $A_{P,4}$ and $S_{H,1}$), while the motion noise observed in two outlier examinations ($S_{P,6}$ and $S_{H,2}$) were not related to the joint status but were rather due to patient movement during acquisition. Hence, a unique fully automatic segmentation model could be developed for bone segmentation in pediatric MR images, regardless of the presence of bone deformity due to musculoskeletal disorders. Furthermore, because of its generic nature, our method could be applied to other anatomical joints such as the knee or the hip, as well as on adult imaging datasets. Such generic framework could provide new perspectives for the management of musculoskeletal disorders, by helping to evaluate treatment response and disease progression as well as being integrated into bio-mechanical models for surgery planning.

\subsection{Latent shape space analysis}
\begin{figure*}[t]
\begin{adjustbox}{width=\textwidth}

\begin{tikzpicture}

\node[inner sep=0pt] at (0,0)
    {\includegraphics[width=.25\textwidth]{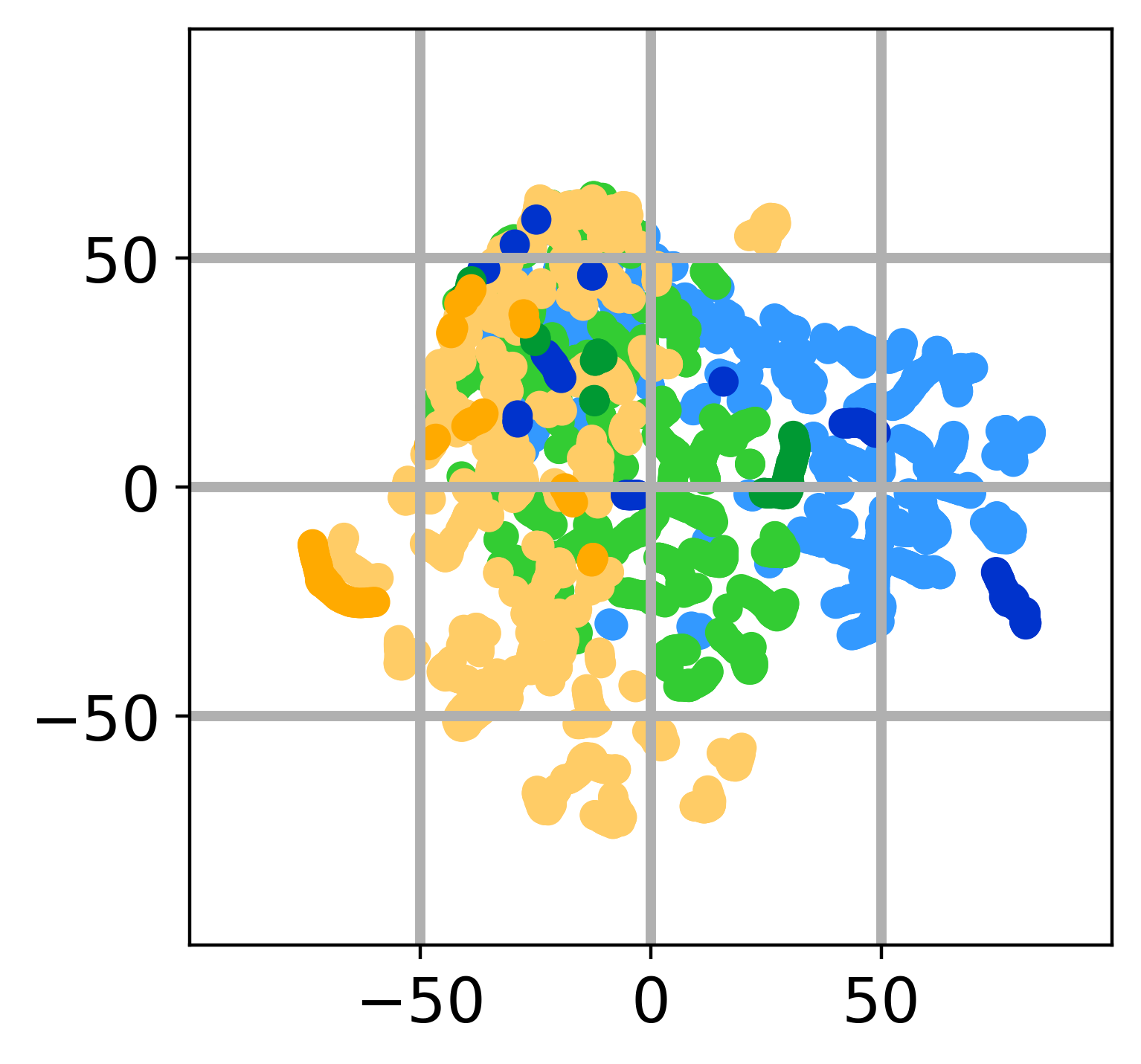}};
\node[inner sep=0pt] at (4.5,0)
    {\includegraphics[width=.25\textwidth]{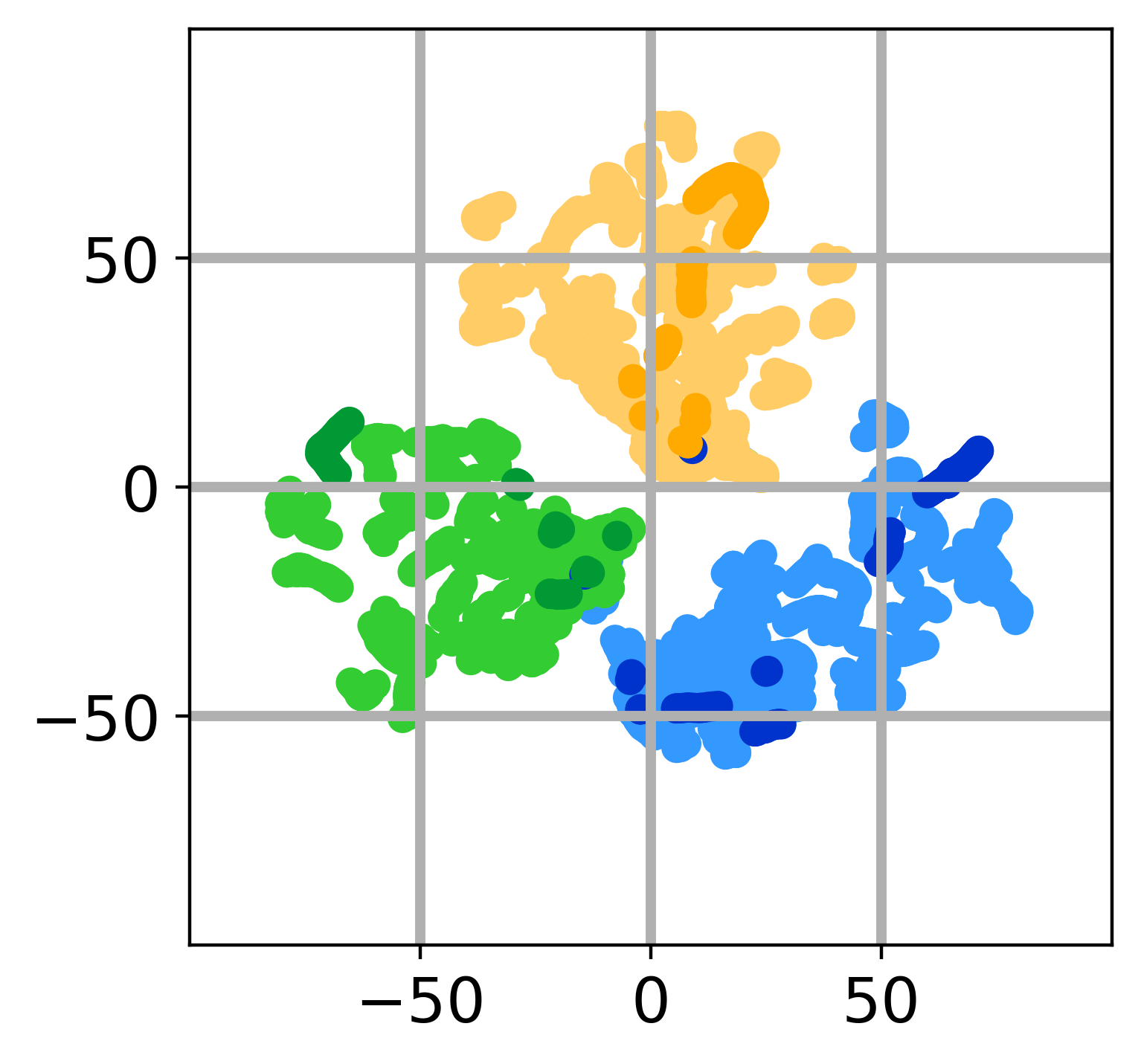}};
    
\node[inner sep=0pt] at (10,0)
    {\includegraphics[width=.25\textwidth]{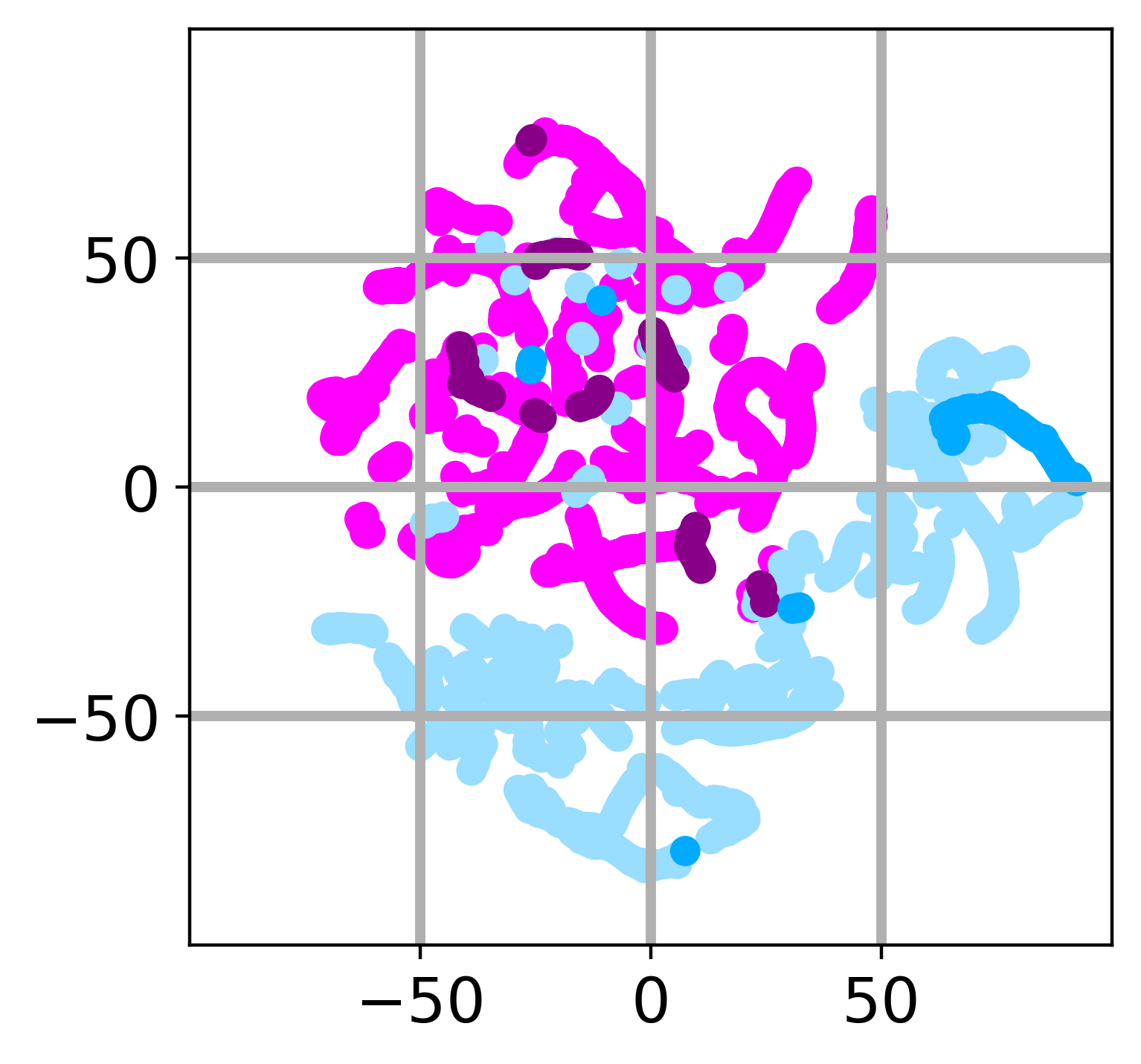}};
\node[inner sep=0pt] at (14.5,0)
    {\includegraphics[width=.25\textwidth]{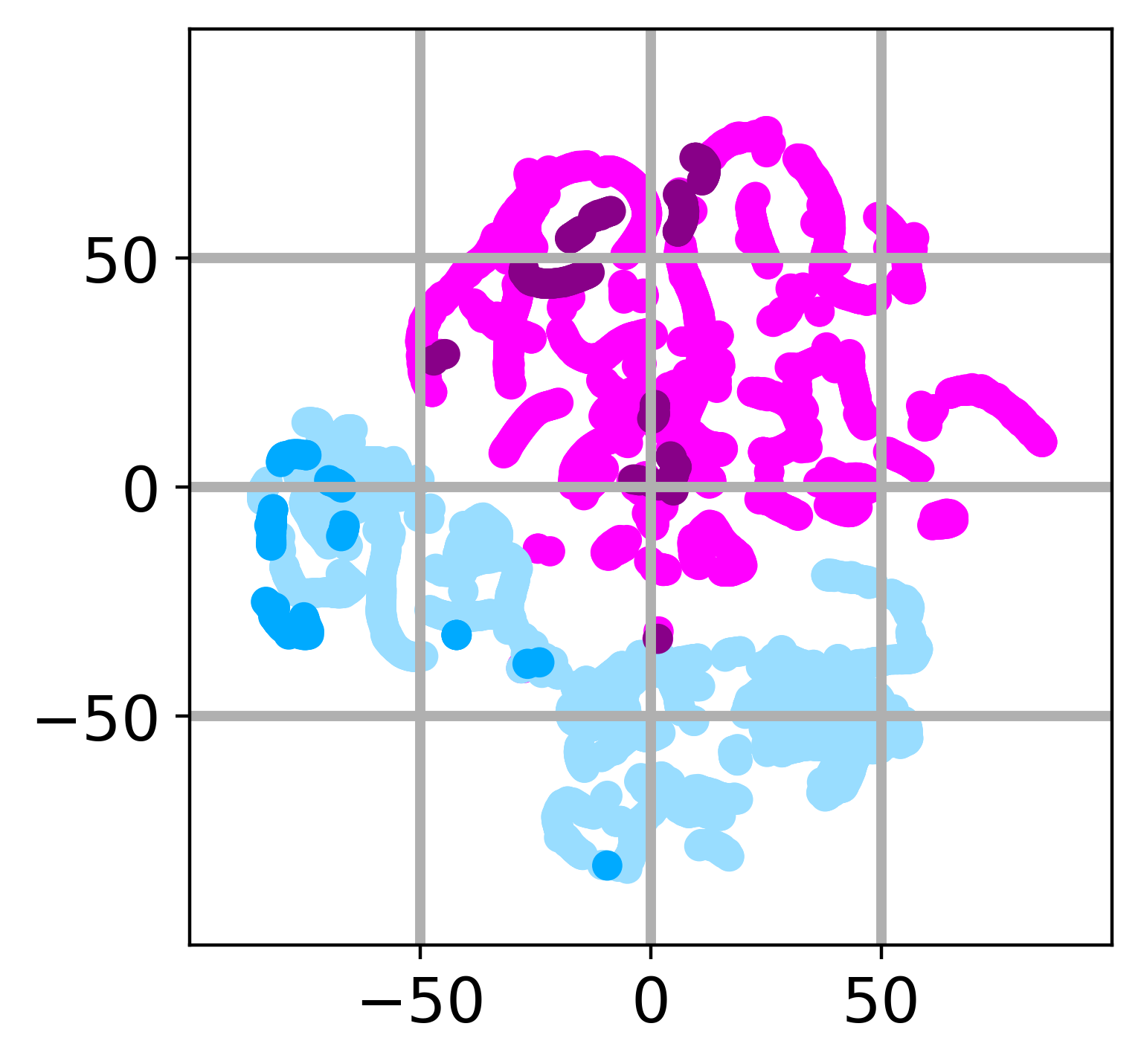}};
    
\node[inner sep=0pt] at (7.55,-3.1)
    {\includegraphics[width=.95\textwidth]{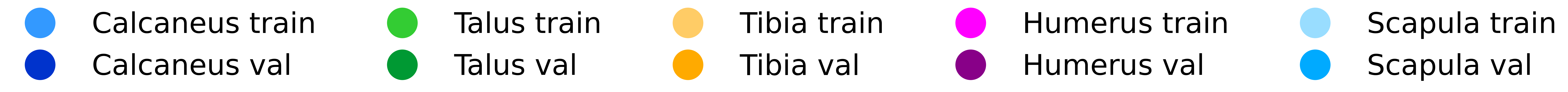}};
    
\node at (2.55,2.25) {\scalebox{1.3}{Ankle}};
\node at (12.55,2.25) {\scalebox{1.3}{Shoulder}};
\node at (0.3,-2.2) {\scalebox{1.2}{Global}};
\node at (4.8,-2.2) {\scalebox{1.2}{Multi}};
\node at (10.3,-2.2) {\scalebox{1.2}{Global}};
\node at (14.8,-2.2) {\scalebox{1.2}{Multi}};

\draw (-.65,-3.6) -- (-.65,-2.6) -- (15.95,-2.6) -- (15.95,-3.6) -- cycle;

\end{tikzpicture}
\end{adjustbox}
    
\caption{Visualization of the latent shape spaces learnt by the global-class and multi-class auto-encoders on ankle and shoulder datasets. The visualization was obtained using the t-SNE algorithm \cite{maaten_visualizing_2008} in which each colored dot corresponds to a 2D binary mask of one of the anatomical objects of interest. The projection included 2D masks originating from the training (train) and validation (val) sets for each joint.}
\label{fig:t-sne_visualization}
\end{figure*}
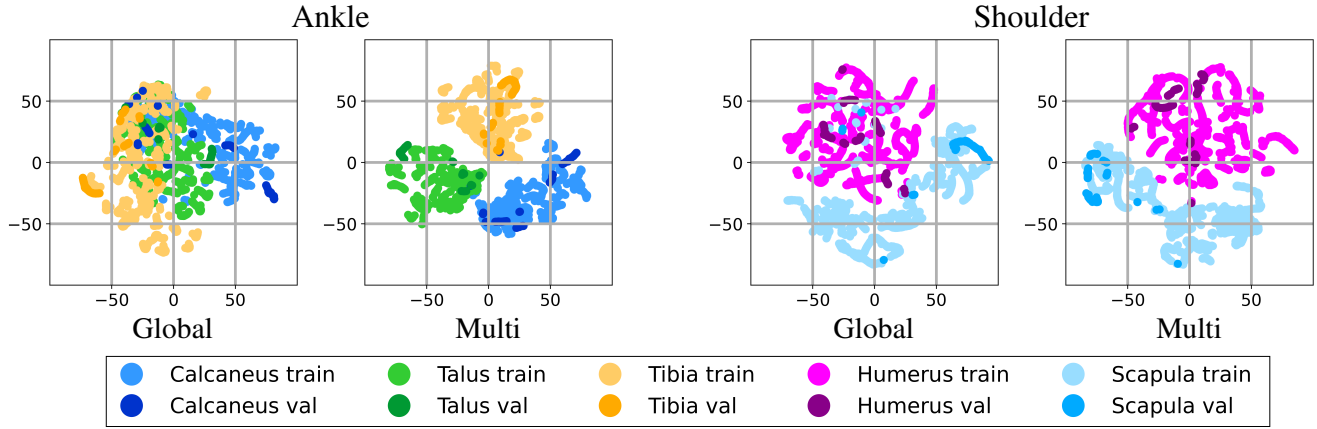{}

While the work of Biffi et al. \cite{biffi_explainable_2020} demonstrated that a deep auto-encoder could learn to differentiate pathological from healthy cardiac shapes, our study focused on the comparison of shape representations arising from two different bone segmentation strategies. The pattern recognition behavior of the deep learning networks can be analyzed by visualizing the compact space learnt during training. We analysed the latent representation learnt by the global-class and multi-class auto-encoders using the t-SNE dimensionality reduction algorithm \cite{maaten_visualizing_2008}. We used the auto-encoders trained on ground truth annotations and employed their encoder components to create latent codes of the bones of the training and validation subjects. We then applied global max pooling and obtained $512$ dimensional codes from 2D bone masks. Finally, in order to visualize the $512$ dimensional feature vectors, we applied a two-step dimensionality reduction as recommended in \cite{maaten_visualizing_2008}. We first employed principal component analysis which reduced the representations to 50 dimensional feature vectors, then the t-SNE algorithm embedded the data into a 2D space (Figure \ref{fig:t-sne_visualization}). The perplexity and learning rate of the t-SNE algorithm were set to 30 and 200 respectively.

For both ankle and shoulder examinations, the latent representation learnt by the global-class auto-encoder did not differentiate anatomical structures, contrary to the shape representation obtained by the multi-class auto-encoder which presented different clusters for each bone (Figure \ref{fig:t-sne_visualization}). Thus, ankle bones were aggregated into a unique cluster in the global-class representation, as opposed to the multi-class one which presented distinctive calcaneus, talus and tibia clusters. The obtained visualizations reinforced our assumption that the global-class auto-encoder imposed the extraction of shared bone features, whereas the multi-class auto-encoder learnt to extract discriminative bone features while complying with inter-bone relationships. 

\subsection{Limited interpretability}

Although incorporating regularization through the loss function successfully constrains the network's parameters and promotes the desired characteristics for robust bone extraction, it fails to provide a better understanding of the inference process. Additionally, as the computation of the proposed regularization losses is based on deep learning models (an auto-encoder and a discriminator), the interpretability of the respective regularization remains also limited. Hence, it would be beneficial to develop more interpretable models (segmentation network, shape encoder and discriminator) in order to better analysis the internal behaviour of the pipeline. More precisely, such interpretable segmentation model is crucial in medical image analysis applications, as it would allow a better analysis of the network failures.

While attention maps computed by attention gates successfully provide a coarse localization of the anatomical structures of interest \cite{oktay_attention_2018}, these approaches fail to explain the representation learnt by the segmentation models. In this direction, the visualization of the learnt feature maps represents the first step toward understanding the internal behaviour of the "black box" type CNN models. An example is the work of Kamnitsas et al. \cite{kamnitsas_efficient_2017} which shows that CNN learns concepts similar to the ones used by clinical expert. However, the learnt convolutional layer can be activated by a mixture of pattern, hence Zhang et al. \cite{zhang_interpretable_2018} have devised an interpretable CNN in which each filter explicitly memorizes a specific object part without ambiguity and provides a clear semantic representation which could be of great interest for computed-aided musculoskeletal system analysis.

\subsection{Perspectives}

\begin{table*}[t!]
\centering
\caption{Quantitative comparison of SPAR \cite{boutillon_multi-structure_2021}, UNet DSL+$\mathcal{L}_{\text{Contrastive}}$ \cite{boutillon_multi-task_2021}, and proposed CombReg$^{\text{Multi}}_{\text{Res-UNet}}$ models on ankle and shoulder datasets. Metrics encompass Dice (\%), sensitivity (\%), specificity (\%), MSSD (mm), ASSD (mm) and RAVD (\%). Bold results correspond to the best performance for each dataset and for each metric. $\dagger$ indicates that reported results based on multi-class masks have been transformed into global-class scores.}
\begin{tabular}{||P{.3cm}||P{4.4cm}||P{1.5cm}|P{1.5cm}|P{1.5cm}|P{1.5cm}|P{1.5cm}|P{1.5cm}||}
    \hline
    & Method & Dice $\uparrow$ & Sens. $\uparrow$ & Spec. $\uparrow$ & MSSD $\downarrow$ & ASSD $\downarrow$ & RAVD $\downarrow$ \\
    \hline
    \hline
    \multirow{3}{*}{\rotatebox[origin=c]{90}{Ankle}} & SPAR $^{\dagger}$ \cite{boutillon_multi-structure_2021} & 92.9$\pm$1.1 & 92.9$\pm$3.3 & 99.7$\pm$0.2 & 7.7$\pm$2.7 & 0.7$\pm$0.1 & 6.2$\pm$4.2 \\\cline{2-8}
    & UNet DSL+$\mathcal{L}_{\text{Contrastive}}$ $^{\dagger}$ \cite{boutillon_multi-task_2021} & 93.7$\pm$1.1 & 92.6$\pm$3.3 & \textbf{99.8$\pm$0.1} & 11.0$\pm$11.4 & 0.7$\pm$0.3 & 5.0$\pm$3.0 \\\cline{2-8}
    & CombReg$^{\text{Multi}}_{\text{Res-UNet}}$ (Proposed) & \textbf{94.3$\pm$1.1} & \textbf{93.6$\pm$3.1} & \textbf{99.8$\pm$0.1} & \textbf{6.1$\pm$2.8} & \textbf{0.6$\pm$0.1} & \textbf{4.7$\pm$2.9} \\
    \hline
    \hline
   
    \multirow{3}{*}{\rotatebox[origin=c]{90}{Shoulder}} & SPAR $^{\dagger}$ \cite{boutillon_multi-structure_2021} & 85.8$\pm$8.2 & 86.2$\pm$10.7 & 99.8$\pm$0.1 & 30.1$\pm$28.5 & 1.7$\pm$2.0 & 9.8$\pm$12.1 \\\cline{2-8}
    & UNet DSL+$\mathcal{L}_{\text{Contrastive}}$ $^{\dagger}$ \cite{boutillon_multi-task_2021} & 86.7$\pm$7.7 & 85.4$\pm$10.5 & 99.8$\pm$0.1 & 23.1$\pm$13.4 & 1.3$\pm$1.0 & 7.5$\pm$8.1 \\\cline{2-8}
    & CombReg$^{\text{Multi}}_{\text{Res-UNet}}$ (Proposed) & \textbf{90.7$\pm$3.0} & \textbf{90.7$\pm$3.6} & \textbf{99.9$\pm$0.1} & \textbf{19.3$\pm$14.2} & \textbf{0.8$\pm$0.5} & \textbf{3.5$\pm$3.4} \\
    \hline
  
\end{tabular}
\label{tab:quantitative_comparison}
\end{table*}

While multiple techniques exist to integrate shape priors into a segmentation neural network, most regularization schemes based on auto-encoders use a distance measure (i.e. either Manhattan or Euclidean) between ground truth and predicted shape codes derived by the shape encoder \cite{myronenko_3d_2019, oktay_anatomically_2018, pham_deep_2019, ravishankar_learning_2017}. However, these metrics may lose their usefulness in high-dimension due to the curse of dimensionality. Hence, recent works have proposed to employ a discriminator operating on the shape representation at the output of the shape encoder which encourages predicted shape codes to be close to ground truth ones \cite{boutillon_multi-structure_2021}. This shape code discriminator allows learning an adaptive distance metric and has demonstrated equivalent or superior performance than Euclidean-based shape priors regularization on both ankle and shoulder joint datasets. Following this, it could be beneficial to combine the typical adversarial loss operating on segmentation masks and an adversarial loss operating on the learnt shape representation, i.e. shape priors based adversarial regularization (SPAR) \cite{boutillon_multi-structure_2021}. Nevertheless, the simultaneous minimization of losses provided by different discriminators as a multi-objective optimization problem remains a challenge \cite{albuquerque_multi-objective_2019}. In this work, we thus employed an Euclidean-based shape priors regularization in combination with the standard adversarial network to leverage both constraints in a more stable optimization setting.

Avoiding over-fitting is one of the key problem in machine learning, especially when considering pediatric datasets whose small sample size may induce limited generalizability in deep learning models. Models with too much capacity (i.e. number of trainable parameters) are one typical cause of over-fitting as they may learn the dataset and task too well. In practice, it is therefore essential to design models with optimal capacity which depends on the task considered and available imaging resources. In this sense, we observed that the employed Res-UNet model (13.6 million parameters) achieved better performance through the leave-one-out evaluation (Table \ref{tab:quantitative_results}) compared to VGG-UNet (34.7 million parameters). The multi-class segmentation strategy also allowed us to reduce the number of trainable parameters while transfer learning aimed at reducing over-fitting by using weights learnt on a large scale non-medical database. Most importantly, we proposed a combined regularization methodology based on shape priors and adversarial network to enhance the generalization capabilities of the model during optimization. It was observed during experiments (Tables \ref{tab:quantitative_results_ablation_study} and \ref{tab:quantitative_results}) that all these techniques led to progressive improvement in segmentation performance on unseen images. Nevertheless, the obtained results still reflect the difficulties of developing generalizable tools without large scale datasets. In the context of pediatric imaging, recent works have proposed to optimize a single network on multiple data\-sets arising from distinct (but related) anatomical regions \cite{boutillon_multi-task_2021}. This multi-task, multi-domain learning scheme allows to learn more generalizable representations by benefiting from a larger and more diverse dataset. Furthermore, the network described in \cite{boutillon_multi-task_2021} incorporated domain-specific layers (DSL) which enables to adapt the statistics of each domain, while a contrastive loss operating on image domains disentangled the learnt representations.

Quantitative comparison with SPAR \cite{boutillon_multi-structure_2021} and UNet DSL +$\mathcal{L}_{\text{Contrastive}}$ \cite{boutillon_multi-task_2021} approaches (Table \ref{tab:quantitative_comparison}) demonstrated the performance benefits of the proposed CombReg$^{\text{Multi}}_{\text{Res-UNet}}$ model, especially on the ankle dataset. However, it should be emphasized that both SPAR and UNet DSL+$\mathcal{L}_{\text{Contrastive}}$ did not integrate pre-trained encoder. It should also be noted that for comparison purposes, the previously reported results based on multi-class segmentation masks have been transformed into global-class scores. These approaches remain complementary and in this direction, it could be advantageous to integrate the proposed combined regularization into a multi-task, multi-domain learning scheme to further reduce over-fitting.

\section{Conclusion}

In this work, we proposed and evaluated a partially pre-trained convolutional encoder-decoder with combined regularization from shape priors and adversarial network which achieved promising performance for the task of multi-structure bone segmentation on scarce heterogeneous pediatric imaging datasets of the musculoskeletal system. The generalization abilities of the segmentation model were enhanced by exploiting shape priors based regularization which enforced globally consistent shape predictions and an adversarial regularization which encouraged precise delineations. In addition, the proposed method exploited specific as well as shared bone features arising from multi-class annotations in order to improve segmentation performance.

The obtained results bring new perspectives for the management of musculoskeletal disorders in pediatric population. Nevertheless, our framework is currently limited to bone tissues, hence in future we aim at improving our model to detect other anatomical structures such as shoulder muscles or ankle cartilages. The severity of the pathologies will then be computed on the basis of a more complete musculoskeletal modeling. 

\section*{Acknowledgment}

\noindent This work was funded by IMT, Fondation Mines-Télécom and Institut Carnot TSN through the Futur \& Ruptures program. Data were acquired with the support of Fondation motrice (2015/7), Fondation de l’Avenir (AP-RM-16-041), PHRC 2015 (POPB 282) and Innoveo (CHRU Brest). We would like to acknowledge Dr. Christelle Pons from Ildys Foundation, Brest, France for her involvement in clinical data collection and patient enrolment process.

\bibliographystyle{elsarticle-num}
\bibliography{refs}

\setcounter{table}{0}
\setcounter{section}{0}
\setcounter{figure}{0}
\setcounter{equation}{0}
\renewcommand{\thetable}{S\arabic{table}}
\renewcommand{\thesection}{S\arabic{section}}
\renewcommand{\thefigure}{S\arabic{figure}}
\renewcommand{\theequation}{S\arabic{equation}}

\begin{center}
\textbf{\large Supplementary Materials}
\end{center}

\section{Statistical analysis}
\label{sec:statistical_analysis}

\begin{table*}[t!]
\centering
\caption{Statistical analysis between the proposed model and the four backbone architectures: Att-UNet \cite{oktay_attention_2018}, VGG-UNet \cite{simonyan_very_2015}, Dense-UNet \cite{huang_densely_2017} and Res-UNet \cite{he_deep_2016} on ankle and shoulder datasets. Regularization methods include: baseline, shape priors based regularization \cite{oktay_anatomically_2018}, adversarial regularization \cite{singh_breast_2020} and the proposed combined regularization; and bone segmentation strategies comprise: individual, global and multi. Statistical analysis performed through Wilcoxon signed-rank non-parametric test using Dice ($\%$), sensitivity ($\%$) and specificity ($\%$) computed on 2D slices. Bold \textit{p}-values ($<0.05$) highlight statistically significant results for each dataset and for each metric.}
 \begin{tabular}{||P{.3cm}||P{.3cm}|P{.95cm}|P{1.7cm}||P{1.5cm}|P{1.5cm}||P{1.5cm}|P{1.5cm}||P{1.5cm}|P{1.5cm}||}
    \hline
    & \multicolumn{3}{c||}{Method} & Dice 2D & \textit{p}-value & Sens. 2D & \textit{p}-value & Spec. 2D & \textit{p}-value \\
    \hline
    \hline
    \multirow{18}{*}{\rotatebox[origin=c]{90}{Ankle Dataset}} & \multirow{12}{*}{\rotatebox[origin=c]{90}{Att-UNet}} & \multirow{4}{*}{Indiv} & Base & $70.9^{+22.1}_{-28.0}$ & \textbf{$<$1$\times$10$^{-6}$} & $72.3^{+22.1}_{-26.3}$ & \textbf{$<$1$\times$10$^{-6}$} & $98.4^{+1.5}_{-1.6}$ & \textbf{$<$1$\times$10$^{-6}$} \\\cline{4-10}
    & & & ShapeReg & $74.3^{+19.6}_{-18.2}$ & \textbf{$<$1$\times$10$^{-6}$} & $75.1^{+21.2}_{-20.3}$ & \textbf{$<$1$\times$10$^{-6}$} & $98.7^{+1.2}_{-1.5}$ & \textbf{$<$1$\times$10$^{-6}$} \\\cline{4-10}
    & & & AdvReg & $74.6^{+18.6}_{-20.6}$ & \textbf{$<$1$\times$10$^{-6}$} & $74.4^{+20.6}_{-21.5}$ & \textbf{$<$1$\times$10$^{-6}$} & $99.0^{+1.0}_{-1.1}$ & \textbf{$<$1$\times$10$^{-6}$} \\\cline{4-10}
    & & & CombReg & $76.6^{+17.9}_{-19.4}$ & \textbf{$<$1$\times$10$^{-6}$} & $77.1^{+18.8}_{-22.3}$ & \textbf{$<$1$\times$10$^{-6}$} & $99.0^{+0.9}_{-0.7}$ & \textbf{$<$1$\times$10$^{-6}$} \\\cline{3-10}
    & & \multirow{4}{*}{Global} & Base & $77.9^{+16.7}_{-16.0}$ &  \textbf{$<$1$\times$10$^{-6}$} & $85.3^{+13.6}_{-8.2}$ & 8.7$\times$10$^{-2}$ & $97.7^{+2.0}_{-1.8}$ & \textbf{$<$1$\times$10$^{-6}$} \\\cline{4-10}
    & & & ShapeReg & $77.3^{+17.1}_{-16.0}$ & \textbf{$<$1$\times$10$^{-6}$} & $82.1^{+16.2}_{-16.1}$ & \textbf{$<$1$\times$10$^{-6}$} & $98.1^{+1.7}_{-1.7}$ & \textbf{$<$1$\times$10$^{-6}$} \\\cline{4-10}
    & & & AdvReg & $77.9^{+15.9}_{-16.1}$ & \textbf{$<$1$\times$10$^{-6}$} & $85.8^{+12.4}_{-8.0}$ & 6.4$\times$10$^{-1}$ & $97.9^{+1.7}_{-1.5}$ & \textbf{$<$1$\times$10$^{-6}$} \\\cline{4-10}
    & & & CombReg & $79.9^{+14.0}_{-10.3}$ & \textbf{$<$1$\times$10$^{-6}$} & $82.3^{+15.4}_{-7.9}$ & \textbf{$<$1$\times$10$^{-6}$} & $98.7^{+1.1}_{-1.2}$ & \textbf{$<$1$\times$10$^{-6}$} \\\cline{3-10}
    & & \multirow{4}{*}{Multi} & Base & $76.2^{+19.3}_{-23.0}$ & \textbf{$<$1$\times$10$^{-6}$} & $76.7^{+19.9}_{-22.8}$ & \textbf{$<$1$\times$10$^{-6}$} & $99.1^{+0.8}_{-0.7}$ & \textbf{$<$1$\times$10$^{-6}$} \\\cline{4-10}
    & & & ShapeReg & $80.5^{+15.4}_{-11.2}$ & \textbf{$<$1$\times$10$^{-6}$} & $81.4^{+16.3}_{-17.2}$ & \textbf{$<$1$\times$10$^{-6}$} & $99.2^{+0.8}_{-0.7}$ &  \textbf{$<$1$\times$10$^{-6}$} \\\cline{4-10}
    & & & AdvReg & $79.0^{+16.3}_{-13.6}$ & \textbf{$<$1$\times$10$^{-6}$} & $79.5^{+18.3}_{-16.6}$ & \textbf{$<$1$\times$10$^{-6}$} & $99.0^{+0.9}_{-1.1}$ & \textbf{$<$1$\times$10$^{-6}$} \\\cline{4-10}
    & & & CombReg & $78.1^{+17.4}_{-15.9}$ & \textbf{$<$1$\times$10$^{-6}$} & $77.0^{+19.7}_{-19.7}$ & \textbf{$<$1$\times$10$^{-6}$} & $99.3^{+0.7}_{-0.6}$ & \textbf{$<$1$\times$10$^{-6}$} \\\cline{2-10}
    & \multirow{2}{*}{\rotatebox[origin=c]{90}{VGG}} & \multirow{2}{*}{Multi} & Base & $81.7^{+14.9}_{-10.7}$ & \textbf{$<$1$\times$10$^{-6}$} & $83.7^{+14.9}_{-11.8}$ &  \textbf{1.9$\times$10$^{-3}$} & $99.2^{+0.7}_{-0.7}$ & \textbf{$<$1$\times$10$^{-6}$} \\\cline{4-10}
    & & & CombReg & $82.9^{+13.6}_{-9.7}$ & \textbf{$<$1$\times$10$^{-6}$} & $84.2^{+14.4}_{-12.3}$ & \textbf{2.0$\times$10$^{-4}$} & $99.2^{+0.7}_{-0.6}$ & \textbf{$<$1$\times$10$^{-6}$} \\\cline{2-10}
    & \multirow{2}{*}{\rotatebox[origin=c]{90}{\small{Dense}}} & \multirow{2}{*}{Multi} & Base & $84.8^{+12.1}_{-7.8}$ & \textbf{$<$1$\times$10$^{-6}$} & $83.7^{+13.8}_{-10.1}$ & \textbf{$<$1$\times$10$^{-6}$} & $\mathbf{99.6^{+0.4}_{-0.3}}$ & \textbf{$<$1$\times$10$^{-6}$} \\\cline{4-10}
    & & & CombReg & $84.9^{+12.0}_{-5.4}$ & \textbf{1.2$\times$10$^{-4}$} & $84.2^{+13.4}_{-7.2}$ & \textbf{$<$1$\times$10$^{-6}$} & $\mathbf{99.6^{+0.4}_{-0.4}}$ & \textbf{$<$1$\times$10$^{-6}$} \\\cline{2-10}
    & \multirow{2}{*}{\rotatebox[origin=c]{90}{Res}} & \multirow{2}{*}{Multi} & Base & $83.9^{+13.1}_{-6.8}$ & \textbf{$<$1$\times$10$^{-6}$} & $83.5^{+14.1}_{-8.2}$ & \textbf{$<$1$\times$10$^{-6}$} & $99.5^{+0.4}_{-0.3}$ & \textbf{$<$1$\times$10$^{-6}$} \\\cline{4-10}
    & & & CombReg & $\mathbf{86.2^{+10.8}_{-6.0}}$ & \raisebox{1.8pt}{\rule{5pt}{1pt}} & $\mathbf{86.2^{+11.5}_{-6.5}}$ & \raisebox{1.8pt}{\rule{5pt}{1pt}} & $99.5^{+0.4}_{-0.3}$ & \raisebox{1.8pt}{\rule{5pt}{1pt}} \\
    \hline
    \hline
   
    \multirow{18}{*}{\rotatebox[origin=c]{90}{Shoulder Dataset}} & \multirow{12}{*}{\rotatebox[origin=c]{90}{Att-UNet}} & \multirow{4}{*}{Indiv} & Base & $82.6^{+13.0}_{-12.2}$ & \textbf{$<$1$\times$10$^{-6}$} & $83.6^{+12.8}_{-9.3}$ & \textbf{$<$1$\times$10$^{-6}$} & $99.8^{+0.2}_{-0.2}$ & \textbf{3.3$\times$10$^{-2}$} \\\cline{4-10}
    & & & ShapeReg & $82.6^{+13.4}_{-8.9}$ & \textbf{$<$1$\times$10$^{-6}$} & $81.4^{+15.2}_{-15.2}$ & \textbf{$<$1$\times$10$^{-6}$} & $\mathbf{99.9^{+0.1}_{-0.1}}$ & \textbf{$<$1$\times$10$^{-6}$} \\\cline{4-10}
    & & & AdvReg & $83.8^{+11.8}_{-7.7}$ & \textbf{$<$1$\times$10$^{-6}$} & $83.1^{+13.9}_{-11.23}$ & \textbf{$<$1$\times$10$^{-6}$} & $99.8^{+0.2}_{-0.1}$ & \textbf{$<$1$\times$10$^{-6}$} \\\cline{4-10}
    & & & CombReg & $83.6^{+11.9}_{-7.9}$ & \textbf{$<$1$\times$10$^{-6}$} & $83.2^{+13.4}_{-10.7}$ & \textbf{$<$1$\times$10$^{-6}$} & $99.8^{+0.2}_{-0.1}$ & \textbf{1.4$\times$10$^{-5}$} \\\cline{3-10}
    & & \multirow{4}{*}{Global} & Base & $82.8^{+13.1}_{-13.0}$ & \textbf{$<$1$\times$10$^{-6}$} & $82.7^{+14.3}_{-14.7}$ & \textbf{$<$1$\times$10$^{-6}$} & $99.8^{+0.2}_{-0.1}$ & \textbf{3.5$\times$10$^{-2}$} \\\cline{4-10}
    & & & ShapeReg & $84.7^{+11.3}_{-6.2}$ & \textbf{$<$1$\times$10$^{-6}$} & $85.1^{+12.6}_{-8.1}$ & \textbf{$<$1$\times$10$^{-6}$} & $99.8^{+0.2}_{-0.2}$ & \textbf{$<$1$\times$10$^{-6}$} \\\cline{4-10}
    & & & AdvReg & $84.6^{+11.3}_{-9.3}$ & \textbf{$<$1$\times$10$^{-6}$} & $85.7^{+11.8}_{-8.1}$ & \textbf{$<$1$\times$10$^{-6}$} & $99.8^{+0.2}_{-0.1}$ & \textbf{$<$1$\times$10$^{-6}$} \\\cline{4-10}
    & & & CombReg & $86.4^{+9.4}_{-8.0}$ & \textbf{$<$1$\times$10$^{-6}$} & $87.2^{+10.1}_{-7.3}$ & \textbf{$<$1$\times$10$^{-6}$} & $99.8^{+0.2}_{-0.1}$ & \textbf{$<$1$\times$10$^{-6}$} \\\cline{3-10}
    & & \multirow{4}{*}{Multi} & Base & $84.8^{+11.5}_{-6.4}$ & \textbf{$<$1$\times$10$^{-6}$} & $84.7^{+13.0}_{-9.2}$ & \textbf{$<$1$\times$10$^{-6}$} & $99.8^{+0.2}_{-0.1}$ & \textbf{$<$1$\times$10$^{-6}$} \\\cline{4-10}
    & & & ShapeReg & $86.5^{+9.7}_{-7.1}$ & \textbf{$<$1$\times$10$^{-6}$} & $85.9^{+11.7}_{-8.7}$ & \textbf{$<$1$\times$10$^{-6}$} & $\mathbf{99.9^{+0.1}_{-0.1}}$ & 5.5$\times$10$^{-1}$ \\\cline{4-10}
    & & & AdvReg & $85.9^{+10.2}_{-8.6}$ & \textbf{$<$1$\times$10$^{-6}$} & $87.6^{+10.2}_{-8.0}$ & \textbf{$<$1$\times$10$^{-6}$} & $99.8^{+0.2}_{-0.2}$ & \textbf{$<$1$\times$10$^{-6}$} \\\cline{4-10}
    & & & CombReg & $87.1^{+9.2}_{-6.3}$ & \textbf{$<$1$\times$10$^{-6}$} & $87.1^{+10.4}_{-7.7}$ & \textbf{$<$1$\times$10$^{-6}$} & $99.8^{+0.2}_{-0.1}$ & \textbf{$<$1$\times$10$^{-6}$} \\\cline{2-10}
    & \multirow{2}{*}{\rotatebox[origin=c]{90}{VGG}} & \multirow{2}{*}{Multi} & Base & $89.1^{+7.0}_{-3.6}$ & \textbf{$<$1$\times$10$^{-6}$} & $91.4^{+6.5}_{-4.0}$ & \textbf{$<$1$\times$10$^{-6}$} & $99.8^{+0.2}_{-0.2}$ & \textbf{$<$1$\times$10$^{-6}$} \\\cline{4-10}
    & & & CombReg & $89.6^{+6.2}_{-4.1}$ & \textbf{$<$1$\times$10$^{-6}$} & $\mathbf{92.3^{+6.1}_{-4.2}}$ & \textbf{$<$1$\times$10$^{-6}$} & $99.8^{+0.2}_{-0.2}$ & \textbf{$<$1$\times$10$^{-6}$} \\\cline{2-10}
    & \multirow{2}{*}{\rotatebox[origin=c]{90}{\small{Dense}}} & \multirow{2}{*}{Multi} & Base & $90.4^{+5.5}_{-4.0}$ & \textbf{$<$1$\times$10$^{-6}$} & $91.1^{+6.0}_{-3.8}$ & \textbf{9.0$\times$10$^{-4}$} & $\mathbf{99.9^{+0.1}_{-0.1}}$ & \textbf{1.4$\times$10$^{-4}$} \\\cline{4-10}
    & & & CombReg & $90.4^{+5.3}_{-3.7}$ & 1.6$\times$10$^{-1}$ & $90.2^{+6.9}_{-4.5}$ & \textbf{1.2$\times$10$^{-5}$} & $\mathbf{99.9^{+0.1}_{-0.1}}$ & \textbf{$<$1$\times$10$^{-6}$} \\\cline{2-10}
    & \multirow{2}{*}{\rotatebox[origin=c]{90}{Res}} & \multirow{2}{*}{Multi} & Base & $89.9^{+5.8}_{-4.0}$ & \textbf{7.3$\times$10$^{-5}$} & $90.5^{+6.8}_{-4.2}$ & 8.5$\times$10$^{-1}$ & $\mathbf{99.9^{+0.1}_{-0.1}}$ & \textbf{2.3$\times$10$^{-2}$} \\\cline{4-10}
    & & & CombReg & $\mathbf{90.5^{+5.4}_{-3.3}}$ & \raisebox{1.8pt}{\rule{5pt}{1pt}} & $90.8^{+6.6}_{-5.1}$ & \raisebox{1.8pt}{\rule{5pt}{1pt}} & $\mathbf{99.9^{+0.1}_{-0.1}}$ & \raisebox{1.8pt}{\rule{5pt}{1pt}} \\\hline
  
\end{tabular}
\label{tab:statistical analysis}
\end{table*}

To perform the statistical analysis, we employed the Wil\-coxon signed-rank non-parametric test using Dice, sensitivity and specificity scores obtained from the 1446 ankle (respectively 3357 shoulder) 2D slices corresponding to 17 ankle (respectively 15 shoulder) 3D MR images. The statistical tests were conducted using only the 2D slices containing at least one of the bone of interest. We preliminary verified the non-normality of the 2D results distributions using D’Agostino and Pearson normality test. We then performed the statistical analysis between methods and the obtained $p$-values are summarized in supplementary Table \ref{tab:statistical analysis}. Due to the skew of the non-normal distributions of 2D scores, we reported their mean and the distances from the mean to the upper and lower bound of the $68\%$ confidence interval, which correspond to the 16 and 84 percentiles (as in \cite{schnider_3d_2020}). For ankle datasets, our proposed model CombReg$^{\text{Multi}}_{\text{Res-UNet}}$ ranked best in 2D Dice ($86.2\%$) and 2D sensitivity ($86.2\%$) metrics while remaining $0.1\%$ lower than the best method in specificity 2D metric. For shoulder datasets, CombReg$^{\text{Multi}}_{\text{Res-UNet}}$ outperformed other approaches in 2D Dice ($90.5\%$) and 2D specificity ($99.9\%$) metrics, and ranked $1.5\%$ lower than the best model in 2D sensitivity metric.

The obtained $p$-values indicated that CombReg$^{\text{Multi}}_{\text{Res-UNet}}$ produced statistically significant different results ($p$-values $< 0.05$), except compared with: Base$^{\text{Global}}_{\text{Att-UNet}}$ and AdvReg$^{\text{Global}}_{\text{Att-UNet}}$ on ankle datasets using sensitivity 2D metrics; as well as CombReg$^{\text{Multi}}_{\text{Dense-UNet}}$, Base$^{\text{Multi}}_{\text{Res-UNet}}$ and ShapeReg$^{\text{Multi}}_{\text{Att-UNet}}$ on shoulder datasets using 2D Dice, 2D sensitivity and 2D specificity metrics respectively. In these particular cases, the difference between results obtained by our model and compared methods was not statistically significant. However, in each case CombReg$^{\text{Multi}}_{\text{Res-UNet}}$ produced statistically significant improvements on the remaining 2D metrics. Hence, we considered the overall improvements achieved by our model to be statistically significant.

\section{Ranking robustness}
\label{sec:ranking_robustness}

\begin{table*}[t!]
\centering
\caption{Transformed rankings of the four backbone architectures: Att-UNet \cite{oktay_attention_2018}, VGG-UNet \cite{simonyan_very_2015}, Dense-UNet \cite{huang_densely_2017} and Res-UNet \cite{he_deep_2016} on ankle and shoulder datasets. Regularization methods include: baseline, shape priors based regularization \cite{oktay_anatomically_2018}, adversarial regularization \cite{singh_breast_2020} and the proposed combined regularization; and bone segmentation strategies comprise: individual, global and multi. Rankings were computed using different threshold values: Dice = 75 or 85$\%$, Sensitivity = 75 or 85$\%$, MSSD = 20 or 40 mm, ASSD = 3 or 5 mm and RAVD = 5 or 15$\%$. Modified ranks are in bold.}
\begin{tabular}{||P{.3cm}||P{.3cm}|P{.95cm}|P{1.7cm}||P{.8cm}|P{.8cm}|P{.8cm}|P{.8cm}|P{.8cm}|P{.8cm}|P{.8cm}|P{.8cm}|P{.8cm}|P{.8cm}||}

    \hline
    & \multicolumn{3}{c||}{\multirow{2}{*}{Method}} & \multicolumn{10}{c||}{Rankings} \\\cline{5-14}
    & \multicolumn{3}{c||}{} & Dice$_{75}$ & Dice$_{85}$ & Sens$_{75}$ & Sens$_{85}$ & \footnotesize{MSSD$_{20}$} & \footnotesize{MSSD$_{40}$} & \small{ASSD$_{3}$} & \small{ASSD$_{5}$} & \small{RAVD$_{5}$} & \footnotesize{RAVD$_{15}$} \\
    \hline
    \hline
    
    \multirow{18}{*}{\rotatebox[origin=c]{90}{Ankle Dataset}} &  \multirow{12}{*}{\rotatebox[origin=c]{90}{Att-UNet}} & \multirow{4}{*}{Indiv} & Base & 18 & 18 & 18 & 18 & 18 & 18 & 18 & 18 & 18 & 18 \\ \cline{4-14}
    & & & ShapeReg & 16 & 16 & 16 & 16 & 16 & 16 & 16 & 16 & 16 & 16 \\\cline{4-14}
    & & & AdvReg & 17 & 17 & 17 & 17 & 17 & 17 & 17 & 17 & 17 & 17 \\\cline{4-14}
    & & & CombReg & 15 & 15 & 15 & 15 & 15 & 15 & 15 & 15 & 15 & 15 \\\cline{3-14}
    & & \multirow{4}{*}{Global} & Base & 14 & 14 & 14 & 14 & 14 & 14 & 14 & 14 & 14 & 14 \\ \cline{4-14}
    & & & ShapeReg & 13 & \textbf{12} & 13 & 13 & \textbf{12} & 13 & \textbf{12} & 13 & \textbf{12} & \textbf{12} \\\cline{4-14}
    & & & AdvReg & 12 & \textbf{13} & 12 & 12 & \textbf{13} & \textbf{11} & \textbf{13} & 12 & \textbf{11} & \textbf{13} \\\cline{4-14}
    & & & CombReg & 9 & 9 & 9 & 9 & 9 & 9 & 9 & 9 & 9 & 9 \\\cline{3-14}
    & & \multirow{4}{*}{Multi} & Base & 11 & 11 & 11 & 11 & 11 & \textbf{12} & 11 & 11 & \textbf{13} & 11 \\\cline{4-14}
    & & & ShapeReg & 8 & 8 & 8 & 8 & 8 & 8 & 8 & 8 & 8 & 8 \\\cline{4-14}
    & & & AdvReg & 10 & 10 & 10 & 10 & 10 & 10 & 10 & 10 & 10 & 10 \\\cline{4-14}
    & & & CombReg & 7 & 7 & 7 & 7 & 7 & 7 & 7 & 7 & 7 & 7 \\\cline{2-14}
    & \multirow{2}{*}{\rotatebox[origin=c]{90}{VGG}} & \multirow{2}{*}{Multi} & Base & 6 & 6 & 6 & 6 & 6 & 6 & 6 & 6 & 6 & 6 \\ \cline{4-14}
    & & & CombReg & 4 & 4 & 4 & 4 & 4 & 4 & 4 & 4 & 4 & 4 \\\cline{2-14}
    & \multirow{2}{*}{\rotatebox[origin=c]{90}{\small{Dense}}} & \multirow{2}{*}{Multi} & Base & 5 & 5 & 5 & 5 & 5 & 5 & 5 & 5 & 5 & 5 \\ \cline{4-14}
    & & & CombReg & 3 & 3 & 3 & 3 & 3 & 3 & 3 & 3 & 3 & 3 \\\cline{2-14}
    & \multirow{2}{*}{\rotatebox[origin=c]{90}{Res}} & \multirow{2}{*}{Multi} & Base & 2 & 2 & 2 & 2 & 2 & 2 & 2 & 2 & 2 & 2 \\ \cline{4-14}
    & & & CombReg & 1 & 1 & 1 & 1 & 1 & 1 & 1 & 1 & 1 & 1 \\
    
    \hline
    \hline
   
   \multirow{18}{*}{\rotatebox[origin=c]{90}{Shoulder Dataset}} &  \multirow{12}{*}{\rotatebox[origin=c]{90}{Att-UNet}} & \multirow{4}{*}{Indiv} & Base & 18 & 18 & 18 & 18 & 18 & 18 & 18 & 18 & 18 & 18 \\ \cline{4-14}
    & & & ShapeReg & 15 & 15 & 15 & 15 & 15 & \textbf{16} & 15 & 15 & \textbf{16} & 15 \\\cline{4-14}
    & & & AdvReg & 16 & 16 & 16 & 16 & 16 & \textbf{15} & 16 & 16 & \textbf{15} & 16 \\\cline{4-14}
    & & & CombReg & \textbf{13} & 14 & \textbf{13} & \textbf{13} & \textbf{12} & 14 & 14 & \textbf{12} & 14 & 14 \\\cline{3-14}
    & & \multirow{4}{*}{Global} & Base & 17 & 17 & 17 & 17 & 17 & 17 & 17 & 17 & 17 & 17 \\ \cline{4-14}
    & & & ShapeReg & 12 & 12 & \textbf{14} & 12 & \textbf{14} & \textbf{13} & \textbf{13} & \textbf{14} & 12 & \textbf{13} \\\cline{4-14}
    & & & AdvReg & \textbf{14} & 13 & \textbf{12} & \textbf{14} & 13 & \textbf{12} & \textbf{12} & 13 & 13 & \textbf{12} \\\cline{4-14}
    & & & CombReg & \textbf{10} & 11 & 11 & 11 & 11 & 11 & 11 & 11 & \textbf{10} & 11 \\\cline{3-14}
    & & \multirow{4}{*}{Multi} & Base & \textbf{11} & 10 & 10 & 10 & 10 & 10 & 10 & 10 & \textbf{11} & 10 \\\cline{4-14}
    & & & ShapeReg & \textbf{8} & 9 & \textbf{8} & 9 & 9 & 9 & 9 & 9 & \textbf{8} & \textbf{8} \\\cline{4-14}
    & & & AdvReg & \textbf{9} & 8 & \textbf{9} & 8 & 8 & 8 & 8 & 8 & \textbf{9} & \textbf{9} \\\cline{4-14}
    & & & CombReg & 7 & 7 & 7 & 7 & 7 & 7 & 7 & 7 & 7 & 7 \\\cline{2-14}
    & \multirow{2}{*}{\rotatebox[origin=c]{90}{VGG}} & \multirow{2}{*}{Multi} & Base & 6 & 6 & 6 & 6 & 6 & 6 & 6 & 6 & 6 & 6 \\ \cline{4-14}
    & & & CombReg & 5 & 5 & 5 & 5 & 5 & 5 & 5 & 5 & 5 & 5 \\\cline{2-14}
    & \multirow{2}{*}{\rotatebox[origin=c]{90}{\small{Dense}}} & \multirow{2}{*}{Multi} & Base & 4 & 4 & 4 & 4 & 4 & 4 & 4 & 4 & 4 & 4 \\ \cline{4-14}
    & & & CombReg & 3 & 3 & 3 & 3 & 3 & 3 & 3 & 3 & 3 & \textbf{2} \\\cline{2-14}
    & \multirow{2}{*}{\rotatebox[origin=c]{90}{Res}} & \multirow{2}{*}{Multi} & Base & 2 & 2 & 2 & 2 & 2 & 2 & 2 & 2 & 2 & \textbf{3} \\ \cline{4-14}
    & & & CombReg & 1 & 1 & 1 & 1 & 1 & 1 & 1 & 1 & 1 & 1 \\
    \hline
 
\end{tabular}
\label{tab:ranking_robustness}
\end{table*}

To assess the robustness of our ranking, we tested different threshold values for each metric: Dice ($75-85\%$), sensitivity ($75-85\%$), MSSD ($20-40$ mm), ASSD ($3-5$ mm) and RAVD ($5-15\%$). Thresholds were modified independently. Metric value between the corresponding best value and the modified threshold was mapped to the interval $[0,100]$. Supplementary Table \ref{tab:ranking_robustness} summarizes the obtained transformed rankings with modified ranks in bold. For instance, Dice threshold modification to $85\%$ (Dice$_{85}$) led to a permutation of ShapeReg$_{\text{Att-UNet}}^{\text{Global}}$ and AdvReg$_{\text{Att-UNet}}^{\text{Global}}$ ranks ($12^{th}$ and $13^{th}$) on ankle dataset. More importantly, CombReg$_{\text{Res-UNet}}^{\text{Multi}}$ ranked first on both datasets, whatever the selected threshold values.

\section{Metrics definition}
\label{sec:metrics_definition}

Let $GT$ and $P$ be the ground truth and predicted 3D segmentation masks and let $S_{GT}$ and $S_P$ be the surface voxels of the corresponding sets. The metrics were defined as follows:
\begin{align}
    &\textnormal{Dice} = \dfrac{2 \vert GT \cdot P\vert}{\vert GT \vert + \vert P \vert}\\
    &\textnormal{Sensitivity} = \dfrac{\vert GT \cdot P \vert}{\vert GT \vert}\\
    &\textnormal{Specificity} = \dfrac{\vert \overline{GT} \cdot \overline{P} \vert}{\vert \overline{GT} \vert}\\
    &\textnormal{MSSD} = \max ( h(S_{GT}, S_P), h(S_P,S_{GT}) ) \\
    &\textnormal{with} \enspace h(S,S') = \max_{s \in S} \min_{s' \in S'} \norm{s-s'}_2\\
    &\textnormal{ASSD} = \dfrac{1}{\vert S_{GT} \vert + \vert S_P \vert} (\sum_{s \in S_{GT}} d(s,S_P) \nonumber\\
    & \hspace{8.9em} + \sum_{s \in S_{P}} d(s,S_{GT}) ) \\
    &\textnormal{with} \enspace d(s,S') = \min_{s' \in S'} \norm{s-s'}_2\\
    &\textnormal{RAVD} = \dfrac{\left| \vert GT \vert - \vert P \vert \right|}{\vert GT \vert}
\end{align}
\noindent Distance measures were transformed to millimeters using voxel size information extracted from DICOM metadata.

\end{document}